\documentclass[]{jfm} % lineno to number lines

\usepackage{graphicx}
\usepackage{newtxtext}
\usepackage{newtxmath}
\usepackage{natbib}
\usepackage{hyperref}
\hypersetup{
    colorlinks = true,
    linkcolor = blue,
    urlcolor   = blue,
    citecolor  = black,
}

\newcommand{\RomanNumeralCaps}[1]

\usepackage{lineno}
\usepackage{subcaption}

\providecommand\bnabla{\boldsymbol{\nabla}}

\providecommand\btau{\boldsymbol{\tau}}
\providecommand\bcdot{\boldsymbol{\cdot}}
\newcommand\bS{\boldsymbol{S}}

\newcommand{\Reynd}{Re} 

\newcommand{\deltazero}{\delta_0^*}

\newcommand{\tento}[1]{10^{#1}}

\newcommand{\StL}{St_{L}}
\newcommand{\Lsep}{L_{\mathrm{sep}}}

\newcommand{\Jac}{\B{\cal{J}}}
\newcommand{\Reso}{\B{\mathcal{R}}}

\newcommand{\qoline}{\B{\oline{q}}}
\newcommand{\qprime}{\B{q'}}

\newcommand{\qhat}{\B{\what{q}}}

\newcommand{\fhat}{\B{\what{f}}}

\newcommand {\what}   [1] { \widehat{#1} }
\newcommand {\oline}  [1] { \overline{#1} }
\newcommand{\B}       [1] {{\bf {#1}}}

\newcommand {\derp}   [2] { \partial #1 / \partial #2  }

\newcommand {\Derp}   [2] { \frac { \partial #1 } { \partial #2 } }

\newcommand{\DPS}{\displaystyle}
\newcommand{\Cal}       [1] {{\cal{#1}}}

\usepackage{xcolor}

\shortauthor{B. Bugeat et al.}

\title[Resolvent analysis of the OSWBLI]{Low-frequency resolvent analysis of the laminar oblique shock wave / boundary layer interaction}

\author{B. Bugeat\aff{1} 
  \corresp{\email{b.bugeat@tudelft.nl}},
 J.-Ch. Robinet\aff{2}
 J.-C. Chassaing\aff{3}
\and P. Sagaut\aff{4}
}

\affiliation{\aff{1}Process and Energy Department, Delft University of Technology, Leeghwaterstraat 39, 2628 CB Delft, The Netherlands
\aff{3}Sorbonne Universit\'e, CNRS, Institut Jean Le Rond d'Alembert, UMR 7190, F-75005 Paris, France
\aff{2}DynFluid Lab. - Arts \& M\'etiers Paris - 151, Bd. de l'H\^opital - 75013 - Paris - France
\aff{4}Aix Marseille Univ., CNRS, Centrale Marseille, M2P2 UMR 7340, 13451 Marseille, France
}

\begin{document}

\maketitle

\begin{abstract}

Resolvent analysis is used to study the low-frequency behaviour of the laminar oblique shock wave / boundary layer interaction (SWBLI).
It is shown that the computed optimal gain, which can be seen as a transfer function of the system, follows a first-order low-pass filter equation, recovering the results of \cite{TS2011}.
This behaviour is understood as proceeding from the excitation of a single stable, steady global mode whose damping rate sets the time scale of the filter.
Different Mach and Reynolds numbers are studied, covering different recirculation lengths $L$.
This damping rate is found to scale as $1/L$, leading to a constant Strouhal number $\StL$ as observed in the literature.
It is associated with a breathing motion of the recirculation bubble.
This analysis furthermore supports the idea that the low-frequency dynamics of the SWBLI is a forced dynamics, in which background perturbations continuously excite the flow.
The investigation is then carried out for 3D perturbations for which two regimes are identified.
At low wave numbers of the order of $L$, a modal mechanism similar to that of 2D perturbations is found and exhibits larger values of the optimal gain.
At larger wave numbers of the order of the boundary layer thickness, the growth of streaks, which results from a non-modal mechanism, is detected.
No interaction with the recirculation region is observed.
Based on these results, the potential prevalence of 3D effects in the low-frequency dynamics of the SWBLI is discussed.

\end{abstract}


%%%%%%%%%%%%%%%%%%%%%%%%%%%%%%%%%%%%%%%%%%%%%%%%%%%%%%%%%%%%%%%%%%%%%%%%%%%%%%%%%%%%%%%%%%%%%%%%%%%%%%%%%%%%%%%%%%%%%%%%%%%%%%%%
%%%%%%%%%%%%%%%%%%%%%%%%%%%%%%%%%%%%%%%%%%%%%%%%%%%%%%%%%%%%%%%%%%%%%%%%%%%%%%%%%%%%%%%%%%%%%%%%%%%%%%%%%%%%%%%%%%%%%%%%%%%%%%%%
\section{Introduction}

% Context
An accurate understanding of the dynamics of the shock wave / boundary layer interaction (SWBLI) is needed in many aerospace and aeronautical applications to predict, for example, flows around transonic airfoils, supersonic air intakes, or deflected control surfaces of vehicles at transonic or supersonic speed. 
These interactions can lead to an increase of drag, to separation and to loss of performances. 
Moreover, shock wave / turbulent boundary layer interaction (SWTBLI) generally produces a low-frequency unsteadiness of the shock system. 
This unsteadiness can for example modify the thermal load or induce fatigue of the structure.
It is well known that the flow separation may be at the origin of this unsteadiness in any regime, from the incompressible \citep{WMTS2015,MTW2016,FWMTD2020} to the hypersonic \citep{G2015,PM2021}, taking a particular form for compressible flows when the separation is induced by a shock wave.
A considerable amount of work has been carried out to investigate the steady and unsteady aspects (see \cite{G2015} for a review). 
Experimental research of the SWBLI started in the mid-1940s with the work of \citet{AFR47} and has shown a continuing interest since then.
At this time, most of the experiments only measure steady aerodynamic quantities such as pressure distribution, skin friction, and heat transfer rates. 
Detailed investigations of the phenomenon and its dependence on flow and boundary layer parameters have been already published \citet{AM1980,DM86,S88}. 

% General presentation
It was only at the end of the 90s that the study of the dynamics of a SWTBLI was more systematically undertaken \citep{D01,KYPZ03}. 
It has been shown that the dynamics is rich, spanning over several frequency decades. 
Two main length scales are at play: the vorticity thickness of the shear layer coming from the separated region and linked to the disturbances developing along the shear layer and, on a more global scale, the separation length $L$ that is related to lower frequencies and is associated with the dynamics of the separation bubble \citep{D2012,G2015}. 
Three main frequency ranges can be identified.
First, high frequencies, characterised by Strouhal numbers based on $L$ typically in the range of $St_L \geq 1 $, are associated with the transitional or turbulent boundary layer dynamics \citep{BBGBMSJ2019}.
Secondly, a vortex shedding resulting from the shear layer dynamics usually takes place at mid-frequency ($ 0.1 \leq St_L \leq 1 $).
This can be observed either in incompressible recirculation bubble \citep{WMTS2015,MTW2016} or in the SWBLI \citep{TPC1994,ALDDD2012}.
Finally, a third dynamics is observed in the low-frequency range ($\StL < 0.1$) \citep{clemens2014low}.
This is the frequency domain we will mainly focus on from now on.

%basse freq
The low-frequency range has been intensely studied since the early 2000s in order to unveil its physical origin \citep{dussauge2006unsteadiness}. 
\citet{DHAD05,DHD06} have shown that, for an interaction strong enough, low-frequency unsteadiness is observed around $St_{L} \simeq 0.03$ and is characterised by oscillations of the separated shock \citep{dussauge2006unsteadiness,HSO2009}.
Scenarios attempting to model this low-frequency dynamics have been devised. 
The first of them is related to the interaction of large structures upstream of the boundary layer with the shock leading to low-frequency response of the shock \citep{GCD2007}.  However, \citet{DP2008} have shown that the influence of downstream conditions, especially in the recirculation zone, is more significant than the upstream conditions with respect to low-frequencies when the interaction is strong.
\citet{PDDD09} proposed a model based on a mass balance of the system "shear layer - separated zone" where coherent structures in the shear layer feed the recirculation zone which increases up to a critical size beyond which it empties, causing a breathing of the bubble and consequently the movement of separation of the shock \citep{ALD2015}.
\citet{TS2011} furthermore characterised the system using a combination of theoretical and numerical model.
They showed that the separated shock foot acts as a low-pass filter with respect to white noise. 
This result is in agreement with the linear interaction analysis \citep{R53,RC2001} of the shock response to a harmonic infinitesimal perturbation.

%3D
The three-dimensional dynamics resulting from the development of 3D disturbances, independently of aspects strictly related to the effects of turbulence, has been studied more recently. 
Mainly three types of phenomena have been identified. 
When the upstream boundary layer is turbulent and the SWBLI weak (incipient separated zone), streaks from the upstream boundary layer and associated with longitudinal vortices can interact with the shock system \citep{GCD2007,GCD2009,ROLP2021}. 
However, when the SWBLI is strong enough to generate a significant separated zone, the latter can induce a streamline curvature of the boundary layer and initiate a centrifugal G\"{o}rtler-like instability \citep{PHA2017,ZTLSZ2018}. 
When the SWBLI is sufficiently strong, the separated zone becomes three-dimensional and the emergence of a global non-oscillating instability can be observed. 
This characteristic is very general and can be observed from subsonic \citep{THD00,RT2010,RGS2021} to supersonic regime \citep{R07,HDNJC2018}. 
The role that these 3D effects may play in the low-frequency unsteadiness remains however unclear.

%Num
From a numerical point of view, a substantial body of work has been  done over the years, including direct numerical simulations and large eddy simulations.
\citet{TS2009a,TS2009b,PB2011}'s works made it possible for the first time to characterise the dynamics with reasonable statistical convergence, showing a very good agreement with IUSTI's experiments \citep{DHD06,DPSD2008}. 
In particular, these numerical studies also showed the broadband nature of the low-frequency dynamics.
\citet{PWM2009} for a reflected shock and \citet{WM2008} for a compression ramp detailed the low-frequency dynamics and recovered the scaling proposed in the literature.
They also observed a large amplification of the turbulence through the SWTBLI.
\citet{PM2012} for the compression ramp have studied the physical mechanism that drives the shock motion. 
In their simulations, the flow undergoes low-frequency changes of topology in the interaction region, including the breaking-up of the recirculation bubble and the shedding of vortical structures. 
In addition, the growth of energetic turbulent structures in the shear layer was found to be modulated at low frequency.
This could imply a modulation of the shear layer entrainment rate which is consistent with the scenario of \citet{PDDD09}. 
These authors thus suggest that the low frequency dynamics is related to the dynamics of the separated zone. 
\citet{AGR2013} and \citet{PM2012} then showed that the dynamics of the separated zone occurs at medium frequency and is linked to Kelvin-Helmholtz instabilities in the shear layer and that the low-frequency dynamics is only a modulation of the latter and corresponds to the breathing of the separated zone.

% Resolvent
Resolvent analysis, or resolvent-based modelling, has been intensively used in the fluid dynamics community in the recent years.
This approach is based on the singular value decomposition (SVD) of the resolvent operator that provides an optimal orthonormal basis for an external forcing (input) of the system and its associated linear response (output).
The optimality is here defined as the way for a given energy input, at a given frequency, to force the system in order to trigger the largest possible energy growth.
This is therefore particularly suited to study convective instabilities in noise amplifier flows.
In such flows, ranges of frequency of external perturbations are amplified in space and time through linear mechanisms, while others are damped \citep{huerre1990local}.
Resolvent analysis was successfully applied, for example, to boundary layer \citep{aakervik2008global,sipp2013characterization} and jet \citep{NL2011,GLSH2013} flows.
Because non-modal effects are taken into account \citep{schmid2007nonmodal}, algebraic instabilities like streaks or non-modal mechanisms such as the Orr mechanism can furthermore be modelled via this approach \citep{monokrousos2010global}.
Resolvent analysis has also been intensively used in the modelling of turbulent flow since \cite{mckeon2010critical} noticed it could be used to identify coherent, large scale structures.
While turbulence is a highly non-linear phenomenon, the (linear) resolvent analysis yet captures the growth of these structures.
The interpretation, and consequently the relevance of the resolvent analysis, is that non-linear forcing terms, whose knowledge is not ultimately required here, provide background perturbations that grow into energetic coherent structures through linear mechanisms.
Several authors later discussed the potential of the resolvent approach in reduced-order modelling of turbulent flows \citep{semeraro2016stochastic}, noticing, in particular, the link between resolvent and spectral-POD modes \citep{towne2018spectral}.

% SWBLI and resolvent, objective paper
Resolvent analysis has been recently used to study the dynamics of the SWBLI. 
\citep{sartor2015unsteadiness} used a turbulent mean flow of a transonic SWBLI over a bump and compared it to their experimental results.
They obtained similar features at low and medium frequencies, supporting the idea of a forced dynamics as their mean flow was found globally stable.
\citep{BBGBMSJ2019} studied a transitional case, in which they identified three ranges of frequencies and proposed a scenario for the low-frequency dynamics.
In our paper, we focus exclusively on the low-frequency dynamics of the SWBLI, that we analysed by means of the resolvent analysis.
Unlike the two papers cited above, a laminar base flow will be considered, for which similar features of the low-frequency dynamics can be observed \citep{YS02,boin20063d,SSH2014,SSWWS2014,SSH2016}.
Our objective is first to understand to what extent a resolvent-based approach can model the behaviour of the SWBLI for this range of frequency, where the dynamics does not result from convective or global instabilities.
This will also provide insights on the physics of the system that will be discussed.
In particular, by conducting our analysis for different Mach and Reynolds numbers, we aim at characterising the scales involved in this dynamics.
Besides, despite considering a laminar flow, this analysis could, to some extent, support the idea that the turbulent SWBLI has a forced dynamics at low frequency. 
Indeed, if the intrinsic forcing emanating from non-linear interactions in the turbulent case reasonably projects onto the optimal forcing found in resolvent analysis, it is then possible to assume that a similar response will be found.
All of the above will be carried out considering 2D perturbations only, similarly to the two previously cited papers.
The second objective is to explore the resolvent analysis of the SWBLI for 3D perturbations for which, to our best knowledge, no work has been published.
By 3D perturbations, it is meant that Fourier modes will be considered in the homogeneous spanwise direction, characterised by their wave number $\beta$.
The aim is to investigate the role that streaks or any 3D effects may play in the low-frequency dynamics of the system.

The paper is structured as follows.
The governing equations and the theoretical and numerical approaches are described in $\S$ \ref{sec.theory}.
The methodology to generate the set of base flows understudied is then detailed in $\S$ \ref{sec.baseflowSet}.
In section $\S$ \ref{sec.2Dperturbations}, the resolvent analysis is carried out for 2D perturbations on different base flows.
A link with global stability analysis will be proposed, leading to a model of the optimal and sub-optimal gain obtained from the SVD of the resolvent operator.
The scaling of key quantities will be shown, before discussing these results with respect to the literature.
In section $\S$ \ref{sec.3Dperturbations}, 3D perturbations will be considered in the resolvent analysis. 
Only one base flow will be studied, for which the newly obtained features of the dynamics will be classified into two regimes depending on their wave numbers.
This section ends with a discussion on the 3D effects.
A general conclusion can be found in $\S$ \ref{sec.conclusion}.

%%%%%%%%%%%%%%%%%%%%%%%%%%%%%%%%%%%%%%%%%%%%%%%%%%%%%%%%%%%%%%%%%%%%%%%%%%%%%%%%%%%%%%%%%%%%%%%%%%%%%%%%%%%%%%%%%%%%%%%%%%%%%%%%

\section{Theoretical and numerical approach}\label{sec.theory}

\subsection{Governing equations}

The compressible non-linear Navier-Stokes equations are considered for the conservative variables $\B{q}=\left( \rho, \rho \B{u},\rho E \right)^T$, where $\rho$, $\B{u}$ and $E$ are the density, the velocity vector and the specific total energy, respectively. 
The components of the velocity vector are noted $u$, $v$ and $w$ in the streamwise direction $x$, the wall-normal direction $y$ and the spanwise direction $z$, respectively.
The variables $T$, $p$, $\mu$, and $\kappa$, standing for temperature, pressure, dynamic and kinematic viscosity and thermal conductivity, are also introduced. 
All variables are non-dimensional using their value at infinity, except the pressure that is normalised by $\rho_\infty u_\infty^2$. 
The reference length scale and time scale are $\deltazero$ and $\deltazero/u_\infty$, where $\deltazero$ is the compressible displacement thickness at a chosen streamwise location $x$, defined as $\deltazero = \int_0^\infty \{ 1 - \rho u/ (\rho_\infty u_\infty) \} \mathrm{d}y$.
More details on the choice of $\deltazero$ for our problem will be given in section \ref{sec.baseflow}.
The governing equations read 

\begin{subequations} 
  \begin{equation} \label{eq.mass}
   \Derp{\rho}{t}+\bnabla\bcdot\left(\rho \B{u}\right)=0,
  \end{equation}    
  \begin{equation} \label{eq.momentum}
  \Derp{}{t}\left(\rho \B{u}\right) + \bnabla\bcdot\left[\rho \B{u}\otimes\B{u} + p\B{I} -\frac{1}{Re}\btau\right]=0,
  \end{equation}      
  \begin{equation}\label{eq.energy}
   \Derp{}{t}\left(\rho E\right)+ \bnabla\bcdot\left[\left(\rho E+p\right)\B{u} - \frac{1}{Re}\btau\odot\B{u} -\frac{\kappa}{Pr Re (\gamma-1) M^2}\bnabla T\right]=0
  \end{equation}\label{eq.CNS}      
\end{subequations}

\noindent The fluid is assumed to be a perfect gas whose equation of states and total energy read
\begin{equation} \label{eq.state}
    p = \frac{1}{\gamma M_\infty^2} \rho T \>\>,\>\> E = \frac{p}{\rho (\gamma-1)} + \frac{1}{2}\B  {u}\bcdot\B{u} 
\end{equation}    

\noindent Furthermore assuming the fluid to be Newtonian, the viscous stress tensor verifies
\begin{equation}
\btau = 2\mu\left[\bS-\frac{1}{3}\mbox{Tr}\left(\bS\right)\B{I}\right],\>\>\mbox{with}\>\> \bS = \frac{1}{2}\left[\bnabla\otimes\B{u}+\left(\bnabla\otimes\B{u}\right)^T\right]
\end{equation}  

\noindent Sutherland's law is used for the dynamic viscosity
\begin{equation} \label{eq.sutherland}
\mu(T) = T^{3/2}\frac{1+T_s/T_\infty}{T+T_s/T_\infty},
\end{equation}  

\noindent where $T_s=110.4\>\mbox{K}$ and $T_\infty=288K$.
The thermal conductivity $\kappa$ is also assumed to follow Sutherland's law \citep{toro2013riemann}.

Three non-dimensional numbers have been introduced that control the flow: $Re$, $M$ and $Pr$, the Reynolds, Mach and Prandtl numbers respectively, defined as 
\begin{equation}
  Re = \frac{\rho_\infty u_\infty \deltazero}{\mu_\infty} \>,\> M= \frac{u_\infty}{c_\infty} \>,\> Pr = \frac{\mu_\infty c_p}{\kappa_\infty}
\end{equation}

\noindent where $c_\infty$ is the speed of sound at infinity and $c_p$ the heat capacity.
While $Pr$ is set to $Pr=0.72$ in order to model the fluid as air, the influence of $Re$ and $M$ will be studied.    
Note that a geometrical parameter $\phi$ will also be introduced. 
It is defined as the angle between the streamwise direction and the shock wave impinging the boundary layer.

Finally, equations \eqref{eq.CNS} will be recast in the dynamical system form as
  \begin{equation} \label{eq.DS}
    \DPS\Derp{\B{q}}{t} = \Cal{N}(\B{q})
  \end{equation}

\noindent where $\Cal{N}$ is the non-linear differential operator of the Navier-Stokes equations.

\subsection{Global stability}

In this paper, the shock wave boundary layer interaction is studied via two types of linear analysis.
The linearised Navier-Stokes equations, considered in their semi-discrete form, are written as 
  \begin{align}\label{eq.linearNS}
  \Derp{\qprime}{t} = \Jac  \qprime %+ \fprime
  \end{align}

\noindent where $\Jac$ and $\qprime$ are the Jacobian matrix and the state-vector perturbations around the base flow $\qoline$, respectively.
The base flow $\qoline(x,y)\textbf{}$ corresponds to a steady 2D laminar solution of the non-linear equations \eqref{eq.CNS}.
A Fourier transform in both time and space (in the spanwise direction $z$) can then be performed, allowing the perturbations to be studied as the following global modes
  \begin{align} 
      \qprime(x,y,z,t) &= \qhat(x,y) e^{i (\beta z-\omega t)} + c.c. 
  \end{align} 

\noindent with $\omega$ the angular frequency and $\beta$  and the spanwise wave number.
Considering $\omega \in \mathbb{C}$ and $\beta \in \mathbb{R}$, the global stability \citep{T2011} of the system can be studied by solving the eigenvalue problem
  \begin{align}\label{eq.evpGlobal}
  -i \omega \qhat = \Jac  \qhat %+ \fprime
  \end{align}

\noindent If an eigenvalue is found such that the imaginary part of $\omega$ is positive, then the flow is globally unstable.
In this paper, only globally stable flows will be considered: if not externally sustained, any linear perturbations will eventually decay.

\subsection{Resolvent analysis}

Even though globally stable, flows such as boundary layer or mixing layer can exhibit convective instabilities \citep{huerre1990local}.
In such flows, termed as noise amplifier flows, studying the linear forced dynamics is relevant to understand and model their behaviour.
This can be achieved using resolvent analysis \citep{sipp2010dynamics} which has been shown to provide a good model for convective instabilities over ranges of frequency where large energy growths take place \citep{sipp2013characterization}. 
Besides, this approach gives the location and structure of the optimal harmonic forcing of the flow, i.e. the forcing field that triggers the largest energy growth.
More generally, resolvent analysis provides a transfer function of the system, characterising the response to a forcing for each frequency \citep{schmid2007nonmodal}.

The linear forced dynamics can be studied by introducing a harmonic forcing term $\fhat$ in equation \eqref{eq.evpGlobal} as
  \begin{align}\label{eq.linearForcedNS}
  -i \omega \qhat  = \Jac  \qhat + \fhat
  \end{align}

\noindent with $\omega \in \mathbb{R}$ which now represents the forcing frequency that one wants to study. 
Introducing the resolvent matrix $\Reso = \left( -\omega \mathcal{I} -  \Jac \right)^{-1}$, the relationship between any forcing and its linear response is then given by
  \begin{align}\label{eq.resolvent}
    \qhat  = \Reso \fhat
  \end{align}

\noindent One can now look for the forcing field that triggers the most energetic response by computing the optimal gain $\sigma$ defined as
  \begin{align}\label{eq.gain}
    \sigma^2(\omega, \beta)  =  \max\limits_{\fhat} \frac{|| \qhat ||_E^2}{|| \fhat ||_F^2}
  \end{align}

\noindent Here, two norms must be introduced to measure the energy of the forcing and the response fields.
The norm introduced by \cite{chu1965energy} is used for the response while the $L_2$-norm is used for the forcing. 
More details can be found in \cite{bugeat20193d}.
It can be shown that the optimal gain $\sigma$ corresponds to the largest singular value of the resolvent matrix modified by the norm matrices \citep{sipp2013characterization}.
The associated optimal forcing and response are then the first left and right singular vectors, respectively.
From equation \eqref{eq.gain}, $\sigma$ can be seen as a transfer function as it measures the energy of the response with respect to that of an input forcing.
In the following, the largest (or first) singular value will be called the optimal gain and will be referred to as $\sigma$ or $\sigma_1$ without any differences.
The subsequent singular values $\sigma_2, \sigma_3, ...$ such that $\sigma_1 >\sigma_2 > \sigma_3 > ...$  are called sub-optimal gains and are associated with the subsequent singular vectors which form an orthonormal basis.

\subsection{Numerical methods}

  \subsubsection{Base flow computation}

\begin{figure}
    \centering
      \includegraphics[angle=-0,trim=0 0 0 0, clip,width=0.8\textwidth]{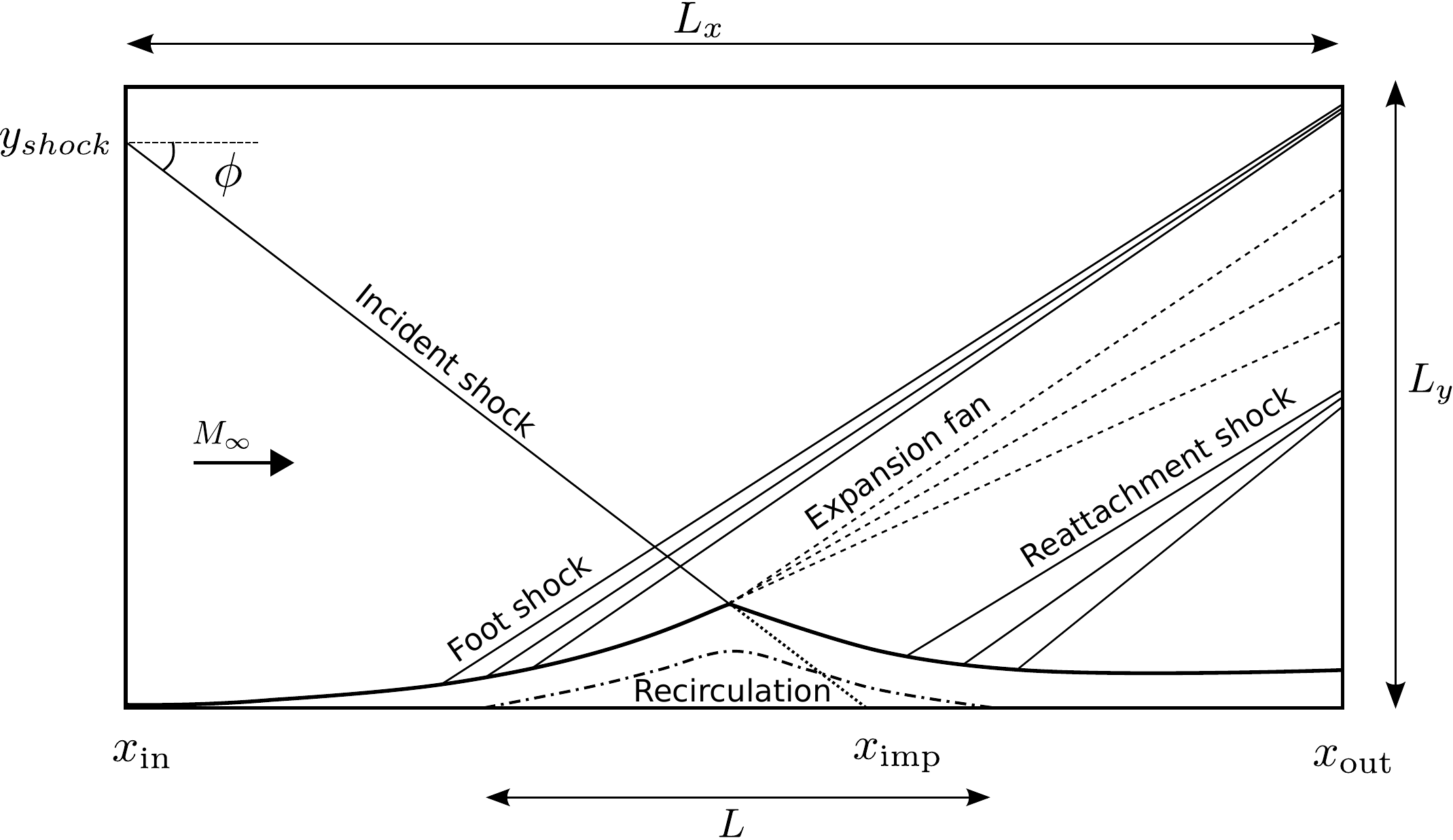}
    \caption{Sketch of the numerical domain and typical structure of the OSWBLI}
     \label{fig.swbli_scheme}
  \end{figure}

In order to get a 2D steady base flow, the non-linear Navier-Stokes equations \eqref{eq.CNS} are solved using a finite volume CFD solver.
Spatial discretisation of convective fluxes is performed using AUSM+ scheme \citep{liou1996sequel} associated with a fifth-order MUSCL extrapolation \citep{boin20063d}. 
Viscous fluxes at cell interfaces are obtained by a second-order centered finite difference scheme. 
The unsteady equations are marched in time until a steady state is reached. 
An implicit dual time stepping method with local time step is used \citep{jameson1991time}. 
More details about this solver and its validation can be found in \citep{boin20063d}.
A rectangular numerical domain is defined as in figure \ref{fig.swbli_scheme}.
The streamwise length $L_x$ is set such that the distance between the impinging shock $x_\mathrm{imp}$ and the outlet of the domain $x_\mathrm{out}$ is $70 \deltazero$.
The height $L_y$ is set such that the reflecting shock leaves the domain at the outlet boundary.
This value depends on the physical case considered.
In any case, the baseflow eventually used in the stability and resolvent calculations is cropped to $L_y=50 \deltazero$. 
A Cartesian mesh is set with a geometrical progression from the wall. 
The number of points is $600 \times 270$ in the $x$ and $y$ directions, respectively.
The independence of the results from the aforementioned numerical parameters is checked in appendix \ref{sec.mesh}.
The following boundary conditions are used.
At the inlet, oblique shock conditions are imposed: for $y<y_{\mathrm{shock}}$, a parallel flow is set using the Mach number at infinity while the corresponding downstream jump conditions are used for $y>y_{\mathrm{shock}}$.
The latter condition is also set at top boundary of the domain.
Note that $y_{\mathrm{shock}}$ is calculated after choosing the angle of the shock $\phi$ and the location of the impinging shock $x_{\mathrm{imp}}$. 
At the outlet, variables are extrapolated from the inside of the domain.
At the bottom, an adiabatic flat plate is considered with no-slip conditions imposed on the velocity.

  \subsubsection{Linear solvers}

The global stability and resolvent analysis are based on the linearised Navier-Stokes equations, which revolves around the computation of the Jacobian matrix $\Jac$.
In order to compute it, the discretised Navier-Stokes equation are linearised via a finite-difference approximation \citep{mettot2014computation}.
This is achieved by using the discrete residual $\mathcal{N}$ (equation \eqref{eq.DS}) of the non-linear Navier-Stokes used in the base flow solver, ensuring numerical consistency between the base flow calculation and the linear solvers.
The adaptation of this framework proposed by \cite{bugeat20193d} for 3D perturbations in base flows featuring one homogeneous direction is used.
Exhaustive details of the resolvent calculation can also be found in \cite{bugeat20193d} where, in particular, the boundary conditions, Chu's norm matrix for conservative variables and the Krylov subspace algorithm used to compute the singular values are fully described.
A sponge zone of length $10 \deltazero$ is also implemented at the top and downstream boundaries of the domain using the approach described by \cite{agarwal2004calculation}.
Finally, PETSc \citep{petsc-user-ref}, SLEPc \citep{Hernandez2005SSF} and MUMPS \citep{amestoy2001fully} open libraries are called to compute the large scale linear systems and eigenvalues problems featuring sparse matrices.  

%%%%%%%%%%%%%%%%%%%%%%%%%%%%%%%%%%%%%%%%%%%%%%%%%%%%%%%%%%%%%%%%%%%%%%%%%%%%%%%%%%%%%%%%%%%%%%%%%%%%%%%%%%%%%%%%%%%%%%%%%%%%%%%%
%%%%%%%%%%%%%%%%%%%%%%%%%%%%%%%%%%%%%%%%%%%%%%%%%%%%%%%%%%%%%%%%%%%%%%%%%%%%%%%%%%%%%%%%%%%%%%%%%%%%%%%%%%%%%%%%%%%%%%%%%%%%%%%%

\section{Base flow} \label{sec.baseflowSet}

\subsection{Validation}

\begin{figure}
  \centering
  \begin{subfigure}[]{0.49\textwidth}  
    \includegraphics[angle=-90,trim=0 0 0 0, clip,width=0.99\textwidth]{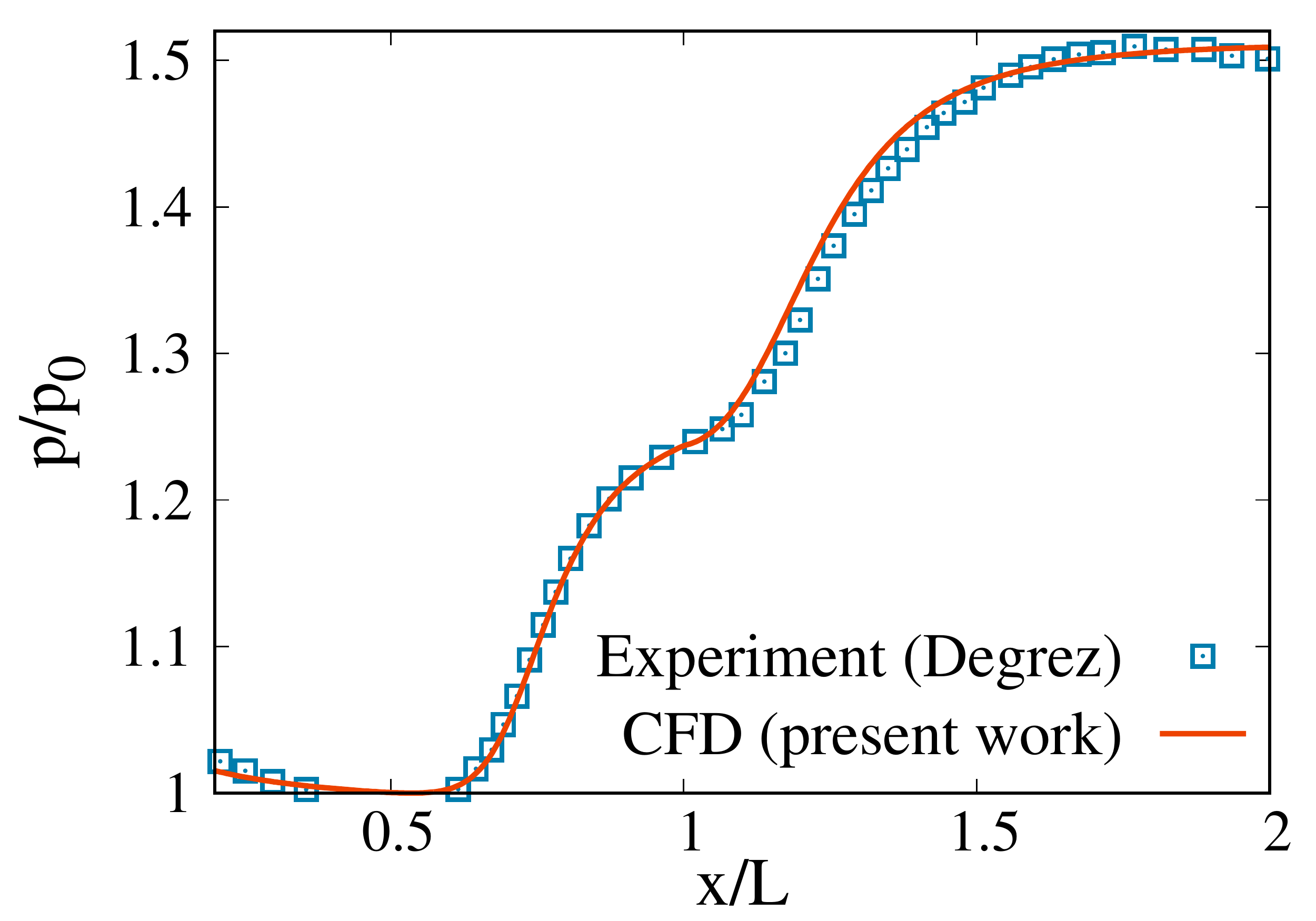}      
  \end{subfigure}
   \begin{subfigure}[]{0.49\textwidth}  
    \includegraphics[angle=-90,trim=0 0 0 0, clip,width=0.99\textwidth]{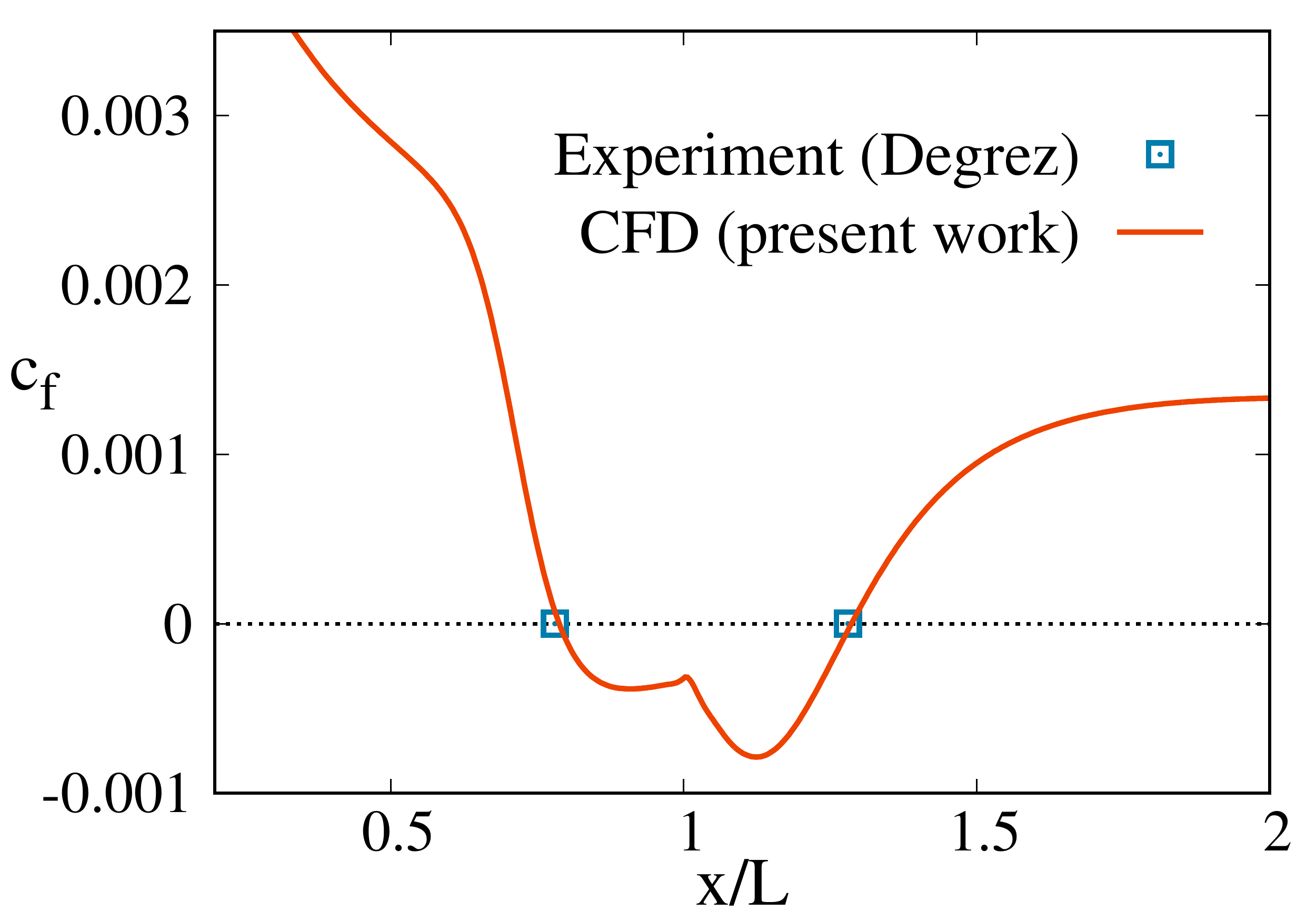}      
  \end{subfigure}
  \caption{Validation against the experimental data from \cite{DBW1987}. Left: wall pressure normalised by the minimum pressure $p_0$ upstream from the shock. Right: Skin friction coefficient $c_f$. The blue squares correspond to the experimental separation and reattachment points.}
   \label{fig.valideDegrez}
\end{figure}

In order to validate the base flow calculation, the flow conditions are matched to those of the experimental investigation of \cite{DBW1987}. 
The freestream Mach number is $M = 2.15$ and the angle of the incident shock measured from the horizontal axis is $\phi = 30.8^\circ$ (corresponding to a flow deflection
angle of $3.81^\circ$). 
The Reynolds number, based on the streamwise length from the leading edge at which the shock impinges the flat plate, is $Re_x = 0.96 \times 10^5$.
Comparisons with the experimental results reported by \cite{DBW1987} are displayed in figure \ref{fig.valideDegrez}. 
This shows that our simulation is in good agreement with experimental data, both in terms of wall pressure measurements and location of the separation bubble.  

\subsection{Set of base flows studied} \label{sec.baseflow}

\begin{figure}
      \centering
      \includegraphics[angle=-0,trim=0 0 0 0, clip,width=0.67\textwidth]{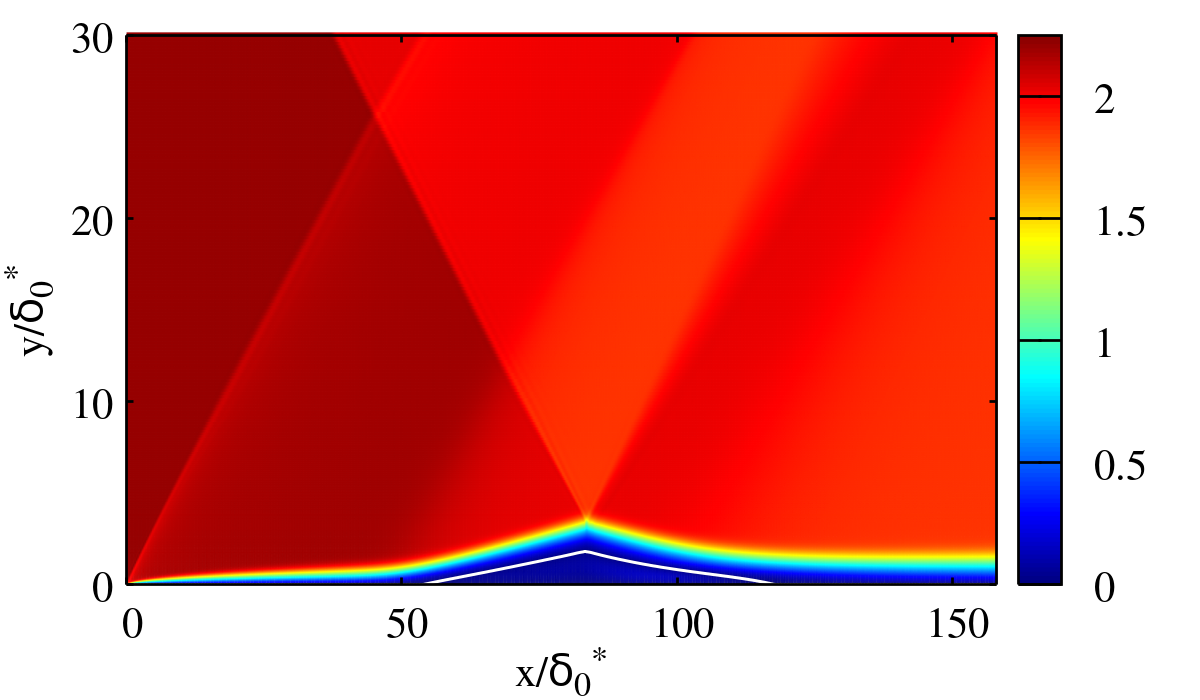}
      \caption{Local Mach number of the base flow at $M=2.20$, $Re=1100$.}
   \label{fig.baseflowCFD}
  \end{figure}

A set of base flows is computed in order to carry out stability and resolvent analysis for different separation lengths $L$.
Different Mach numbers are considered, ranging from $2.00$ (incipient separation) to $2.35$ (large separation length compared to the boundary layer length scale).
First, the Reynolds number $\Reynd$, based on the compressible boundary layer thickness $\delta^*_0$ at which the shock impinges the plate, is kept constant at $\Reynd=1100$ using the following protocol.
For each Mach number, a boundary layer base flow is first computed \textit{without} the impinging shock.
At each $x$-location, $\delta^*$ is then calculated by integrating the quantity $1 - (\rho u) / (\rho_\infty u_\infty)$ in the normal direction.
This allows us to detect the $x$-location where $\Reynd=1100$ and to use it as an input in a second base flow computation $\textit{with}$ a shock impinging at this point.
Thus, the reference length scale $\delta^*_0$ corresponds to the boundary layer thickness at the location of the impinging shock but \textit{prior to} the shock interaction.
In other words, it is not the effective thickness resulting from the shock interaction.
The velocity field obtained at $M=2.20$ and $\Reynd=1100$, which will be used as the reference case, is shown in figure \ref{fig.baseflowCFD}.
By increasing the Mach number, the pressure gradient felt by the boundary layer is increased.
The separation length $L/\deltazero$ is readily computed by detecting the separation and reattachment points along the plate, which are the locations where $\derp{u}{y}$ at the wall becomes negative and positive again, respectively.
The separation length $L/\deltazero$ is found very sensitive to $M$ as its value increases over more than one order of magnitude for the Mach numbers considered in table \ref{tab.baseflows}.
In order to assess the role of viscous effects in the low-frequency receptivity of the SWBLI, different Reynolds numbers are also considered, ranging from 600 to 2200.
For one given Mach number, this corresponds to different $x$-locations at which the shock impinges.
Note that because $\delta^* \sim \sqrt{x}$, this location varies over a wide range of values, e.g. by a factor 10 at $M=2.20$ for which the maximum value of $Re$ considered is 1900.
As shown in table \ref{tab.baseflows}, increasing $\Reynd$ leads to increasing the recirculation bubble as there is less streamwise momentum near the wall to resist the pressure gradient generated by the shock.
Finally, note that the angle of the shock is kept constant at $\phi = 30.8^{\circ}$ for all the computations presented in this paper.

  \begin{table}
    \begin{center}
    \def~{\hphantom{0}}
    \begin{tabular}{ccc}
        $M$  & $Re$ & $L/\deltazero$  \\[3pt]
        $2.00$ & $1100$ & $0.00$\\
        $2.05$ & $1100$ & $2.83$\\
        $2.10$ & $1100$ & $22.0$\\
        $2.15$ & $1100$ & $43.8$\\
        $2.20$ & $1100$ & $63.8$\\
        $2.25$ & $1100$ & $78.2$\\
        $2.30$ & $1100$ & $88.4$\\
        $2.35$ & $1100$ & $94.5$\\
        $2.10$ & $1600$ & $45.4$\\
        $2.10$ & $2200$ & $77.6$\\
        $2.20$ & $600$  & $23.1$\\
        $2.20$ & $850$  & $42.7$\\
        $2.20$ & $1300$ & $80.3$\\
        $2.20$ & $1600$ & $104$\\
        $2.20$ & $1900$ & $128$\\
    \end{tabular}
    \caption{Set of base flows studied with the length of the separation region $L$ obtained in each case.}
    \label{tab.baseflows}
    \end{center}
  \end{table}

%%%%%%%%%%%%%%%%%%%%%%%%%%%%%%%%%%%%%%%%%%%%%%%%%%%%%%%%%%%%%%%%%%%%%%%%%%%%%%%%%%%%%%%%%%%%%%%%%%%%%%%%%%%%%%%%%%%%%%%%%%%%%%%%
%%%%%%%%%%%%%%%%%%%%%%%%%%%%%%%%%%%%%%%%%%%%%%%%%%%%%%%%%%%%%%%%%%%%%%%%%%%%%%%%%%%%%%%%%%%%%%%%%%%%%%%%%%%%%%%%%%%%%%%%%%%%%%%%

\section{Linear dynamics of 2D perturbation at low frequency} \label{sec.2Dperturbations}

    \subsection{Resolvent analysis: optimal forcing and response}

        \subsubsection{Results} \label{sec.resolvent_opt}

\begin{figure}
            \centering
            \includegraphics[angle=-90,trim=0 0 0 0, clip,width=0.625\textwidth]{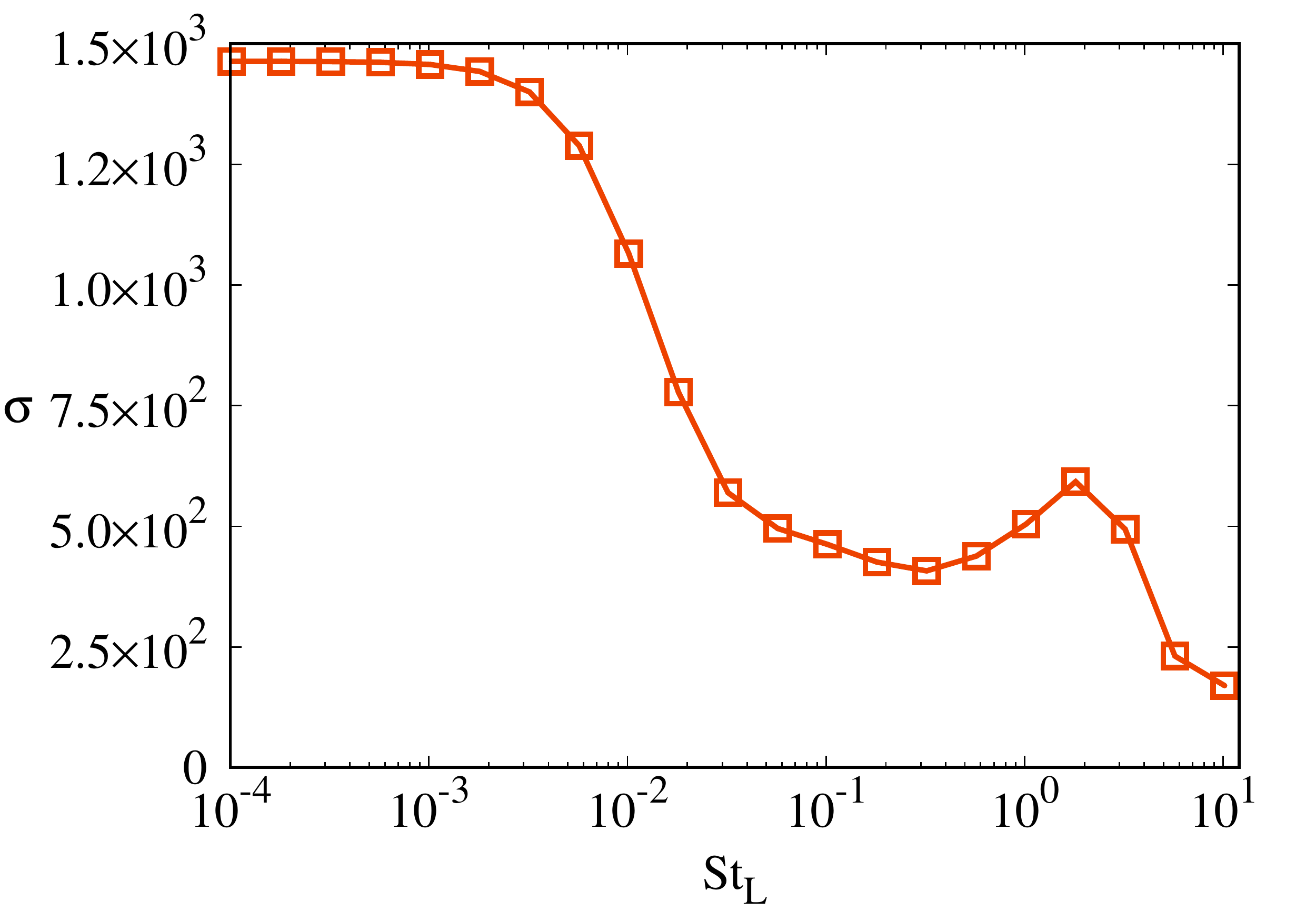}
            % \vspace*{0.3cm}
            \caption{Optimal gain at $M=2.20$, $Re=1100$.}
             \label{fig.optimalGainM220}
        \end{figure}

\begin{figure}
            \centering
            \includegraphics[angle=-0,trim=0 0 0 0, clip,width=0.28\textwidth]{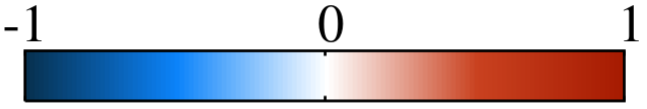}      
            \vspace{0.15cm}
            
            \begin{subfigure}[]{0.49\textwidth}  
                \includegraphics[angle=-0,trim=0 0 0 0, clip,width=0.99\textwidth]{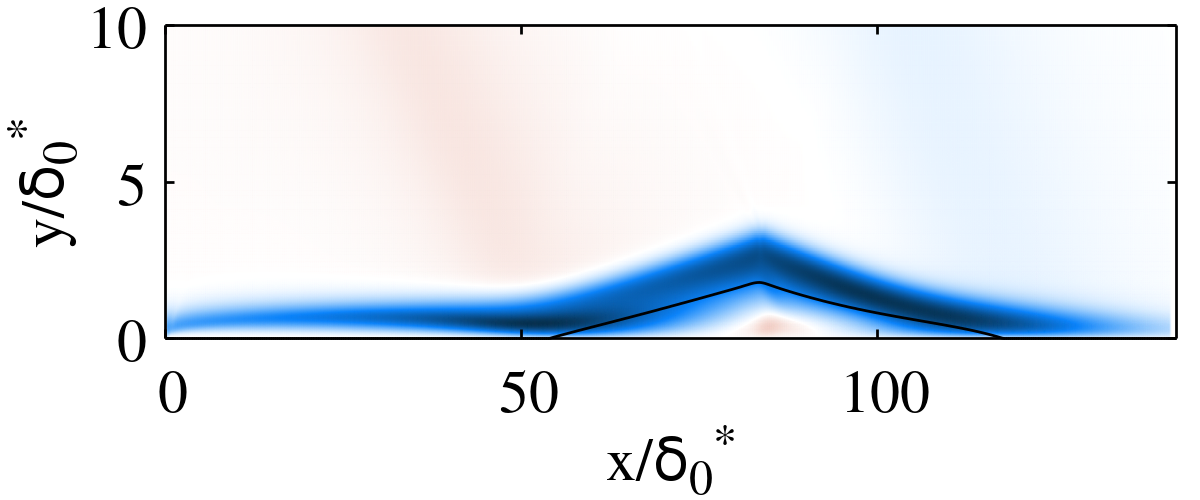}      
            \end{subfigure}
            \begin{subfigure}[]{0.49\textwidth}  
                \includegraphics[angle=-0,trim=0 0 0 0, clip,width=0.99\textwidth]{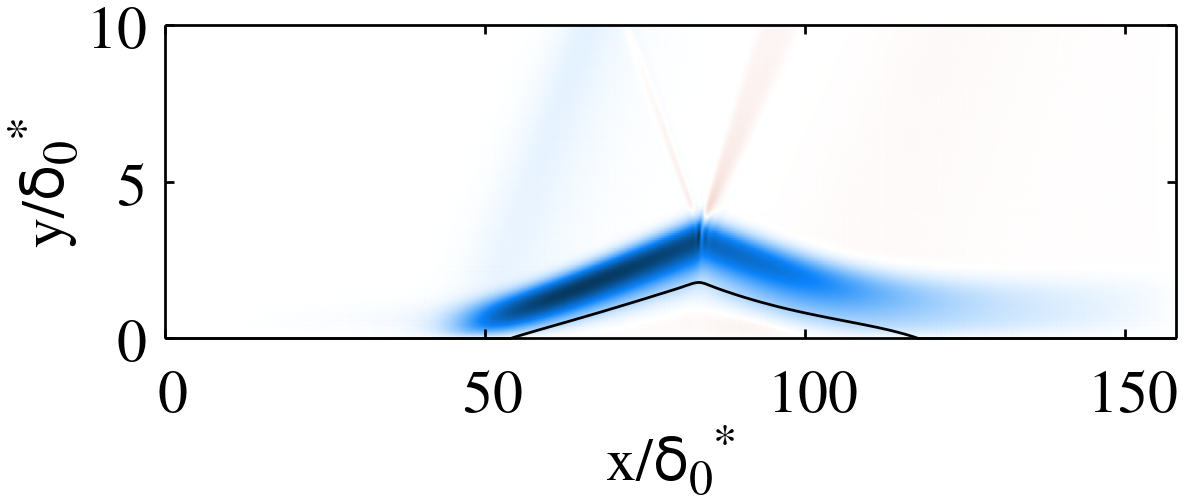}      
            \end{subfigure}
            \caption{Resolvent mode 1 at $St=10^{-4}$, real part of the streamwise velocity. Left: forcing. Right: response.}
            \label{fig.resMode1_1Em4}
        \end{figure}

\begin{figure}
            \centering
            \includegraphics[angle=-0,trim=0 0 0 0, clip,width=0.28\textwidth]{Figures/colorbar1}      
            \vspace{0.15cm}
            
            \begin{subfigure}[]{0.49\textwidth}  
                \includegraphics[angle=-0,trim=0 0 0 0, clip,width=0.99\textwidth]{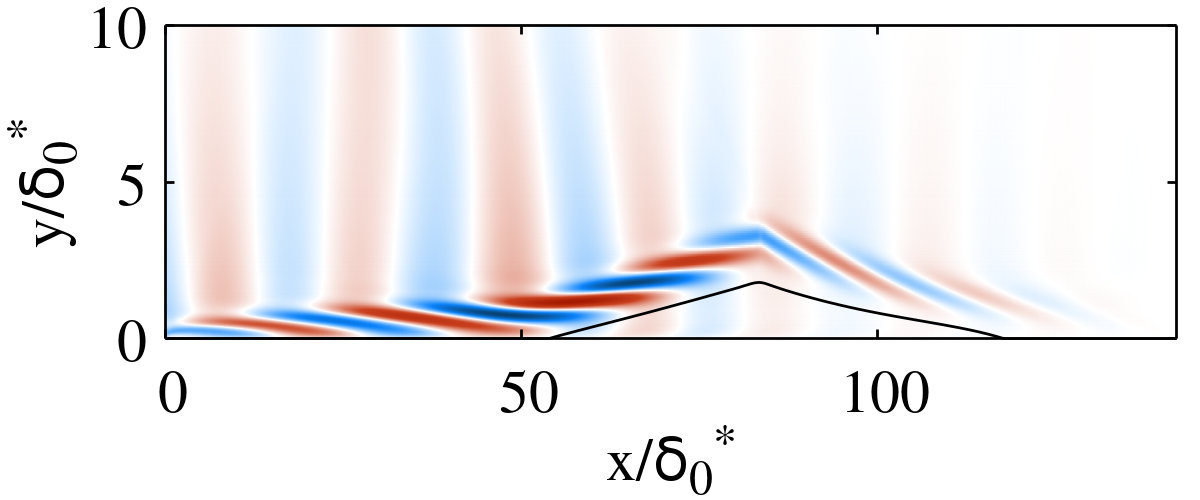}      
            \end{subfigure}
            \begin{subfigure}[]{0.49\textwidth}  
                \includegraphics[angle=-0,trim=0 0 0 0, clip,width=0.99\textwidth]{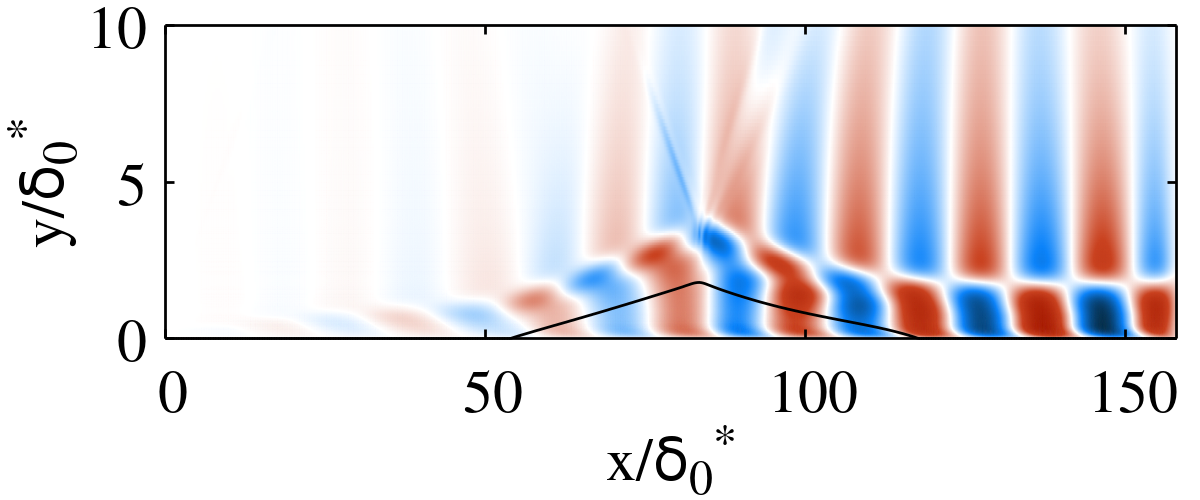}      
            \end{subfigure}
            \caption{Resolvent mode 1 at $St=2$, real part of the streamwise velocity. Left: forcing. Right: response.}
            \label{fig.resMode1_1E0}
        \end{figure}

        The optimal gain is computed for different frequencies at $M=2.20$ and $\Reynd=1100$ (figure \ref{fig.optimalGainM220}). 
        At low frequency ($\StL < 10^{-1}$), the optimal gain monotonically decreases.
        The resolvent modes at $\StL = \tento{-4}$ are shown in figure \ref{fig.resMode1_1Em4}.
        The optimal forcing is spread over a large portion of the domain, admitting maxima near the separation point and along the right part of the recirculation bubble.
        The optimal response is concentrated above the bubble, following its shape.
        Moreover, even though a drop of optimal gain is observed at low frequency, the optimal forcing and response vectors are found independent of the forcing frequency: the same velocity and pressure fields are observed at any low Strouhal numbers.
        Noticeable differences start to appear for frequencies above $\StL \simeq 2 \times \tento{-2}$.
        An interpretation will be proposed in section \ref{sec.model}.
        At higher frequencies, a maximum of optimal gain is detected at $\StL \simeq 2$ (figure \ref{fig.optimalGainM220}).
        It is associated with a convective instability that starts developing in the recirculation region and continues to grow further downstream in the boundary layer (figure \ref{fig.resMode1_1E0}).
        This is reminiscent of the first mode instability of the supersonic boundary layer, which is the compressible counterpart of the Tollmien-Schlichting instability in incompressible boundary layers \citep{mack1984boundary}.
        In order to efficiently trigger this instability, the optimal forcing is located upstream. 
        Its tilted structure shows the concurrent action of the non-modal Orr mechanism, as usually observed in boundary layer flows \citep{EG08,bugeat20193d}.

        \subsubsection{Global stability analysis} \label{sec.stability2D}

\begin{figure}
            \centering
            \includegraphics[angle=-0 ,trim=0 0 0 0, clip,width=0.62\textwidth]{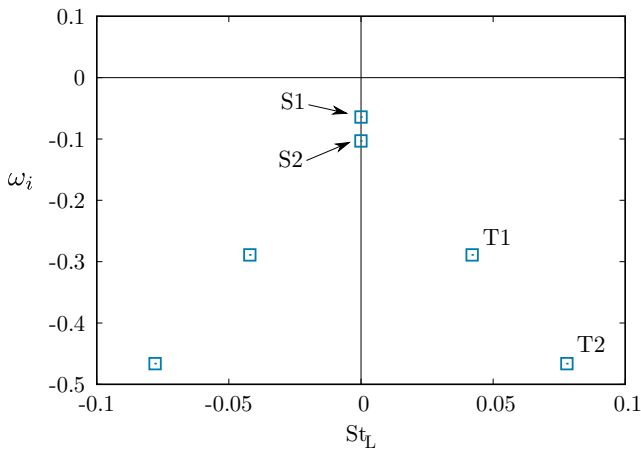}
            \caption{Global stability spectrum at $M=2.20$, $Re=1100$. Steady and unsteady modes are referred to with the letter S and T, respectively.}
         \label{fig.spectrum2D}
        \end{figure}

\begin{figure}
        \centering
        \includegraphics[angle=-0,trim=0 0 0 0, clip,width=0.28\textwidth]{Figures/colorbar1}      
        \vspace{0.15cm}
            
        \begin{subfigure}[]{0.49\textwidth}  
          \includegraphics[angle=-0,trim=0 0 0 0, clip,width=0.99\textwidth]{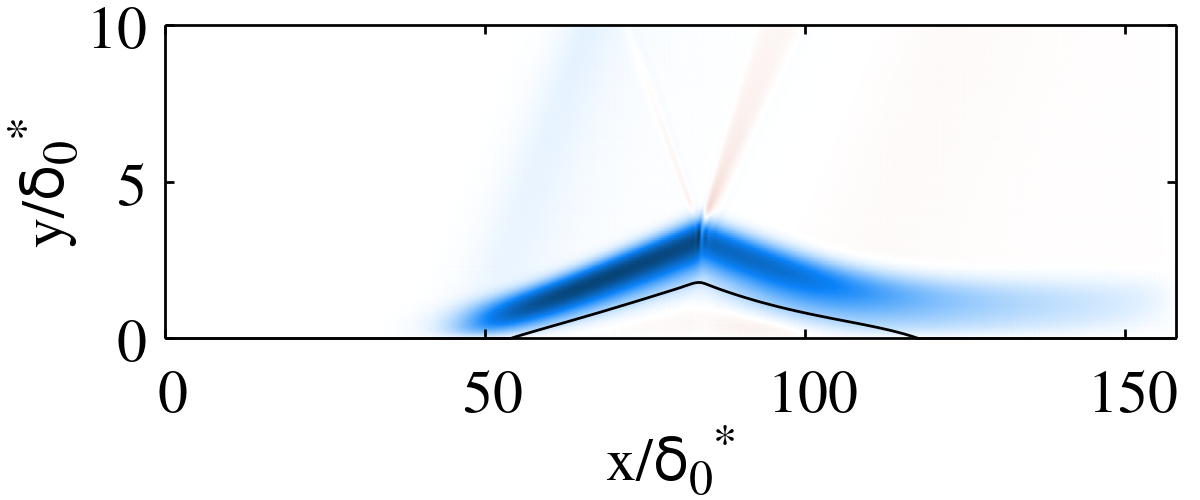}      
        \end{subfigure}
         \begin{subfigure}[]{0.49\textwidth}  
          \includegraphics[angle=-0,trim=0 0 0 0, clip,width=0.99\textwidth]{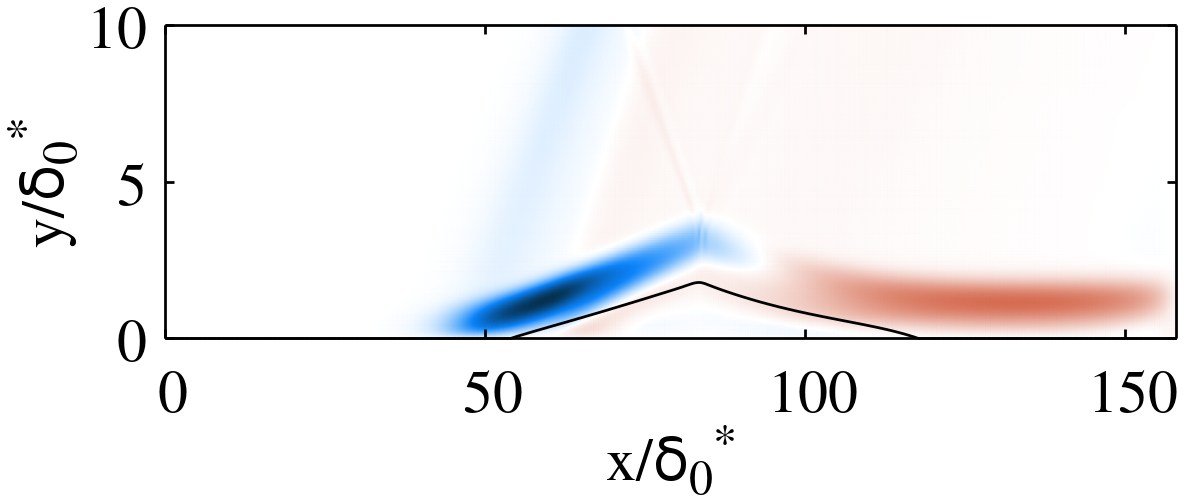}       
        \end{subfigure}
        \caption{Global modes S1 (left) and T1 (right), real part of the streamwise velocity.}
         \label{fig.globalmodes2D}
        \end{figure}

        A global stability computation is performed for the set of base flows described in section \ref{sec.baseflow}.
        The eigenvalue spectrum at $M=2.20$ and $\Reynd=1100$ is presented in figure \ref{fig.spectrum2D}.
        Each eigenvalue has a negative growth rate $\omega_i$ meaning that the system is globally stable.
        This has also been verified for all the base flows considered in this study.  
        The two least stable modes are found to be steady (S1 and S2, $\StL=0$) while the third and fourth modes are unsteady (T1 and T2, $\StL \ne 0$).
        Subsequent modes at higher frequency feature even larger damping rate.
        The velocity field of the mode S1 is mostly localised around the recirculation bubble, following its shape (figure \ref{fig.globalmodes2D}-left).
        The mode T1 shares common characteristics but features a phase opposition between the upstream and downstream region of the bubble (figure \ref{fig.globalmodes2D}-right).
        
        The striking outcome of this global stability analysis is that the least stable global mode S1 is very similar to the optimal response found in the resolvent analysis in the previous section (figure \ref{fig.resMode1_1Em4}-right), which was observed for any low frequencies.
        This means that, at low frequency, the optimal response results from the excitation of the least stable global mode.
        These results suggest that the receptivity at low frequency is a modal phenomenon, as opposed to the non-modal mechanisms usually observed in convective instabilities \citep{CC1997}.
        Note that, because the system is stable, a continuous forcing is required to excite the flow around its base state.
        
        It may appear somewhat unsettling that the receptivity, which pertains to the \textit{unsteady} behaviour of the system, seems to be driven by a \textit{steady} mode. 
        This observation is analysed in the next section.

        \subsubsection{A low-pass filter model} \label{sec.model}

        To clarify how a steady mode could play a role in the low-frequency dynamics of the system, a model based on global stability analysis is proposed to describe the behaviour of the optimal gain. 
        Two sources can cause the singular values of the resolvent operator to increase \citep{schmid2007nonmodal}.
        The first one appears when the forcing frequency is close to the eigenvalues of the Jacobian matrix (as found by a stability analysis).
        The second source is related to the non-normality of the eigenvectors of the Jacobian matrix (i.e. the global modes).
        It is now well-know that the linear superposition of stable modes can lead to energy growth \citep{farrell1988optimal}, termed as non-modal growth \citep{SH01}.
        This idea is translated, in the resolvent framework, into the concept of pseudo-resonance through which increases of optimal gain can occur away from the frequencies of the eigenmodes (see for example \cite{bucci2018roughness}).

        From the insight obtained from the previous stability analysis (section \ref{sec.stability2D}), we would like to test if the optimal gain proceeds from purely modal effects.
        Thus, non-modal phenomena are here discarded from the model.
        Furthermore, if one and only one global mode drives the optimal gain, then the singular value of the resolvent should vary as $\sigma \sim 1/d$, where $d$ is the distance in the complex plane between the forcing frequency $\StL$ and the eigenvalue $\omega / 2\pi = ( \omega_r + i \omega_i )/ 2\pi$ of this global mode.
        This distance is readily obtained as
        \begin{equation} \label{eq.distanceComplex}
            d = \sqrt{\left( \StL - \omega_r/2\pi \right)^2 + \left( \omega_i/ 2\pi \right)^2}
        \end{equation}

        \noindent Because of the resemblance previously observed between the optimal response and the global mode S1, this mode is a natural candidate to test the purely modal receptivity model. 
        Its eigenvalue is then plugged into equation \eqref{eq.distanceComplex}.
        Since this mode is steady ($\omega_r=0$), the low-frequency model of the optimal gain can eventually be recast as
        \begin{equation} \label{eq.lowpassfilter}
            \sigma (\StL) = \frac{\sigma_0}{\sqrt{1+ \left( \frac{\StL}{\omega_i^{(S1)}/2 \pi} \right)^2}}
        \end{equation}

        \noindent where $\sigma_0 = \lim_{\StL \to 0} \sigma(\StL)$ is the value of the optimal gain when $\StL$ goes zero.
        Here, $\sigma_0$ is not predicted \textit{a priori} by the model, but is obtained by one computation of the optimal gain at zero frequency (or any small frequency where the optimal gain experiences a plateau).
        But since the absolute value of the optimal gain ultimately carries little significance \citep{sipp2013characterization}, this value is not essential to this model.
        The aim of this model is rather to detect the drop of optimal gain in order to analyse low-frequency receptivity of the system.
        It can be recognised that equation \eqref{eq.lowpassfilter} is that of a first-order low-pass filter.
        The optimal gain being by definition analogous to a transfer function (see equation \eqref{eq.gain}), describing its behaviour as a filter is consistent.
        
        In equation \eqref{eq.lowpassfilter}, the cut-off frequency is given by the damping rate $\omega_i^{(S1)}$ of the global mode S1, which is the least stable global mode.
        Even though this mode is steady, a harmonic forcing of non-zero frequency is still able to excite it.
        When the forcing frequency $\StL$ goes to zero, the gain does not depend on the frequency as the distance in the complex plane $d$ tends to the constant value $|\omega_i^{(S1)}/2\pi|$.
        However, increasing $\StL$ until the order of $\omega_i^{(S1)}/2\pi$ and above affects the value of the optimal gain that then decreases as the forcing frequency is pushed away from the eigenvalue of the global mode it excites.
        Thus, the time scale appearing in this low-pass filter model is not set by the frequency of the global mode (which would be zero here), but by its damping rate.

        \subsubsection{Test of the model for different Mach and Reynolds numbers} \label{sec.testmodel1}

\begin{figure}
        \centering
        \begin{subfigure}[]{0.49\textwidth}  
          \includegraphics[angle=-90,trim=0 0 0 0, clip,width=0.99\textwidth]{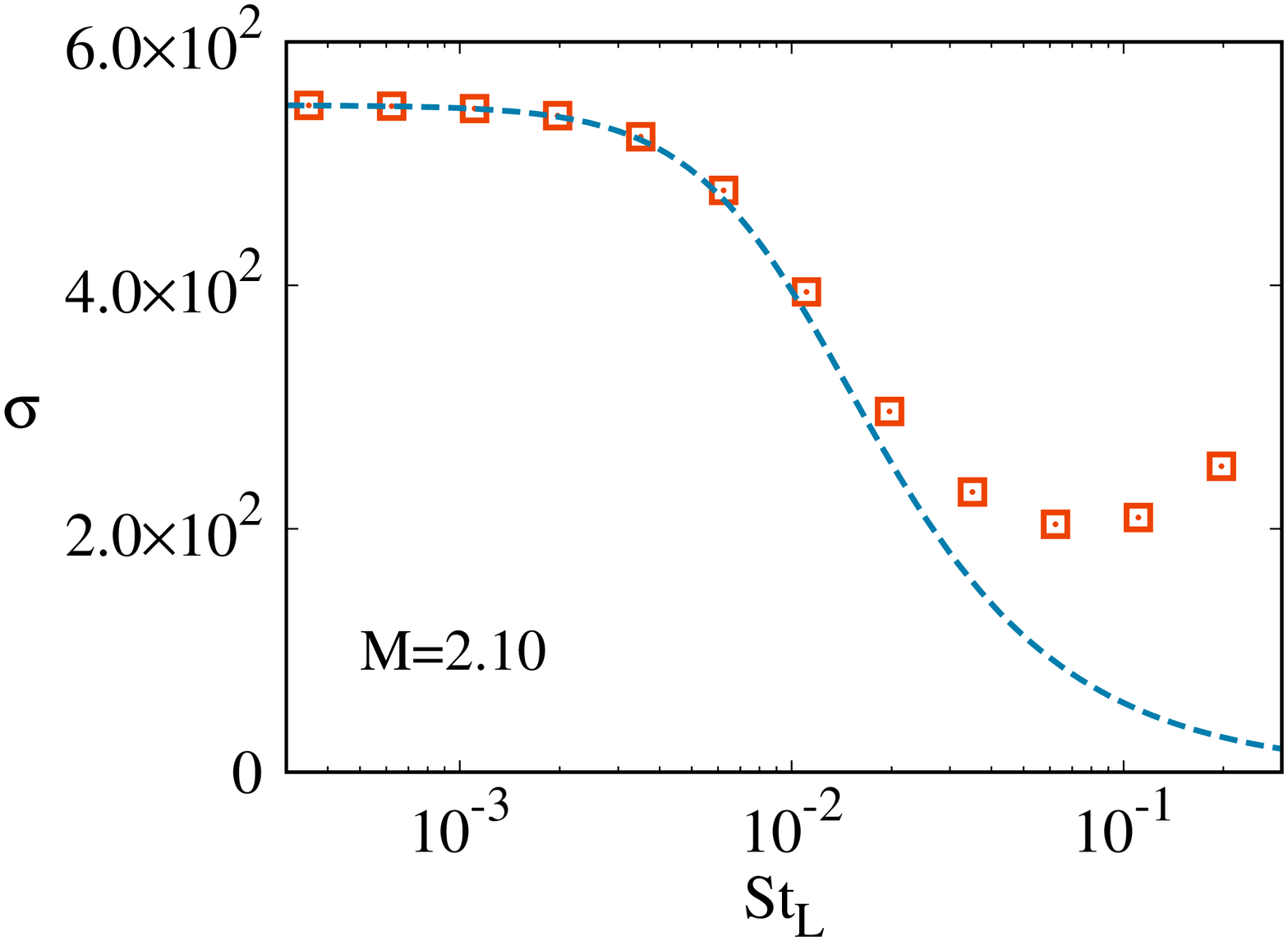}      
        \end{subfigure}

        \begin{subfigure}[]{0.49\textwidth}  
          \includegraphics[angle=-90,trim=0 0 0 0, clip,width=0.99\textwidth]{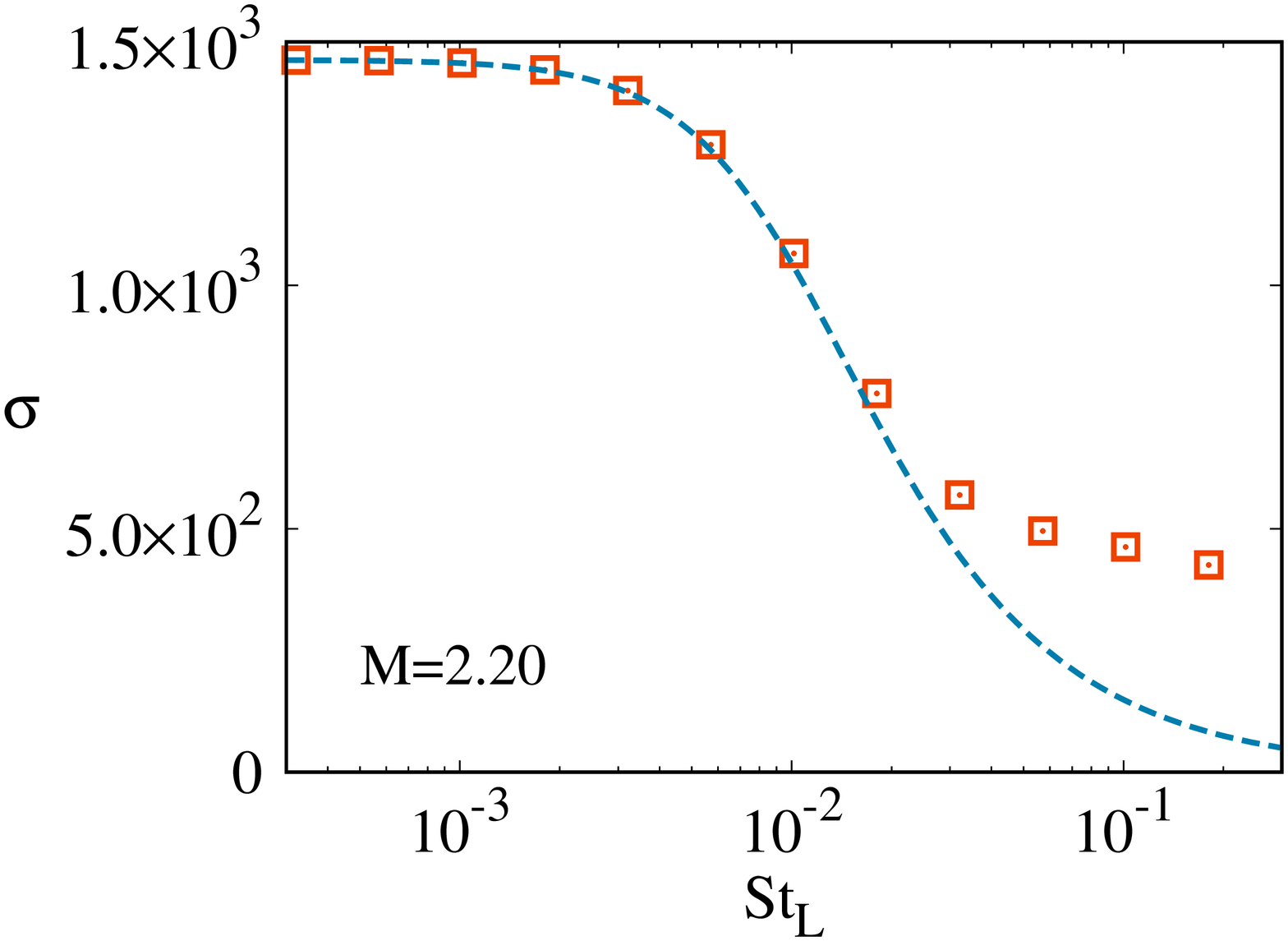}      
        \end{subfigure}        
         \begin{subfigure}[]{0.49\textwidth}  
          \includegraphics[angle=-90,trim=0 0 0 0, clip,width=0.99\textwidth]{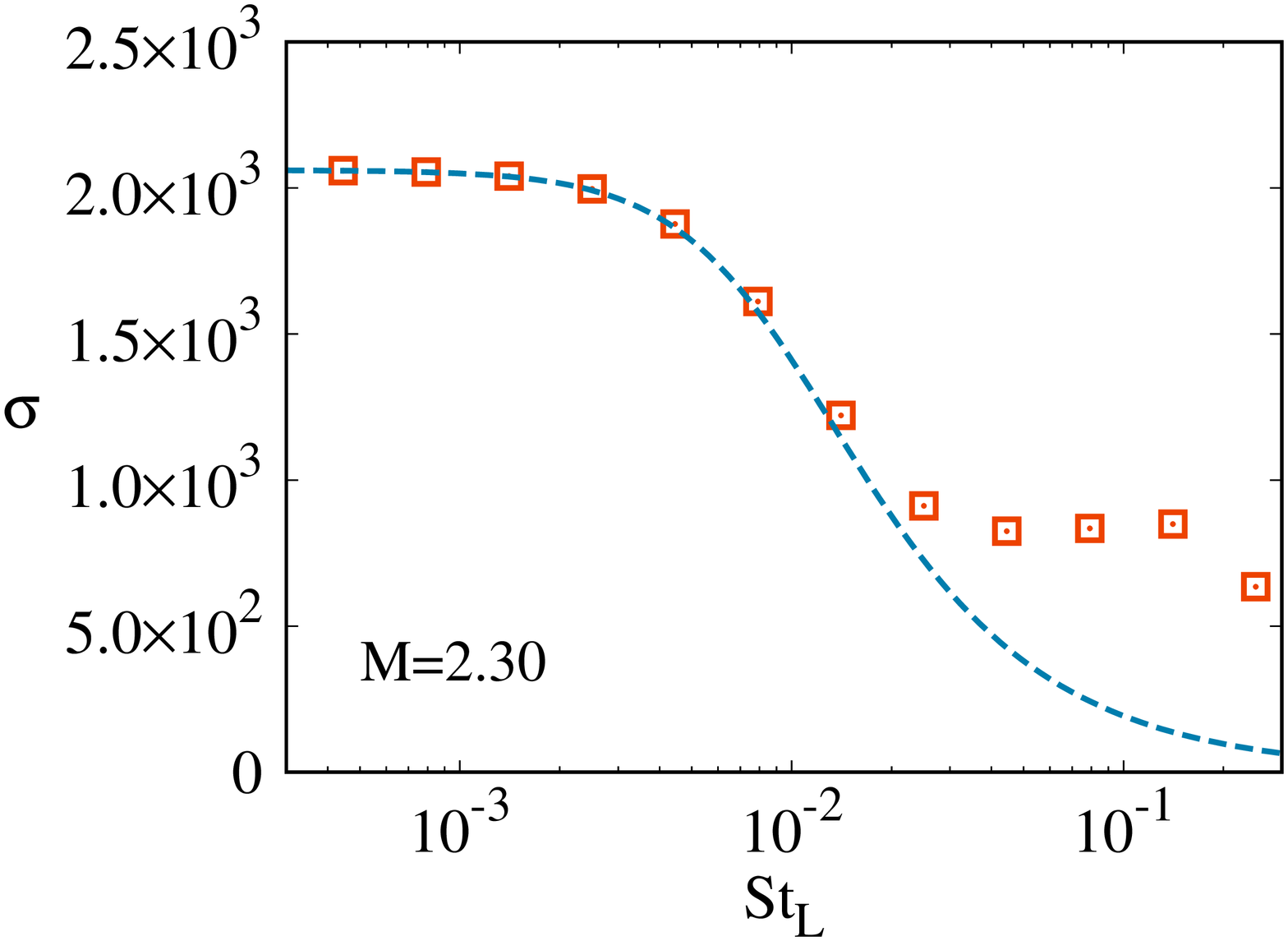}      
        \end{subfigure}
        \vspace{-0.05cm}
        \caption{Comparison between the optimal gain (red squares) and the low-pass filter model from equation \eqref{eq.lowpassfilter} (blue line) for different Mach numbers ($Re=1100$).}
         \label{fig.model1stmode_influenceM}
        \end{figure}

        The optimal gain model proposed in equation \eqref{eq.lowpassfilter} is tested for different Mach and Reynolds numbers.
        Setting $\Reynd=1100$, three Mach numbers are considered: $M=2.10$, $M=2.20$ and $M=2.30$, with separation lengths ranging from $22 \deltazero$ to $88 \deltazero$ (table \ref{tab.baseflows}).
        Figure \ref{fig.model1stmode_influenceM} shows that the low-pass filter model accurately detects the drop of optimal gain that appears between $\StL=\tento{-3}$ and $\tento{-2}$.
        In other words, the cut-off frequency is correctly predicted by the damping rate of the global mode S1.
        This is observed for every Mach numbers.
        Disagreements appear above $\StL \simeq 2 \times \tento{-2}$, where the low-pass filter underestimates the actual gain.
        At these frequencies, the structure of the optimal response starts to change.
        This means that the global mode S1 does not drive the dynamics of the flow alone any longer, but that other modes are getting involved.

\begin{figure}
        \centering
         \begin{subfigure}[]{0.49\textwidth}  
          \includegraphics[angle=-90,trim=0 0 0 0, clip,width=0.99\textwidth]{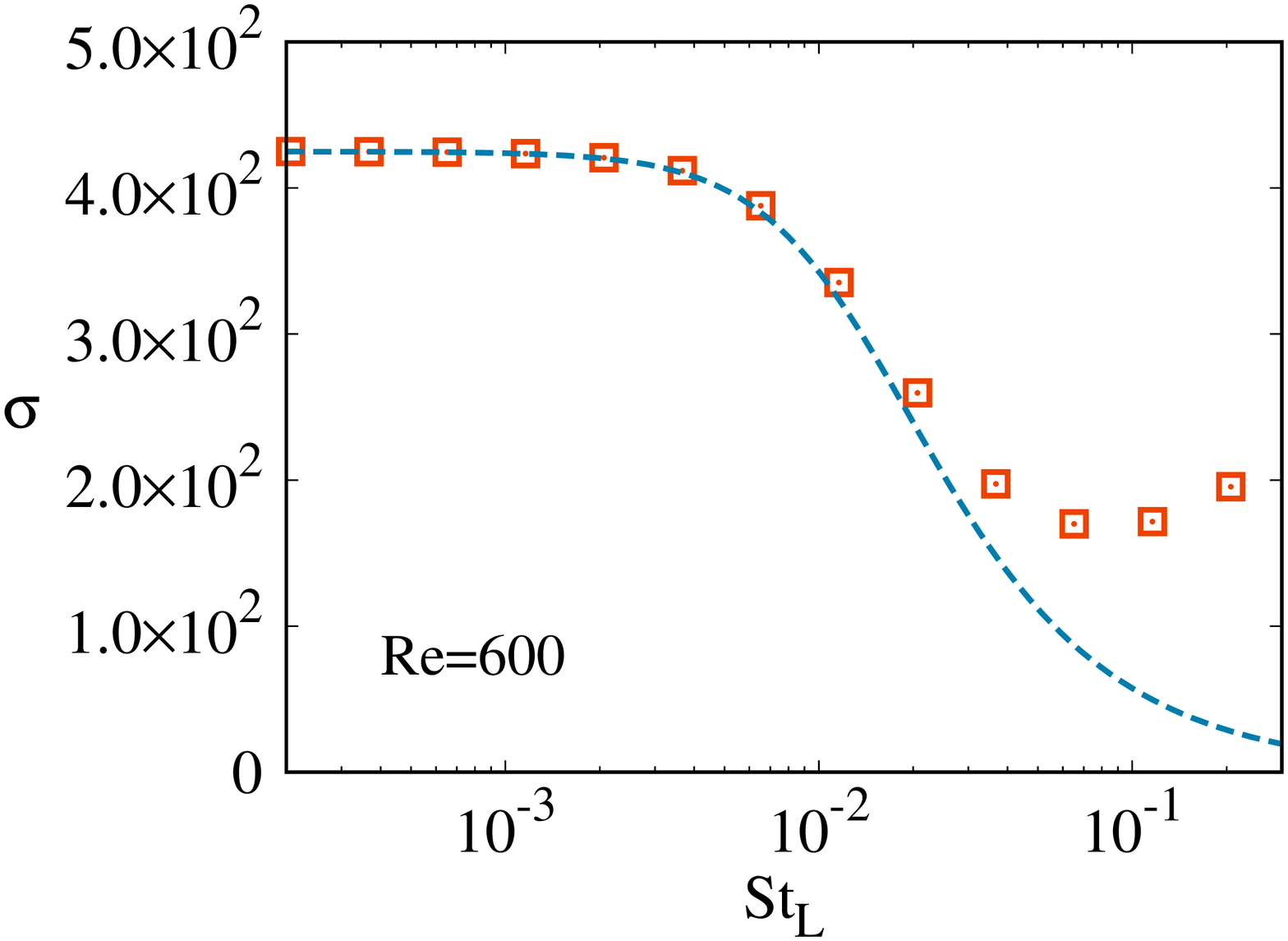}      
        \end{subfigure}
         \begin{subfigure}[]{0.49\textwidth}  
          \includegraphics[angle=-90,trim=0 0 0 0, clip,width=0.99\textwidth]{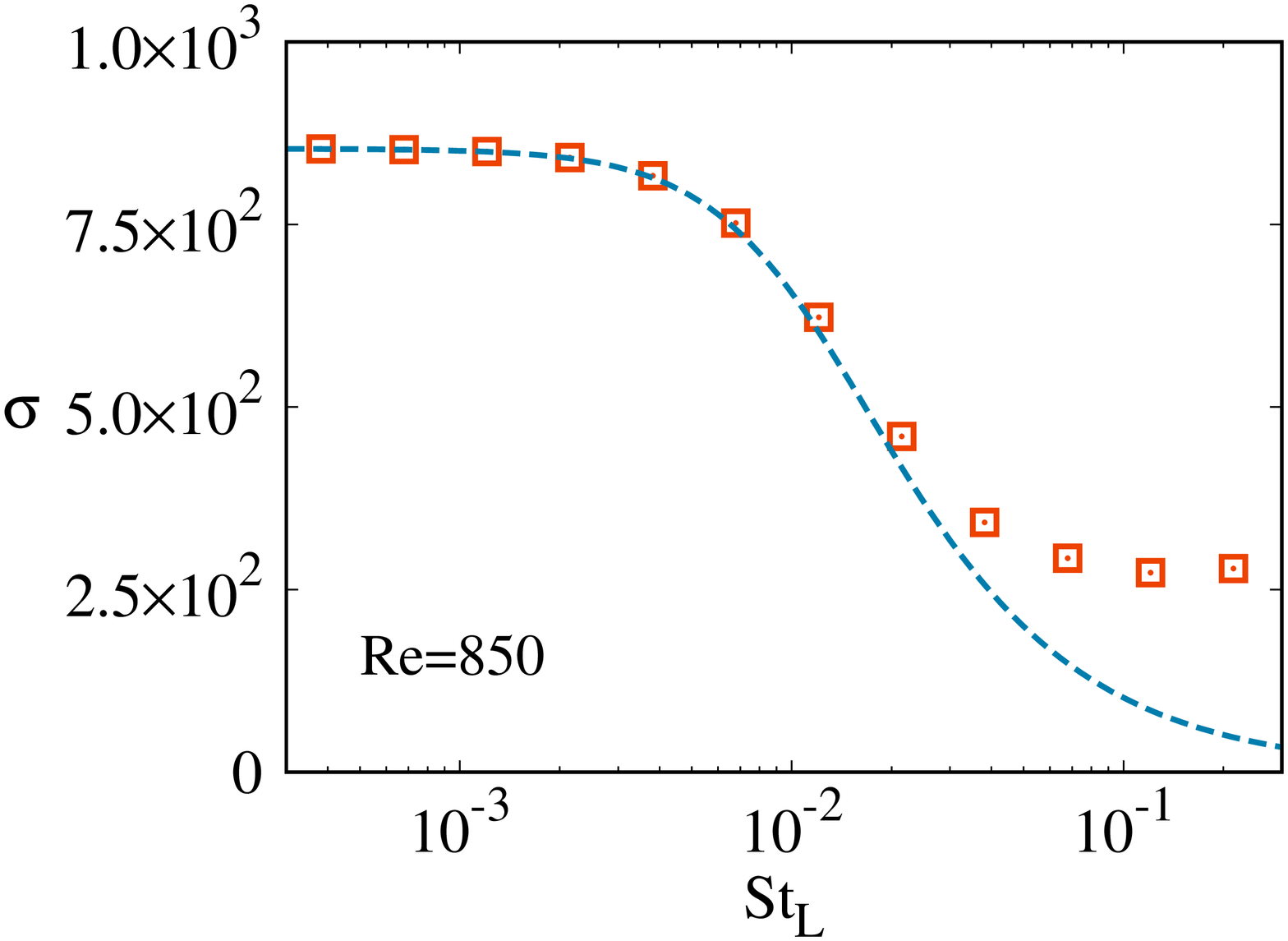}      
        \end{subfigure}

         \begin{subfigure}[]{0.49\textwidth}  
          \includegraphics[angle=-90,trim=0 0 0 0, clip,width=0.99\textwidth]{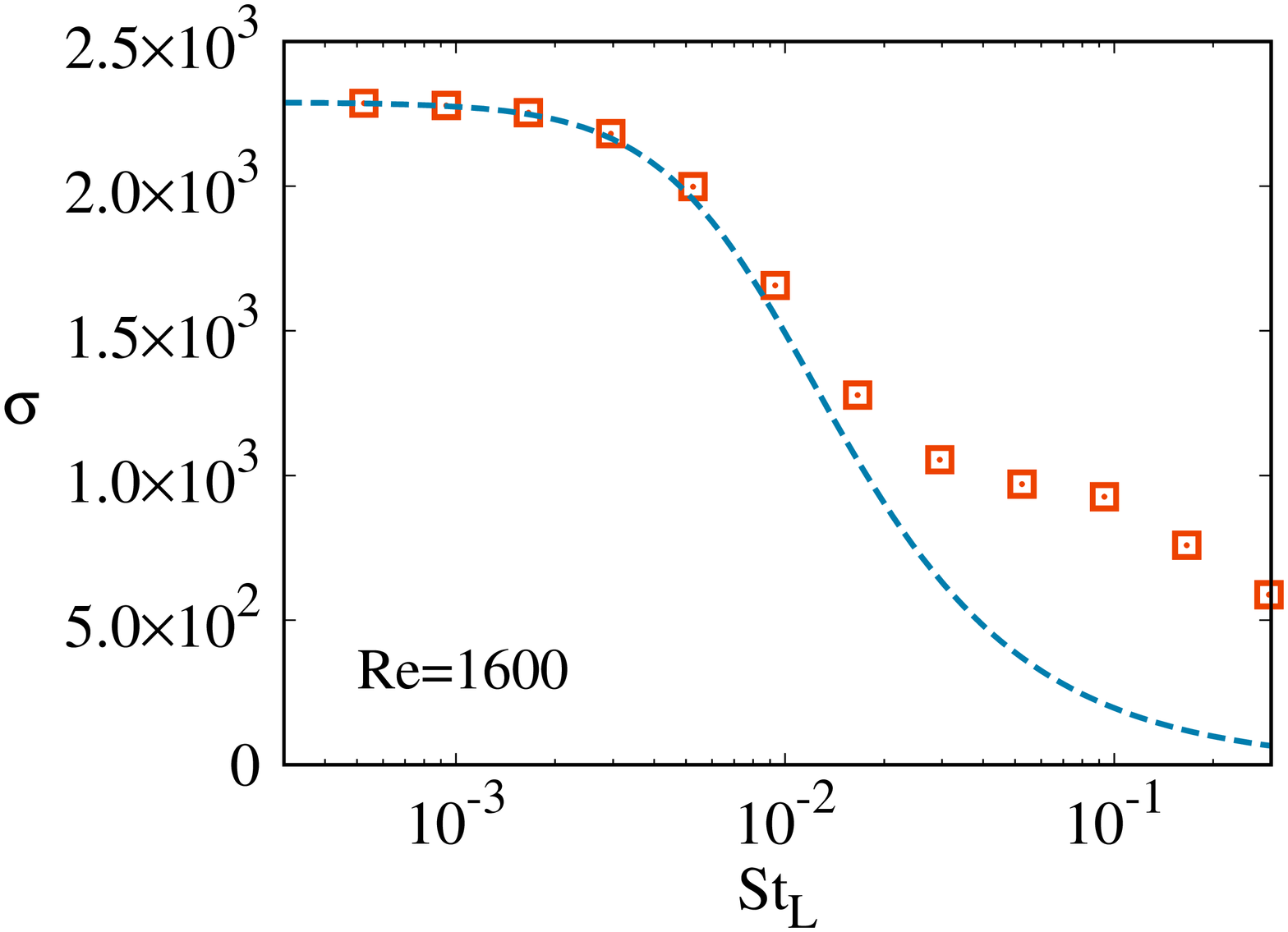}      
        \end{subfigure}
        \begin{subfigure}[]{0.49\textwidth}  
          \includegraphics[angle=-90,trim=0 0 0 0, clip,width=0.99\textwidth]{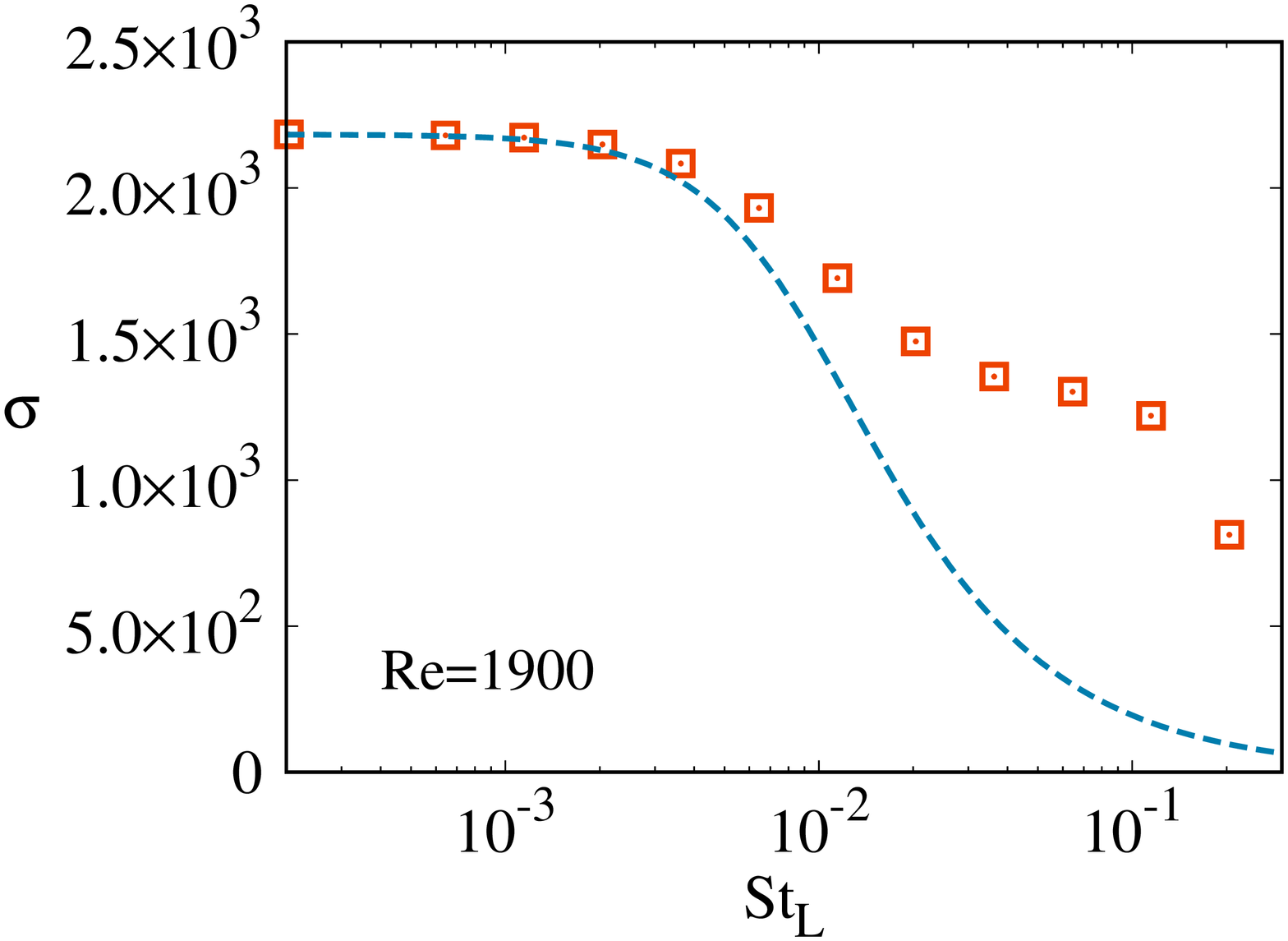}      
        \end{subfigure}
        \vspace{-0.05cm}
        \caption{Comparison between the optimal gain (red squares) and the low-pass filter model from equation \eqref{eq.lowpassfilter} (blue line) for different Reynolds numbers ($M=2.20$).}
         \label{fig.model1stmode_influenceRe}
        \end{figure}

        Other Reynolds numbers, below and above the previous value $\Reynd=1100$, are considered while keeping $M=2.2$.
        Below $\Reynd=1100$, the same agreement as previously described is observed (figure \ref{fig.model1stmode_influenceRe}).
        As the Reynolds number increases ($\Reynd=1600$), discrepancies start to appear for smaller frequencies than $\StL=\tento{-2}$ but the location of the drop of optimal gain is still correctly captured. 
        At $\Reynd=1900$, the model becomes irrelevant. 
        The growing mismatch between the low-pass filter model and the optimal gain as $Re$ is increased can be caused by the action of other modes, previously damped by viscous effects.
        As a result, the model based on the excitation of the global mode S1 alone cannot predict the optimal gain any more, even though this mode does not directly depend on viscous effect (in the next section, it will be shown that the global mode S1 is independent of the viscous length scale $\deltazero$).

        \subsubsection{Scaling} \label{sec.scalingMode1}

\begin{figure}
            \centering
            \includegraphics[angle=-0,trim=0 0 0 0, clip,width=0.67\textwidth]{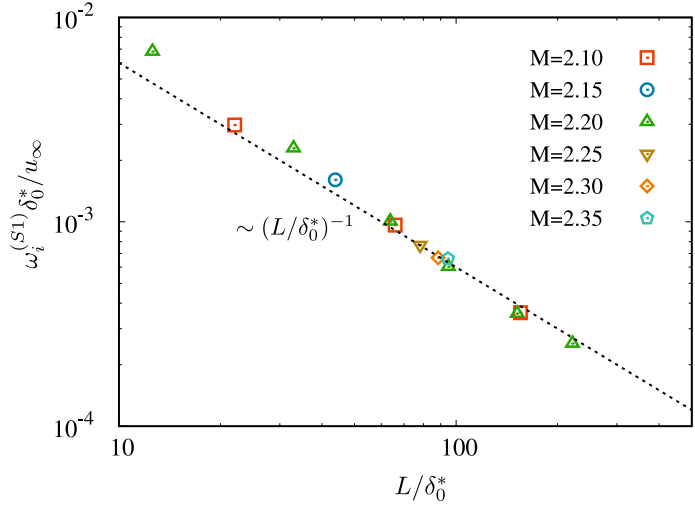}
            \caption{Damping rate of the global mode S1 for different Mach and Reynolds numbers, as a function of the separation length. Here, $\deltazero$ is the boundary layer thickness at $Re=1100$ rather than that at each Reynolds number. 
            %The dashed line, whose slope is -$(L/\deltazero)$, is plotted as a visual reference.
            }
            \label{fig.scaling_mode1}
        \end{figure}

        The low-pass filter model in equation \eqref{eq.lowpassfilter} contains the damping rate $\omega_i^{(S1)}$ of the mode S1.
        This parameter sets the unique time scale of the model.
        Its scaling relative to a reference length scale is now investigated.
        The non-dimensional damping rate $\omega_i^{(S1)}$ is presented in figure \ref{fig.scaling_mode1} as a function of the non-dimensional length of the recirculation bubble $L/\deltazero$.
        Note that each point is associated with a different pair of Reynolds and Mach numbers (see table \ref{tab.baseflows}).
        It is found that $\omega_i^{(S1)} \deltazero / u_\infty \sim (L/\deltazero)^{-1}$, meaning that $\omega_i^{(S1)} \sim u_\infty/L$.
        Because several Reynolds numbers (based on several $\deltazero$, as explained in section \ref{sec.baseflow}) have been tested, this result ensures that $\omega_i^{(S1)}$ is independent of $\deltazero$.
        Instead, the recirculation length $L$ is the relevant length scale.
        The aforementioned scaling furthermore shows that the associated Strouhal number $\StL$, based on the the separation length, is then constant.
        Previous works investigating the low-frequency behaviour of the SWBLI have consistently noted a constant Strouhal number from $\StL \simeq 3 \times \tento{-2}$ to $\StL \simeq 6 \times \tento{-2}$ \citep{dussauge2006unsteadiness}.
        To compare our results to this value, $\omega_i^{(S1)}$ can be translated into a Strouhal number using $\StL = \frac{f L}{u_\infty} = \frac{ \omega^{(S1)} \deltazero }{2 \pi u_\infty} \frac{L}{\deltazero}$, leading to values around $\StL \simeq 1 \times \tento{-2}$.
        While of the same order of magnitude as the data of the literature, the cut-off frequency is slightly underestimated.
        This motivates to push the resolvent analysis a step further by looking at the sub-optimal modes.

        % Scaling

    \subsection{Resolvent analysis: sub-optimal forcing and response}

        \subsubsection{Results}

\begin{figure}
        \centering
          \includegraphics[angle=-90,trim=0 0 0 0, clip,width=0.63\textwidth]{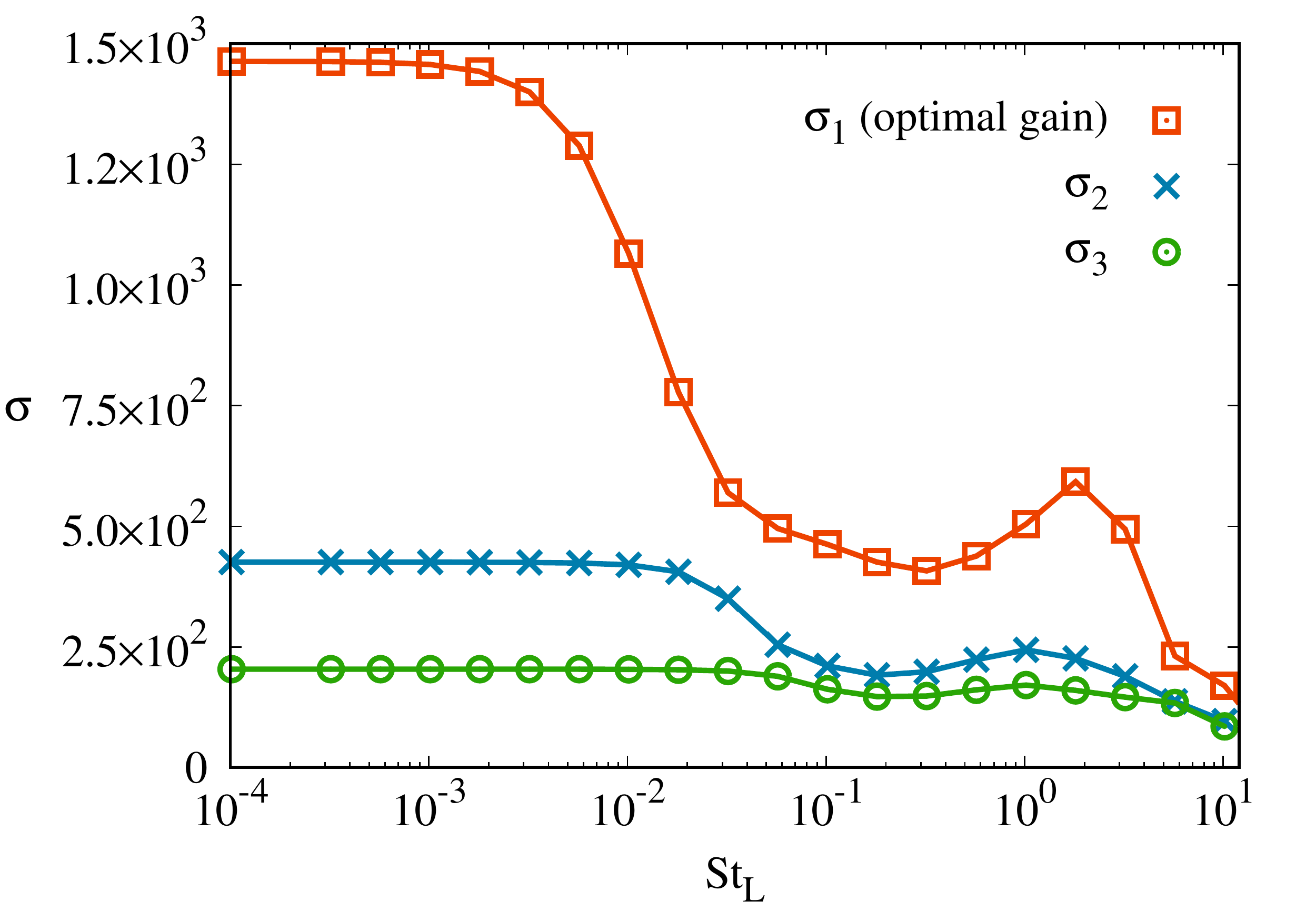}
        \vspace*{-0.05cm}
        \caption{Optimal and sub-optimal gains at $M=2.20$ and $\Reynd=1100$.}
         \label{fig.gain_withSubOpt}
        \end{figure}

\begin{figure}
        \centering
        \includegraphics[angle=-0,trim=0 0 0 0, clip,width=0.28\textwidth]{Figures/colorbar1}      
        \vspace{0.15cm}
            
        \begin{subfigure}[]{0.49\textwidth}  
          \includegraphics[angle=-0,trim=0 0 0 0, clip,width=0.99\textwidth]{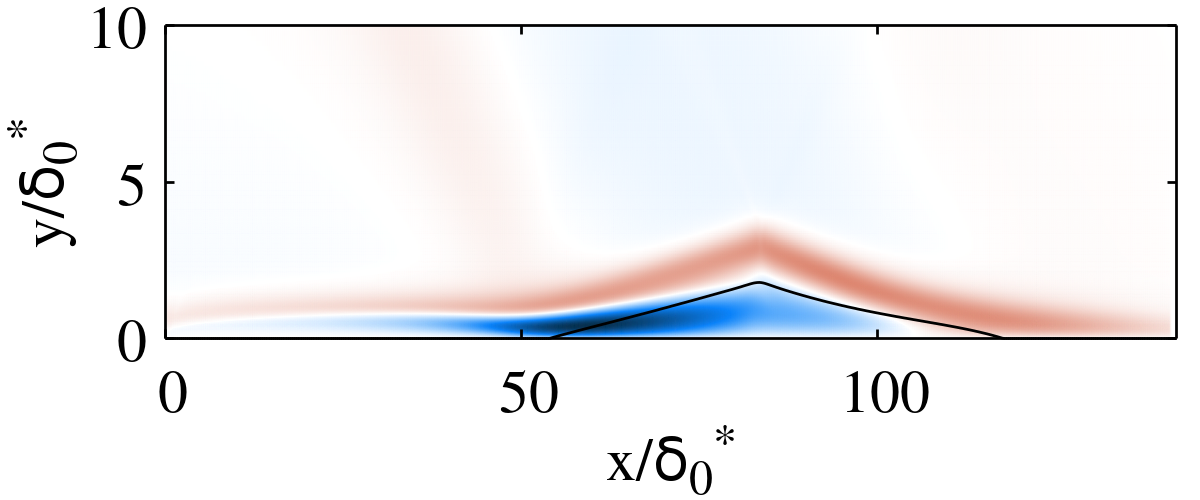}      
        \end{subfigure}
         \begin{subfigure}[]{0.49\textwidth}  
          \includegraphics[angle=-0,trim=0 0 0 0, clip,width=0.99\textwidth]{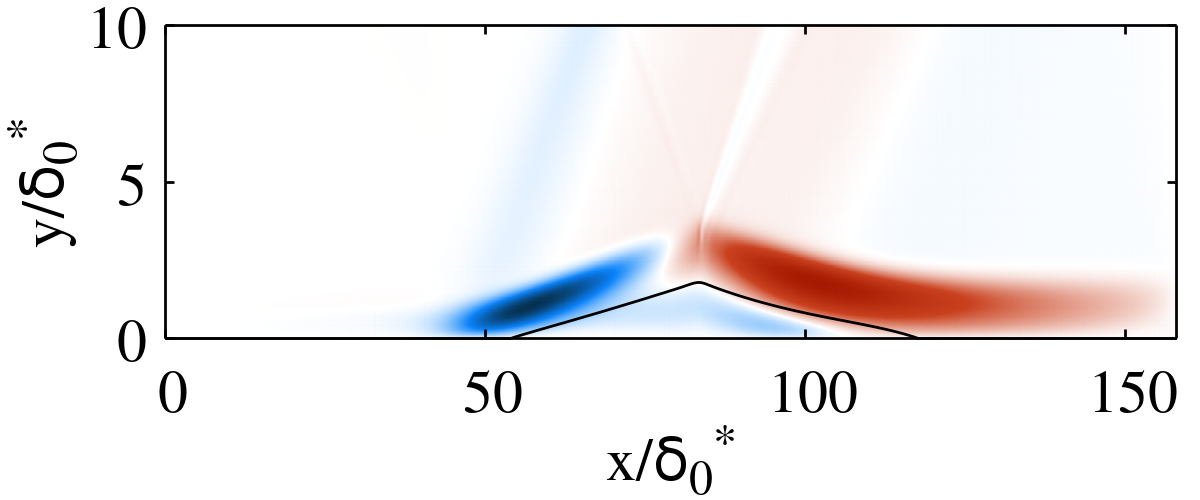}      
        \end{subfigure}
        \caption{Resolvent mode 2 at $\StL=10^{-4}$, real part of the streamwise velocity. Left: forcing. Right: response.}
         \label{fig.resMode2_1Em5}
        \end{figure}

        So far, only the optimal gain $\sigma_1$ and its associated forcing and response modes have been considered.
        Subsequent singular values of the resolvent operator $\sigma_2$ and $\sigma_3$, with $\sigma_1 > \sigma_2 > \sigma_3$ (i.e. sub-optimal gains), are now computed.
        Figure \ref{fig.gain_withSubOpt} compares these three gains as a function of the forcing frequency.
        At low frequency, a factor of about three is observed between $\sigma_1$ and $\sigma_2$, and about two between $\sigma_2$ and $\sigma_3$.
        A drop of sub-optimal gains occurs at higher frequencies that those noted for $\sigma_1$.
        In particular, the decrease of $\sigma_2$ is observed at Strouhal numbers between $\tento{-2}$ and $\tento{-1}$.
        These values lie in the range of the experimental data of the low-frequency unsteadiness more closely than those of $\sigma_1$.
        Our attention will then be focused on the behaviour of $\sigma_2$ and its associated modes at low frequencies.
        The forcing mode follows the shape of the recirculation bubble but is mostly concentrated around the separation point as shown in figure \ref{fig.resMode2_1Em5}-left.
        Unlike the optimal forcing in figure \ref{fig.resMode1_1Em4}-left, the phase varies in the normal direction.
        This is also the case of the associated response (figure \ref{fig.resMode2_1Em5}-right), whose phase varies in the streamwise direction.
        Finally, as previously observed for $\sigma_1$, the sub-optimal modes associated with $\sigma_2$ do not depend on $\StL$ in the low-frequency domain.
        This could suggest that the sub-optimal gain is also based on a modal excitation.
        In the next section, this possibility is discussed by reconsidering the first-order low-pass filter model.

        \subsubsection{Sub-optimal gain: another low-pass filter} \label{sec.modelsubopt}

        The response mode associated with $\sigma_2$ (figure \ref{fig.resMode2_1Em5}-right) shares some features with the unsteady global mode T1 (figure \ref{fig.globalmodes2D}-right) computed in section \ref{sec.stability2D}.
        Besides, the low-frequency behaviour of $\sigma_2$ as a function of $\StL$ is similar to that of $\sigma_1$, the drop of gain being shifted to higher frequencies.
        It is then tempting to build a model of $\sigma_2$ based on the resonance of the global mode T1, derived from the distance in the complex plane between the real forcing frequency and its eigenvalue.
        However, because this mode is unsteady, a resonance model would lead to a peak of gain at the frequency of the mode T1 and a decrease of the gain as the frequency goes to zero.
        These two features are not observed in $\sigma_2$ (figure \ref{fig.gain_withSubOpt}). 
        Such a resonance model is then doomed to fail.

        Instead, equation $\eqref{eq.lowpassfilter}$, which is that of a first-order low-pass filter, satisfies the observation of $\sigma_2$ at low frequency.
        Therefore, we suggest to use this equation to model $\sigma_2$ by plugging the damping rate of  the mode T1 instead of the mode S1 given its resemblance with the response mode associated with $\sigma_2$.
        This equation becomes
        \begin{equation} \label{eq.lowpassfilter2}
            \sigma_2 (\StL) = \frac{\sigma_{2,0}}{\sqrt{1+ \left( \frac{\StL}{\omega_i^{(T1)}/2 \pi} \right)^2}}
        \end{equation}
          
        \noindent As previously mentioned, this approach is different from an approach based on the pure resonance of the global mode T1; instead, we only assume that the damping rate of this mode is the relevant time scale at play.
        The "purely modal dynamics" interpretation elaborated for $\sigma_1$ (see section $\ref{sec.model}$) then no longer holds.
        Other modes, either by normal or non-normal effect, are understood to also contribute to the low-frequency dynamics.
        This is furthermore expected as the response mode of $\sigma_2$ is actually not strictly identical to the mode T1.

\begin{figure}
        \centering
        \begin{subfigure}[]{0.49\textwidth}  
          \includegraphics[angle=-90,trim=0 0 0 0, clip,width=0.99\textwidth]{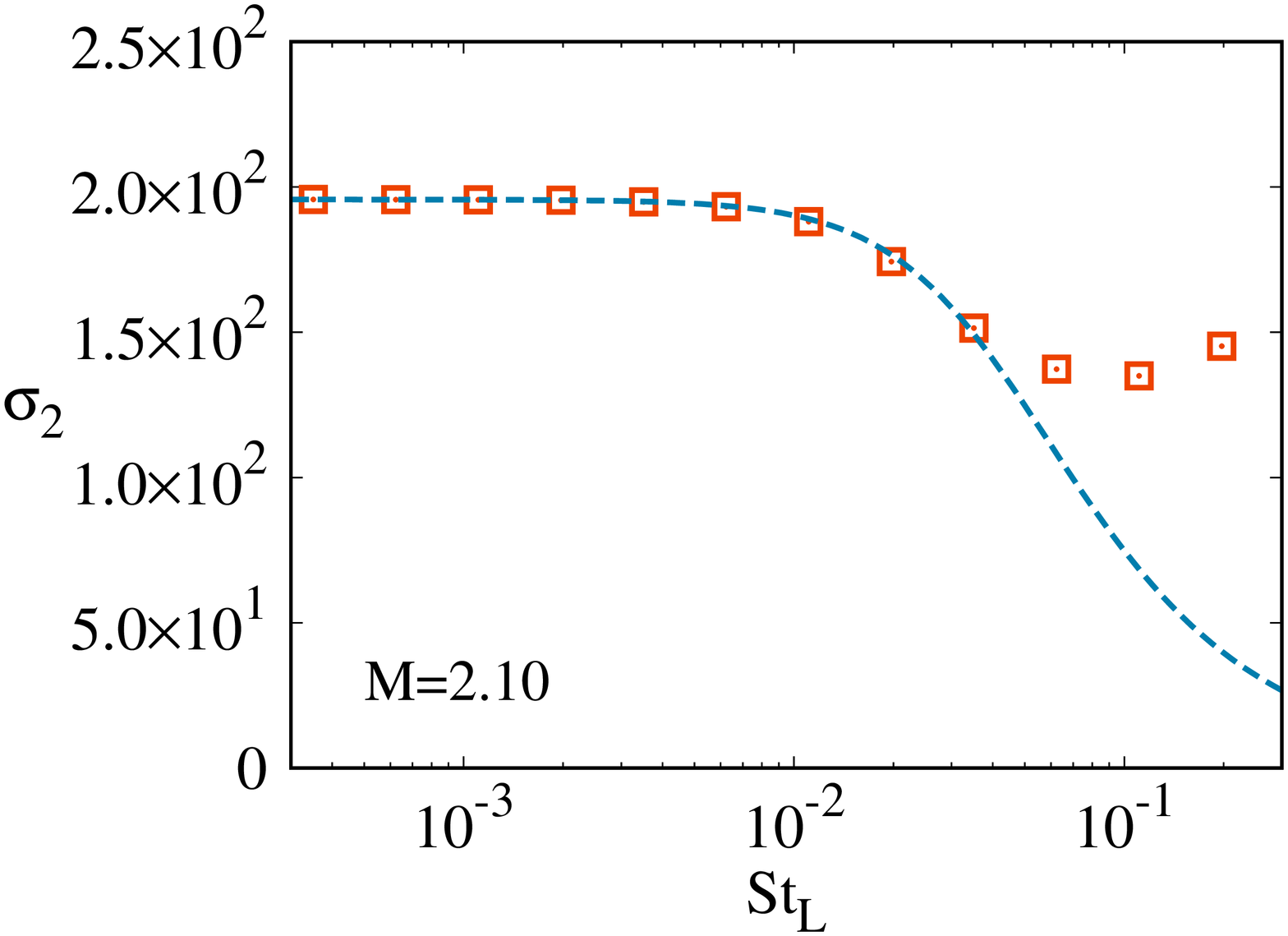}      
        \end{subfigure}

        \begin{subfigure}[]{0.49\textwidth}  
          \includegraphics[angle=-90,trim=0 0 0 0, clip,width=0.99\textwidth]{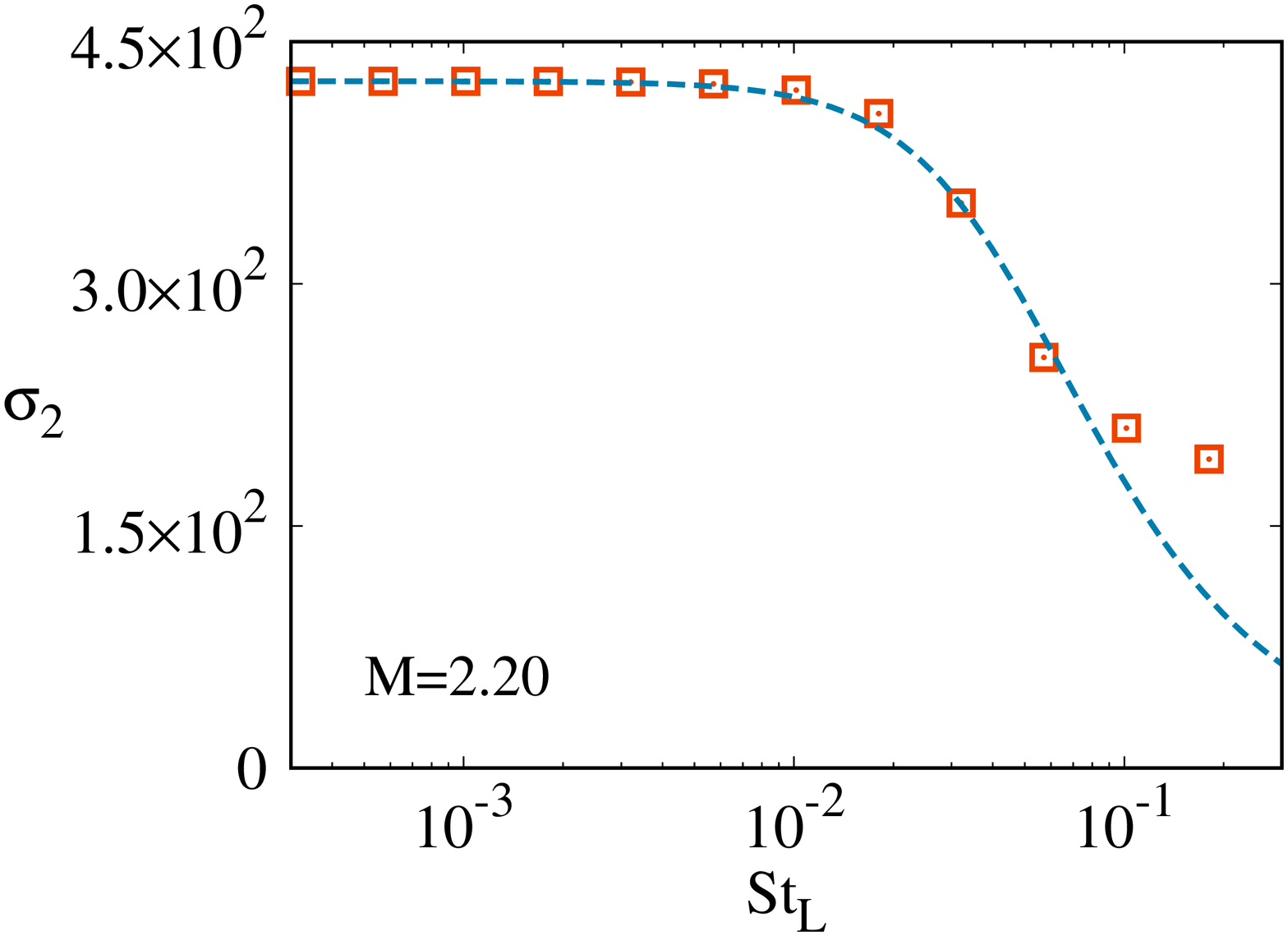}      
        \end{subfigure}
         \begin{subfigure}[]{0.49\textwidth}  
          \includegraphics[angle=-90,trim=0 0 0 0, clip,width=0.99\textwidth]{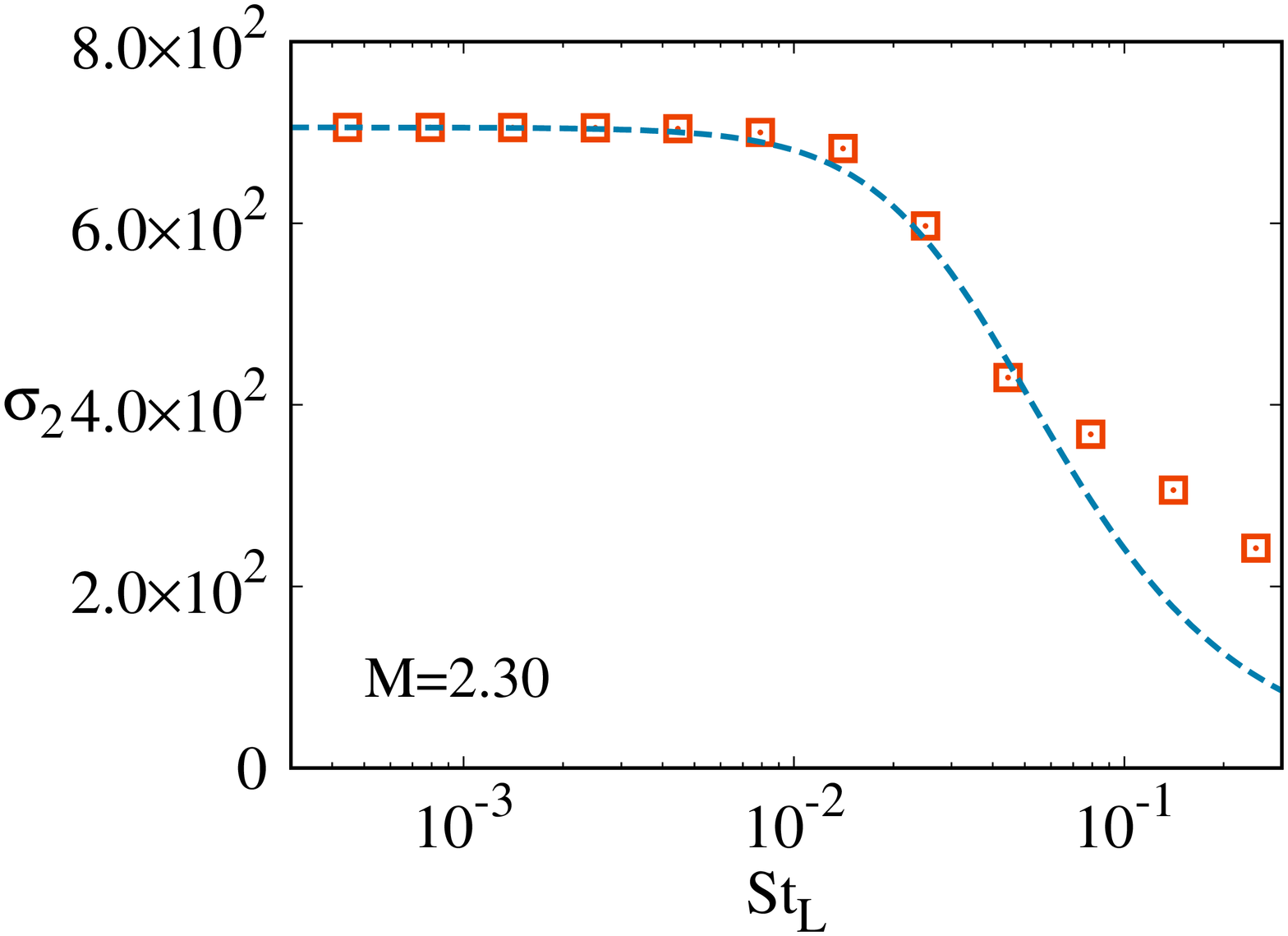}      
        \end{subfigure}
        \vspace{-0.05cm}
        \caption{Comparison between the sub-optimal gain $\sigma_2$ (red squares) and the low-pass filter model in equation \eqref{eq.lowpassfilter2} (blue line) for different Mach numbers ($Re=1100$).}
         \label{fig.model2ndmode_influenceM}
        \end{figure}

\begin{figure}
        \centering
         \begin{subfigure}[]{0.49\textwidth}  
          \includegraphics[angle=-90,trim=0 0 0 0, clip,width=0.99\textwidth]{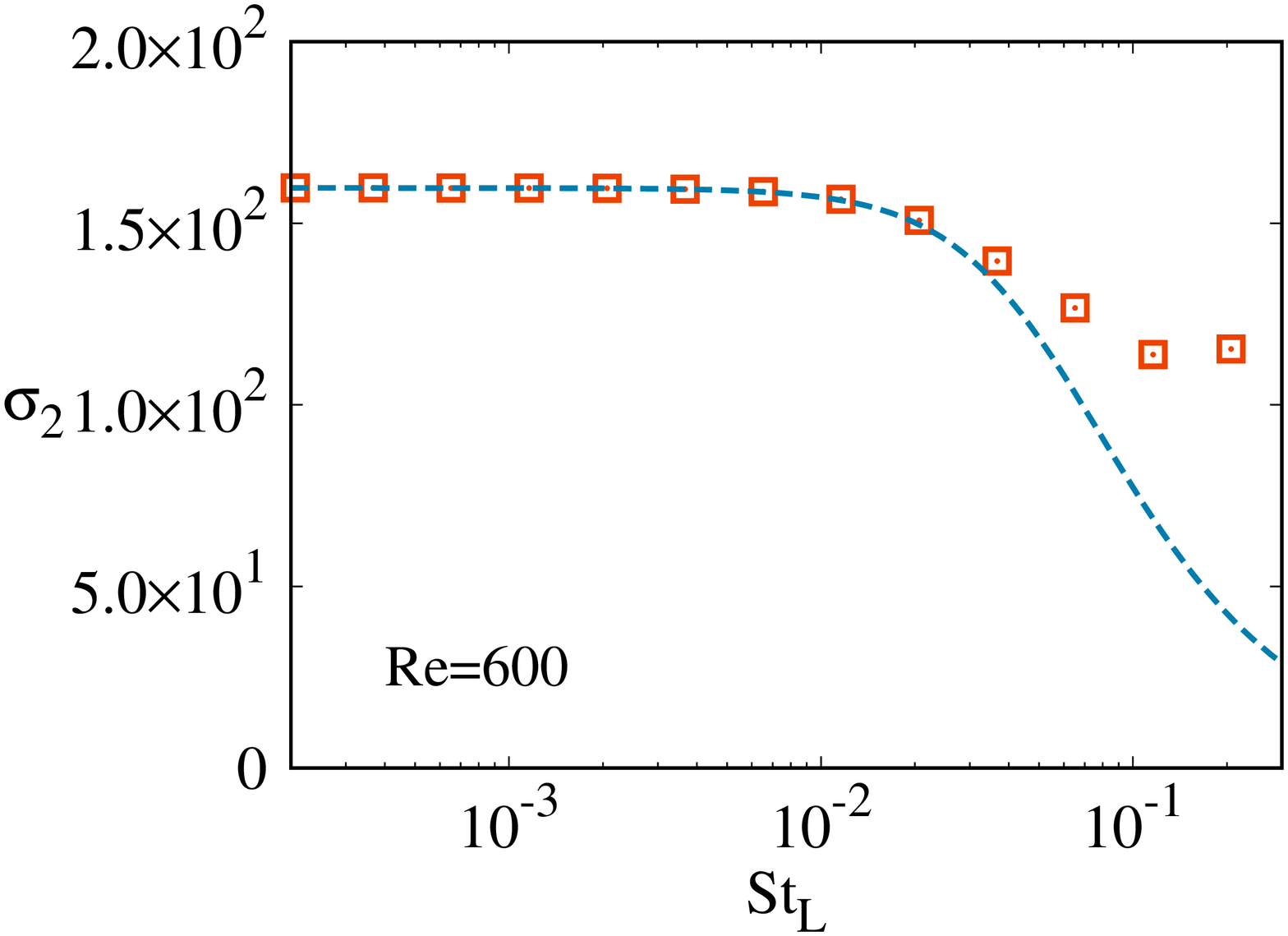}      
        \end{subfigure}
         \begin{subfigure}[]{0.49\textwidth}  
          \includegraphics[angle=-90,trim=0 0 0 0, clip,width=0.99\textwidth]{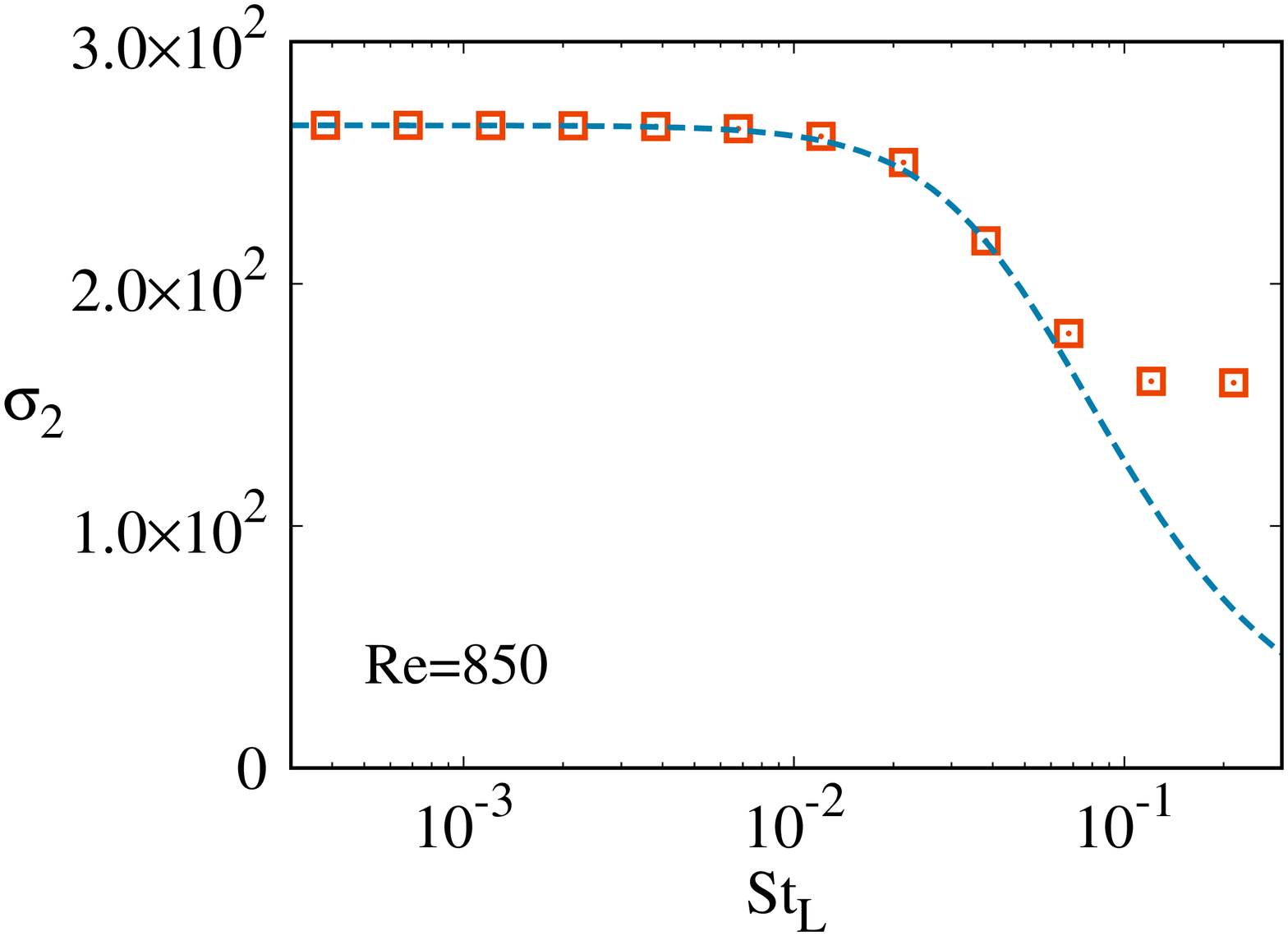}      
        \end{subfigure}

         \begin{subfigure}[]{0.49\textwidth}  
          \includegraphics[angle=-90,trim=0 0 0 0, clip,width=0.99\textwidth]{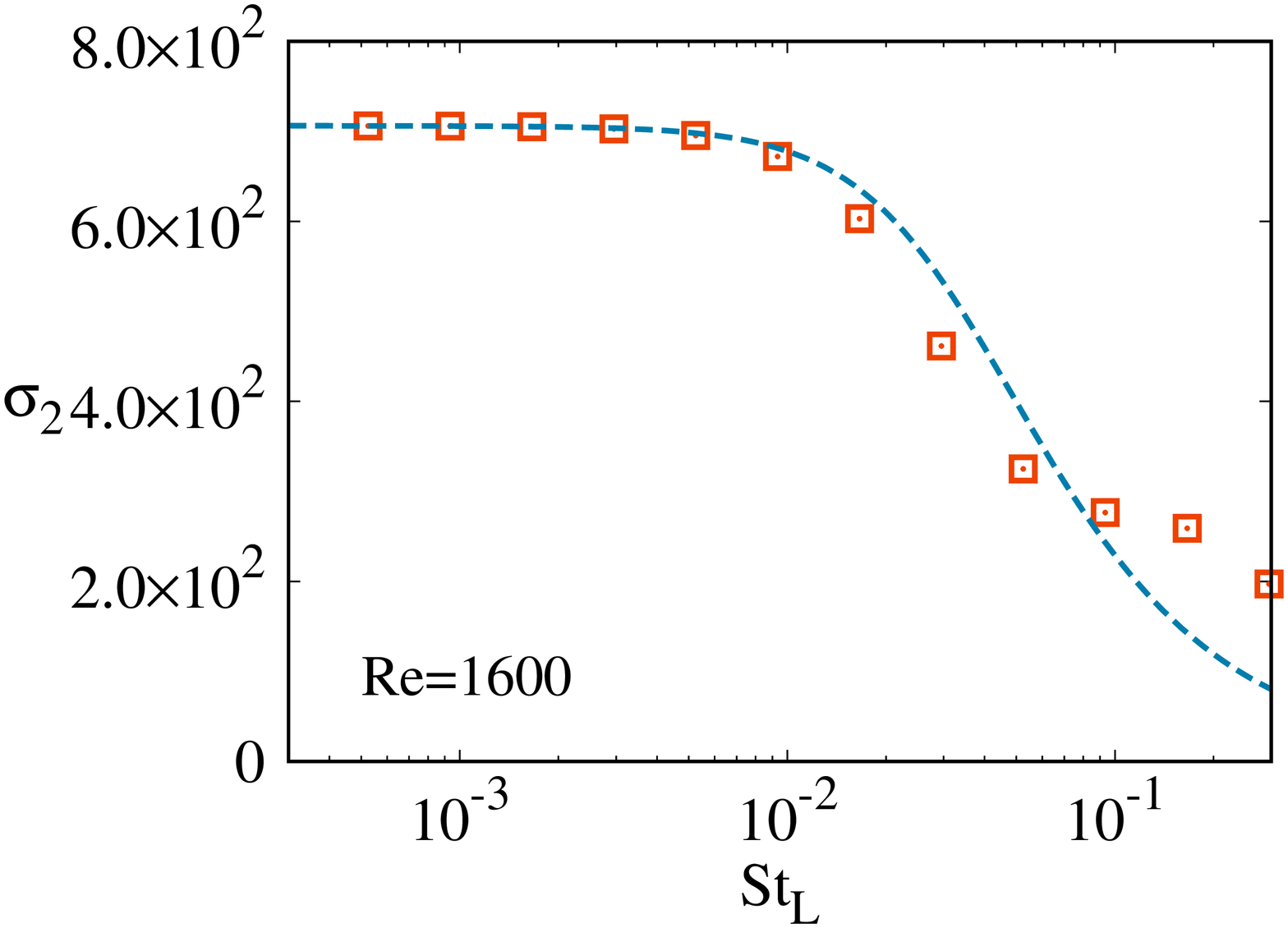}      
        \end{subfigure}
        \begin{subfigure}[]{0.49\textwidth}  
          \includegraphics[angle=-90,trim=0 0 0 0, clip,width=0.99\textwidth]{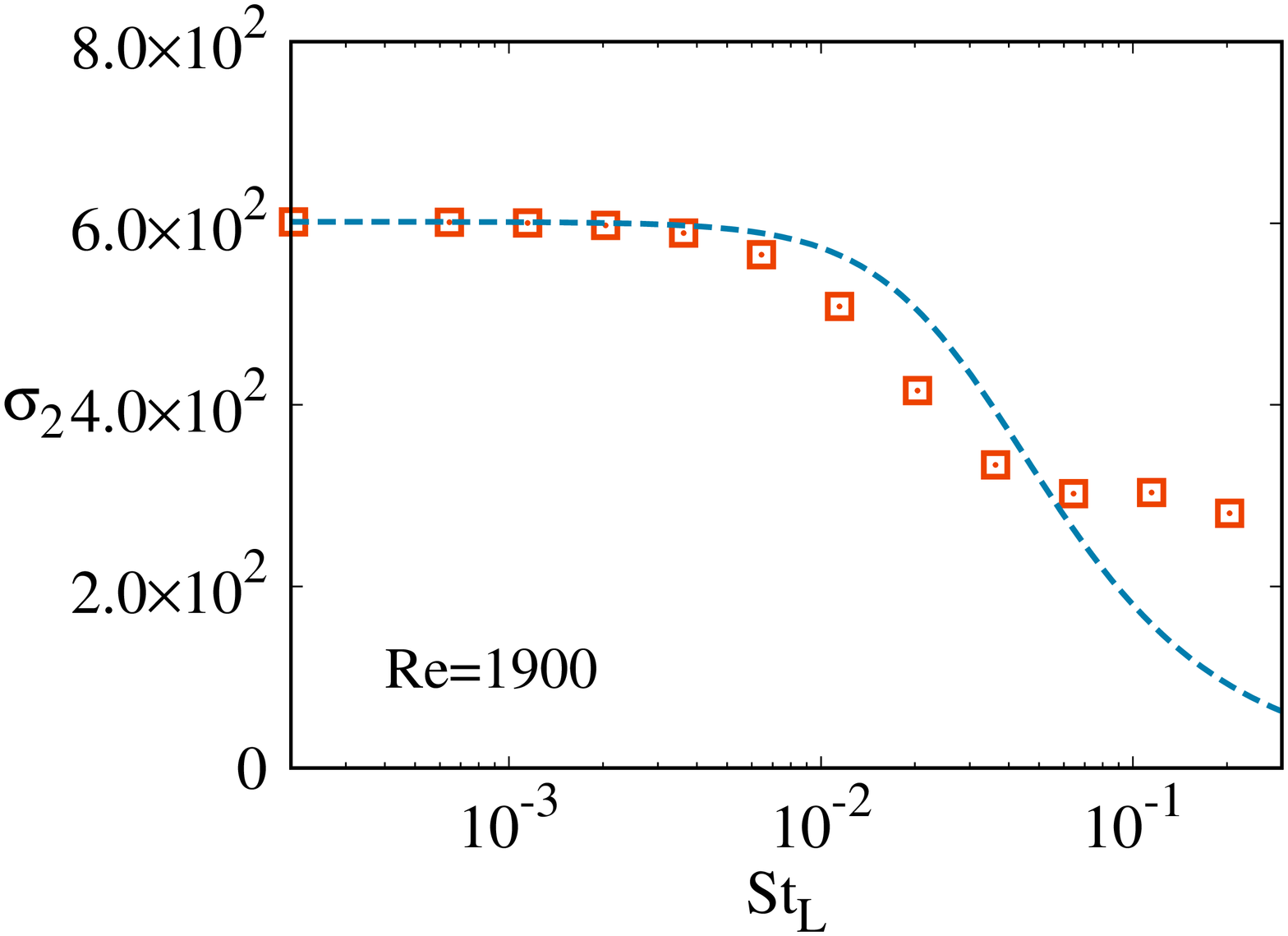}      
        \end{subfigure}
        \vspace{-0.05cm}
        \caption{Comparison between the sub-optimal gain $\sigma_2$ (red squares) and the low-pass filter model in equation \eqref{eq.lowpassfilter2} (blue line) for different Reynolds numbers ($M=2.20$).}
         \label{fig.model2ndmode_influenceRe}
        \end{figure}

        This model is tested for the same set of base flow as previously investigated.
        At $\Reynd=1100$, the location of the drop of the sub-optimal gain $\sigma_2$, observed shortly after $\StL = 10^{-2}$, is captured for the three Mach numbers considered as seen in figure \ref{fig.model2ndmode_influenceM}.
        This means that the cut-off frequency of the low-pass filter, based on the damping rate of the global mode T1, is indeed relevant.
        The slope of the drop is also predicted by the first-order filter until $\StL= 10^{-1}$, except for $M=2.10$ for which the behaviour of $\sigma_2$ changes at lower frequencies.
        Now setting $M=2.20$, the agreement of the cut-off frequency is also correctly captured for Reynolds numbers lower than or equal to $\Reynd=1100$ (figure \ref{fig.model2ndmode_influenceRe}).
        At higher Reynolds number, the model over-estimates this frequency while still reasonably characterising the behaviour at $\Reynd=1600$ .
        For $\Reynd=1900$, the model completely fails (figure \ref{fig.model2ndmode_influenceRe}-bottom), as already noted for $\sigma_1$.

        \subsubsection{Scaling}
\begin{figure}
            \centering
            \includegraphics[angle=-0,trim=0 0 0 0, clip,width=0.67\textwidth]{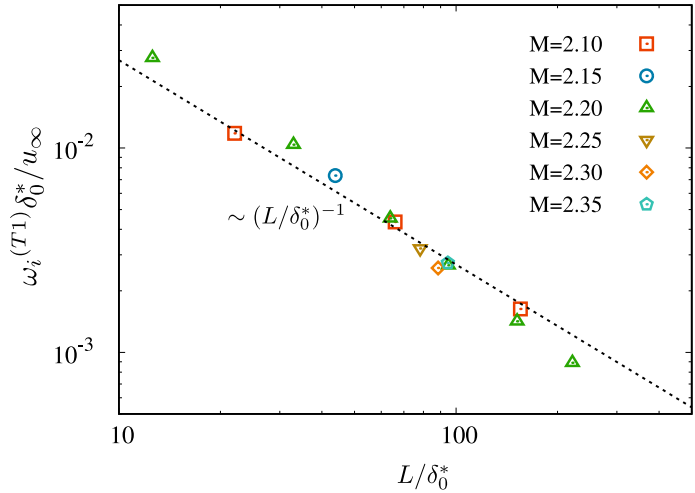}
            \caption{Damping rate of the global mode T1 for different Mach and Reynolds numbers, as a function of the separation length. Here, $\deltazero$ is the boundary layer thickness at $Re=1100$ rather than that at each Reynolds number.}
            \label{fig.scaling_mode3}
            \end{figure}

        The damping rate $\omega_i^{(T1)}$ of the global mode T1 has been shown to capture reasonably well the low-frequency drop of $\sigma_2$.
        Its scaling is considered by plotting its value against the non-dimensional separation length for the set of Reynolds and Mach numbers studied (figure \ref{fig.scaling_mode3}).
        Similarly to section \ref{sec.scalingMode1}, the damping rate is found to scale with $u_\infty/L$ since $\omega_i^{(T1)} \deltazero / u_\infty \sim (L/\deltazero)^{-1}$. 
        This again means that the Strouhal number $\StL$ associated with cut-off frequency is constant for any Mach and Reynolds numbers.
        Its value is found around  $\StL \simeq 4 \times 10^{-2}$, in good agreement with the range of experimental data of the low-frequency unsteadiness.
        Before discussing the implication of this result, the bubble dynamics is studied in the next subsection.

            % Scaling

    \subsection{Bubble dynamics}

\begin{figure}
        \centering
        \begin{subfigure}[]{0.49\textwidth} 
          \includegraphics[angle=-0,trim=0 65 0 0, clip,width=0.99\textwidth]{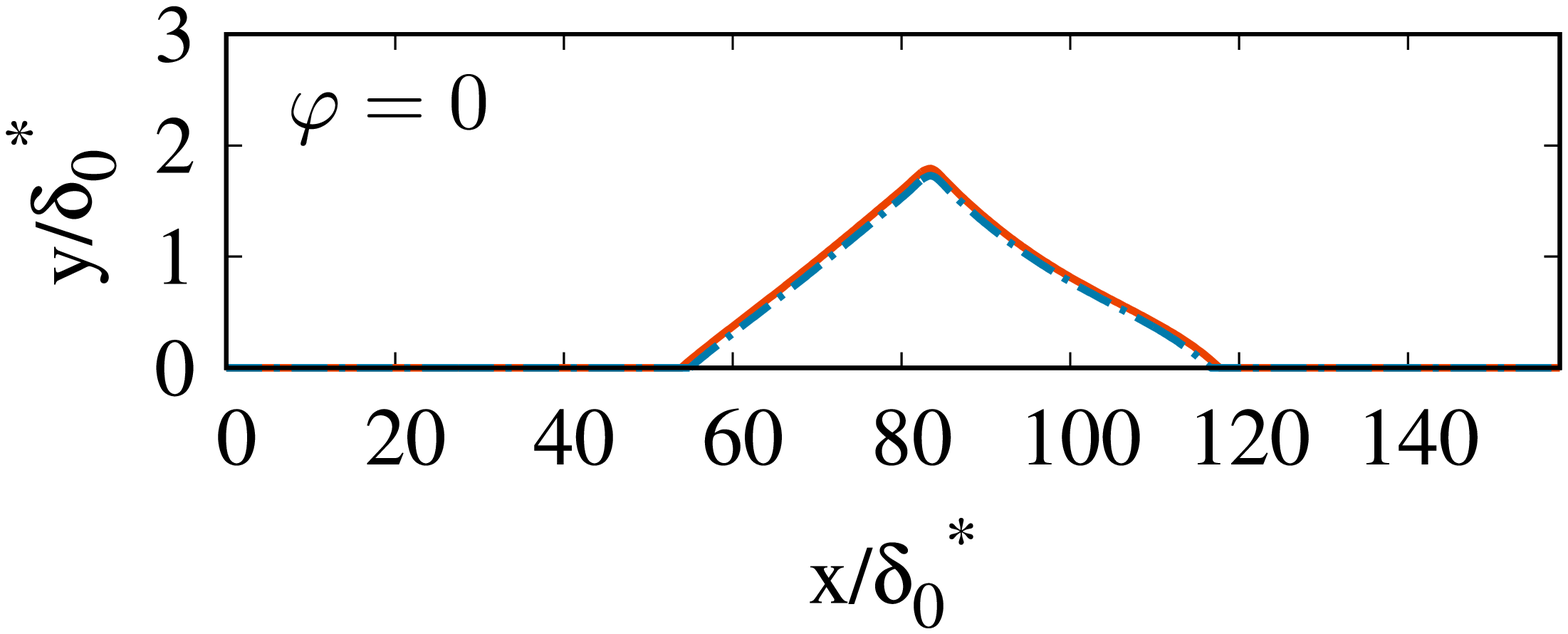}      
        \end{subfigure}        
        \begin{subfigure}[]{0.49\textwidth}  
          \includegraphics[angle=-90,trim=113 63 175 25, clip,width=0.99\textwidth]{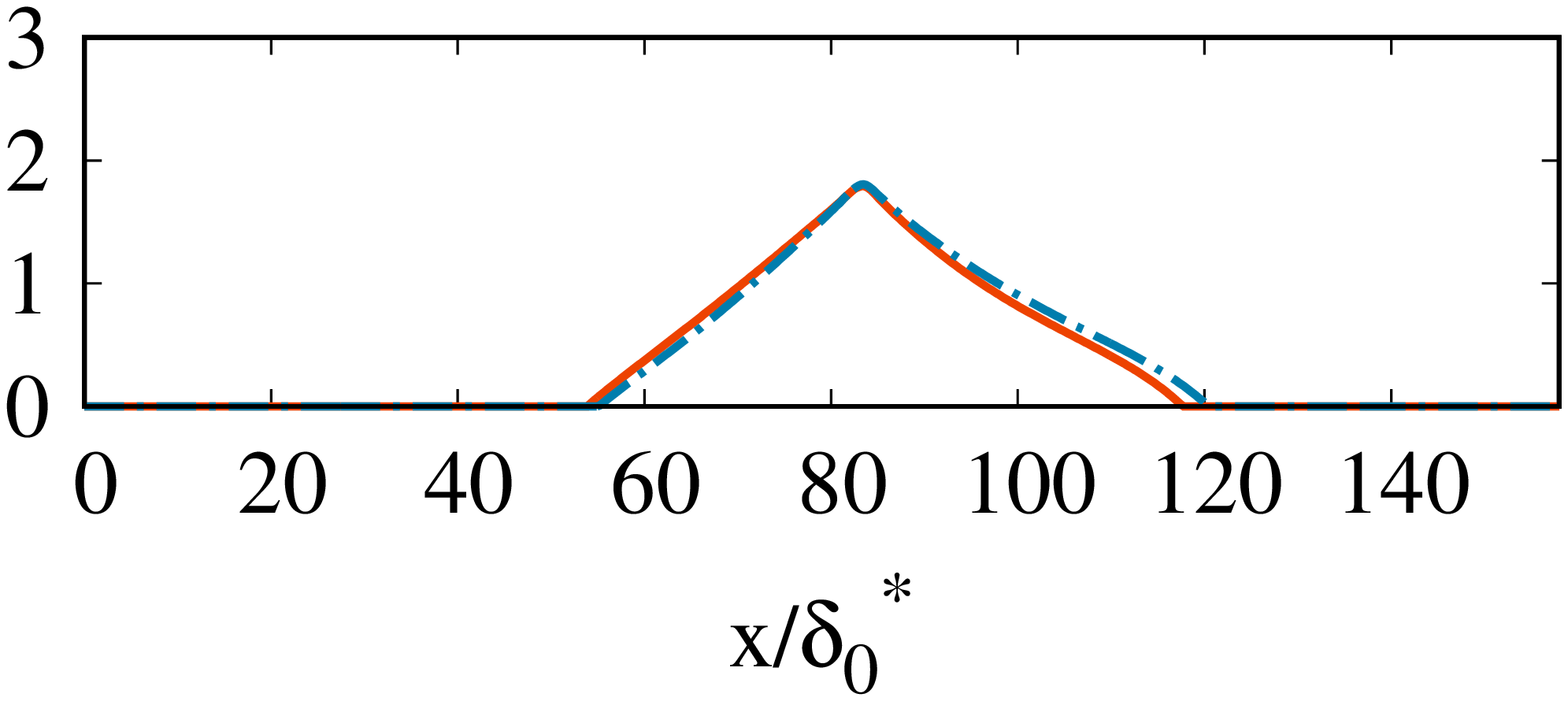}          
        \end{subfigure}

        \begin{subfigure}[]{0.49\textwidth} 
          \includegraphics[angle=-0,trim=0 65 0 0, clip,width=0.99\textwidth]{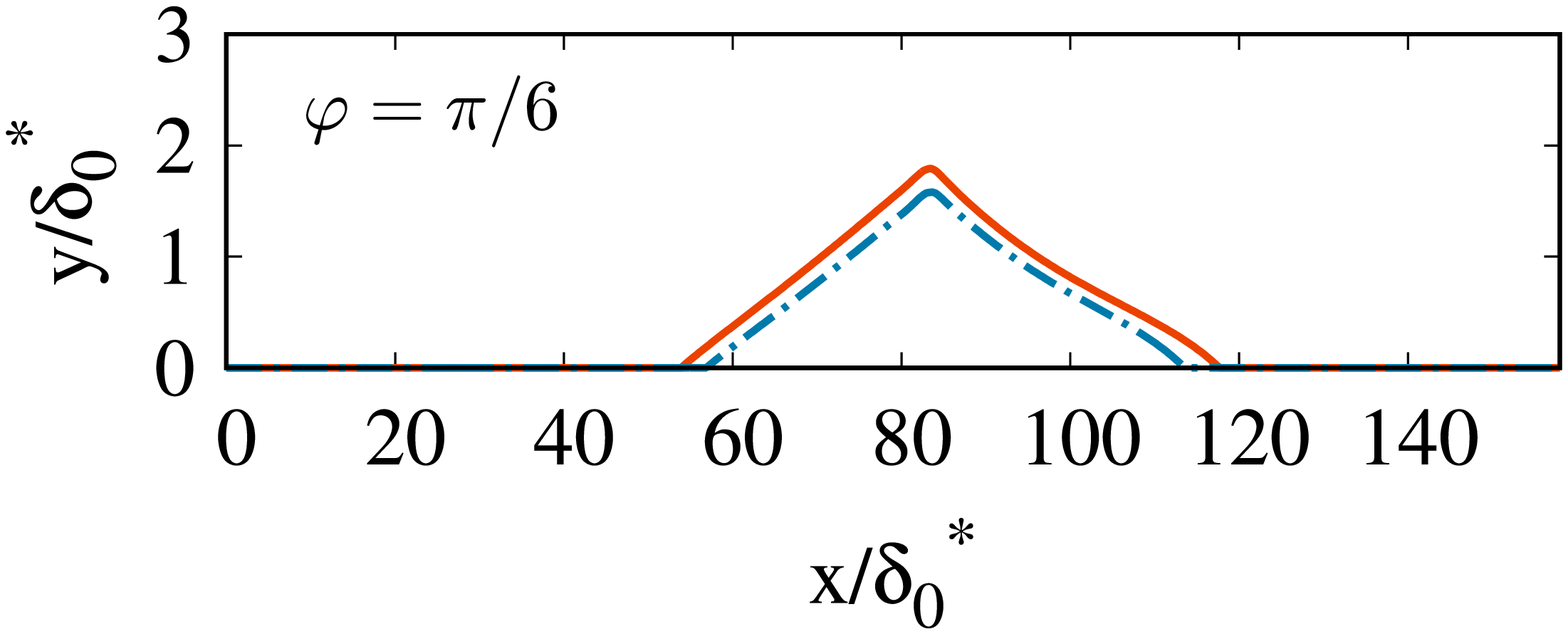}      
        \end{subfigure}        
        \begin{subfigure}[]{0.49\textwidth}  
          \includegraphics[angle=-90,trim=113 63 175 25, clip,width=0.99\textwidth]{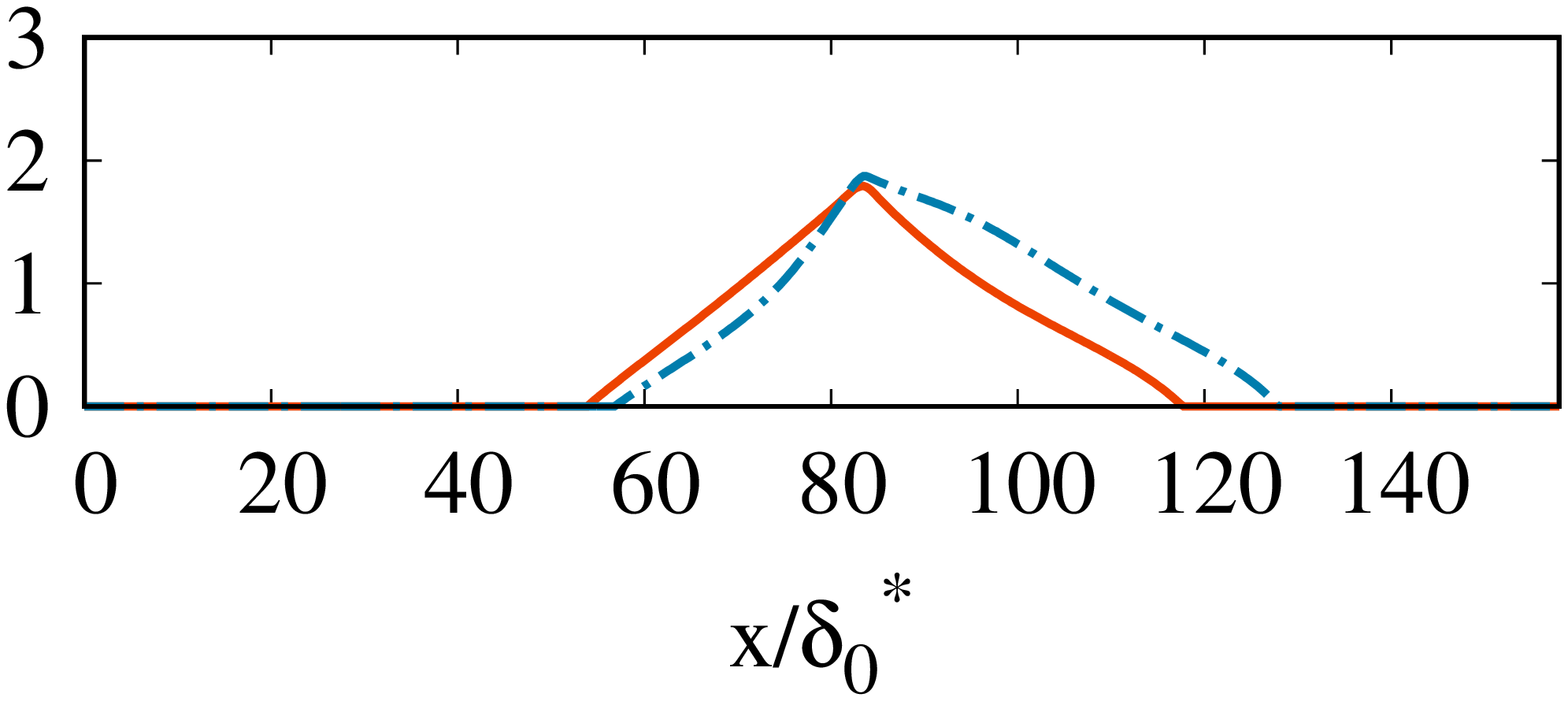}          
        \end{subfigure}

        \begin{subfigure}[]{0.49\textwidth} 
          \includegraphics[angle=-0,trim=0 65 0 0, clip,width=0.99\textwidth]{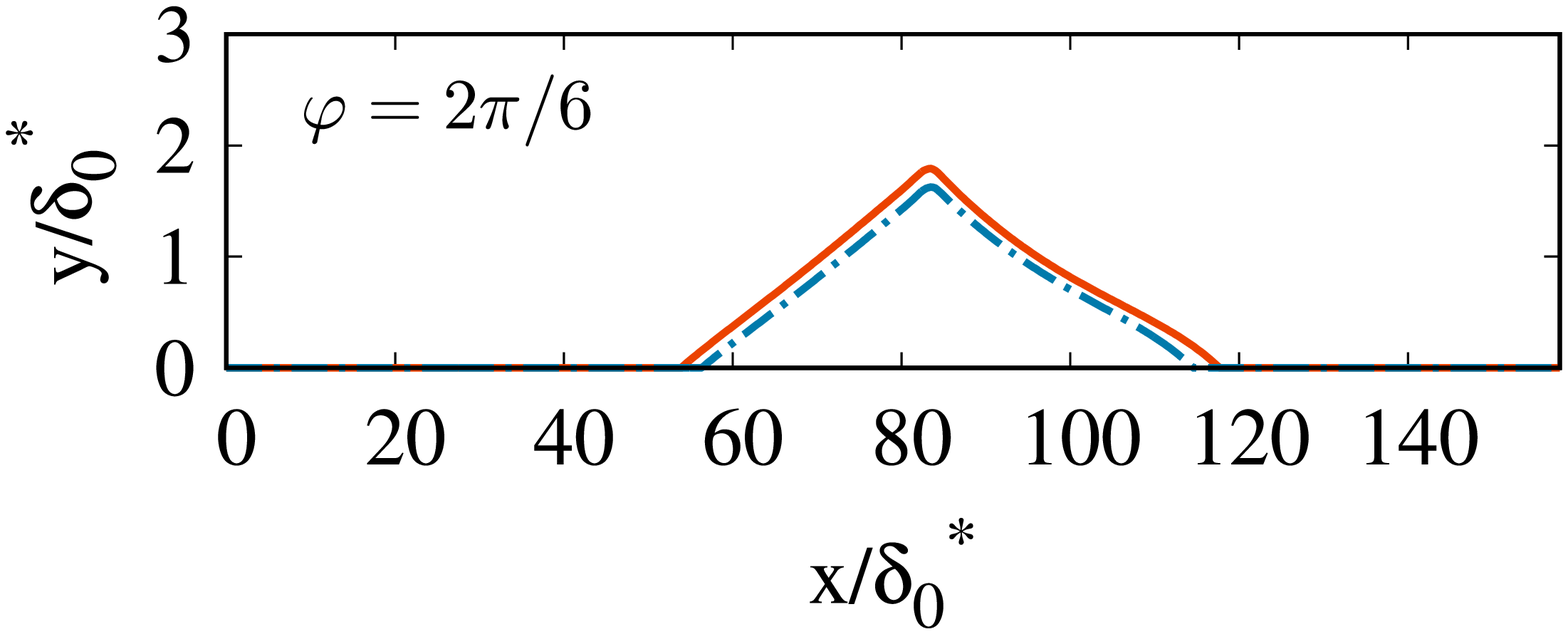}      
        \end{subfigure}        
        \begin{subfigure}[]{0.49\textwidth}  
          \includegraphics[angle=-90,trim=113 63 175 25, clip,width=0.99\textwidth]{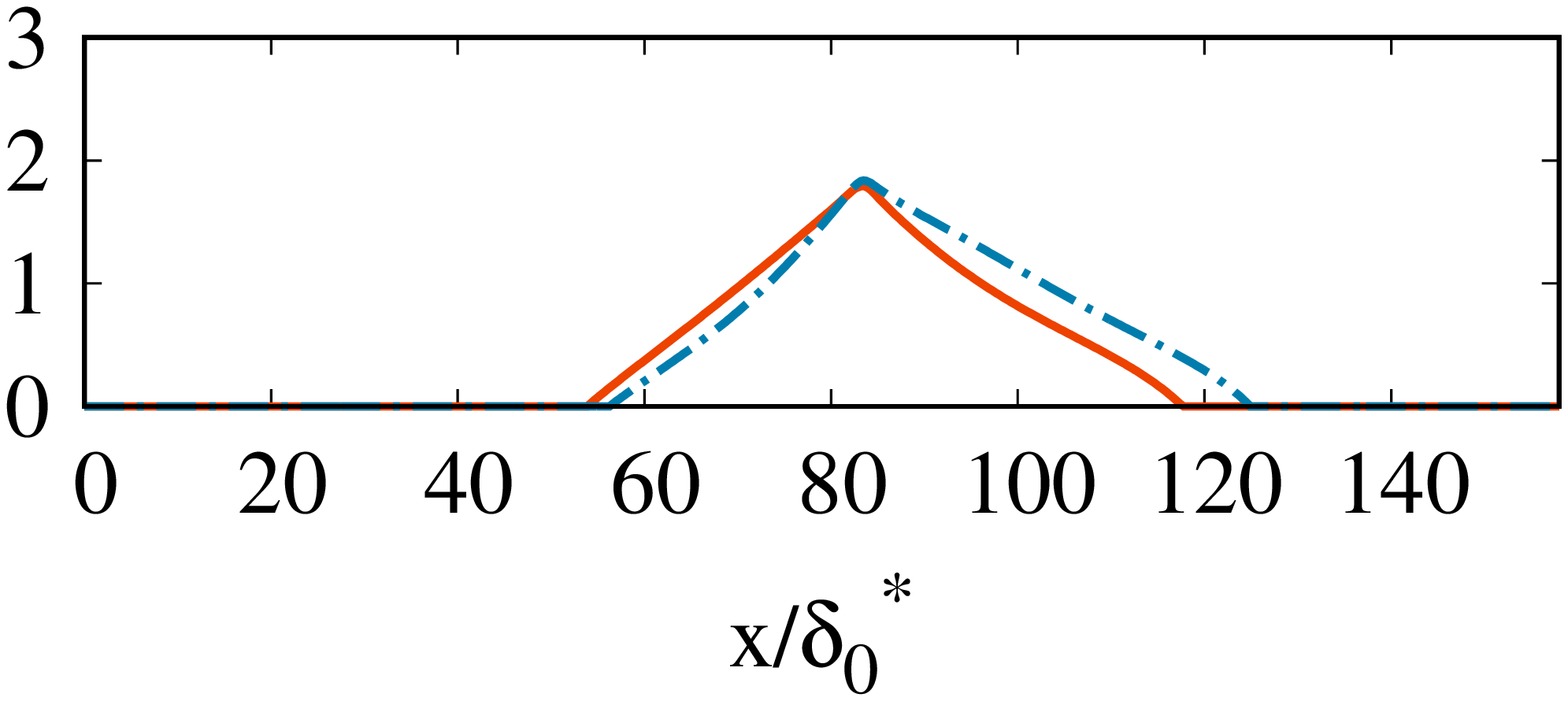}          
        \end{subfigure}

        \begin{subfigure}[]{0.49\textwidth} 
          \includegraphics[angle=-0,trim=0 65 0 0, clip,width=0.99\textwidth]{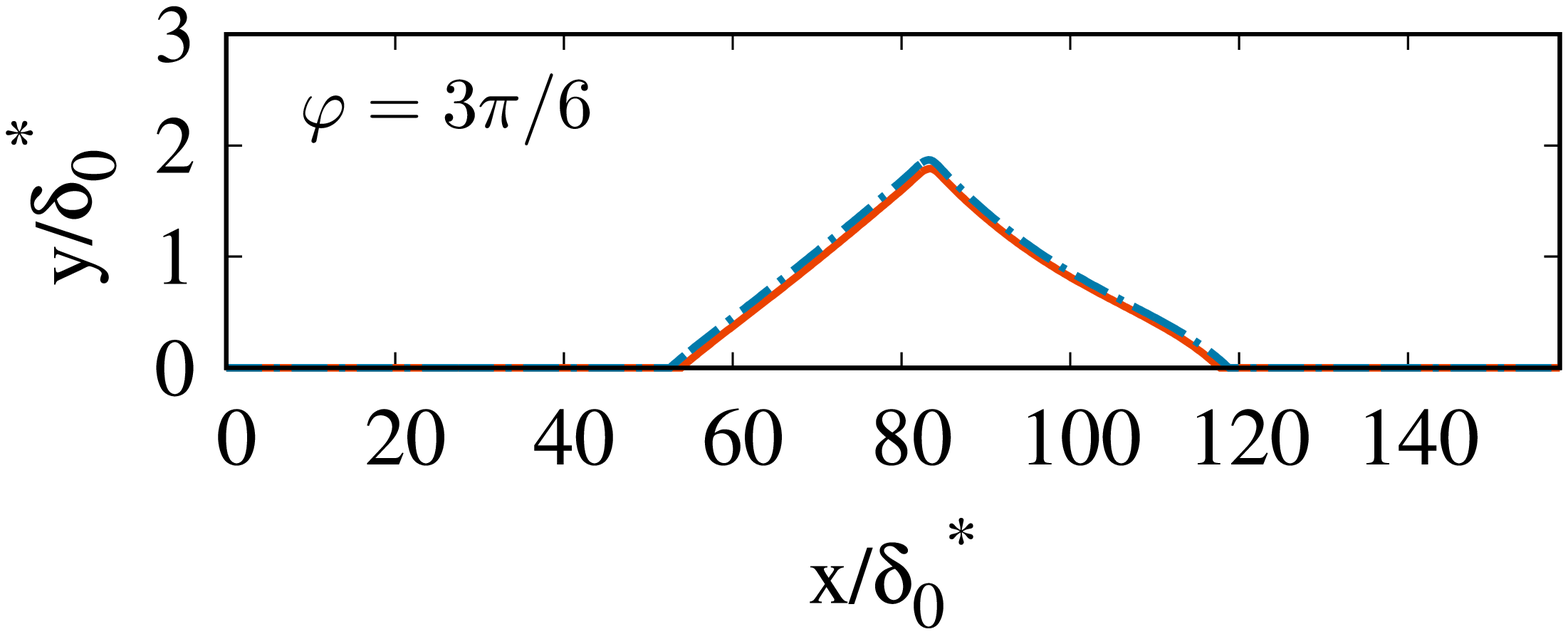}      
        \end{subfigure}        
        \begin{subfigure}[]{0.49\textwidth}  
          \includegraphics[angle=-90,trim=113 63 175 25, clip,width=0.99\textwidth]{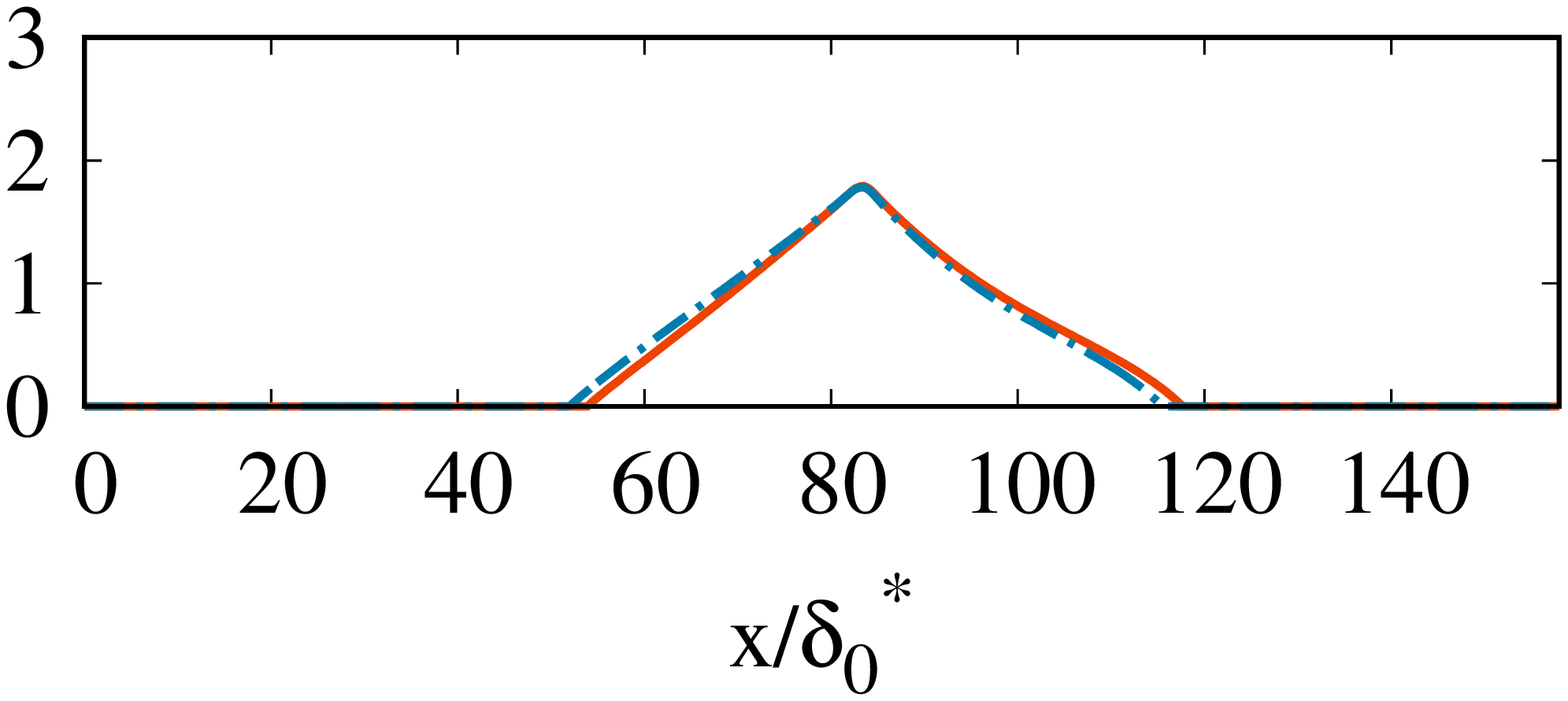}          
        \end{subfigure}

        \begin{subfigure}[]{0.49\textwidth} 
          \includegraphics[angle=-0,trim=0 65 0 0, clip,width=0.99\textwidth]{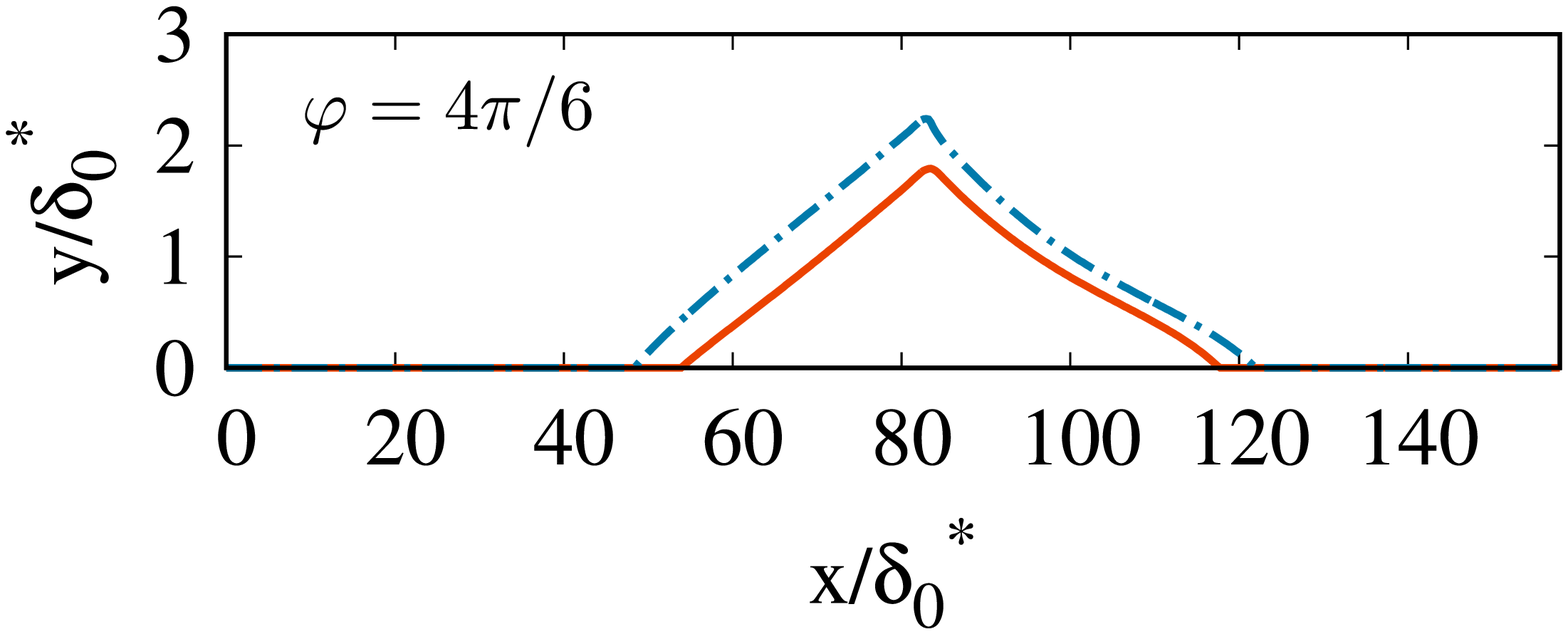}      
        \end{subfigure}        
        \begin{subfigure}[]{0.49\textwidth}  
          \includegraphics[angle=-90,trim=113 63 175 25, clip,width=0.99\textwidth]{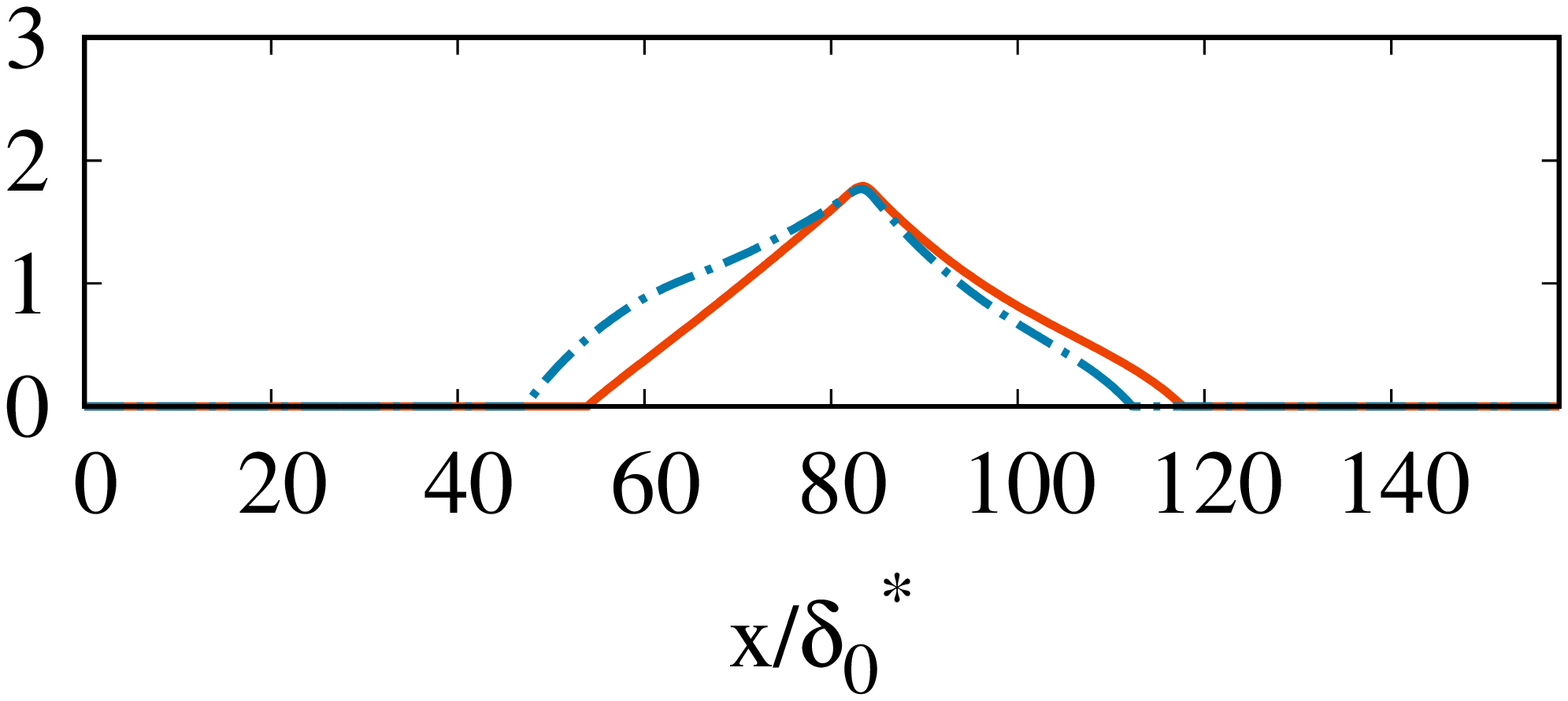}          
        \end{subfigure}

        \begin{subfigure}[]{0.49\textwidth}  
          \includegraphics[angle=-0,trim=0 0 0 0, clip,width=0.99\textwidth]{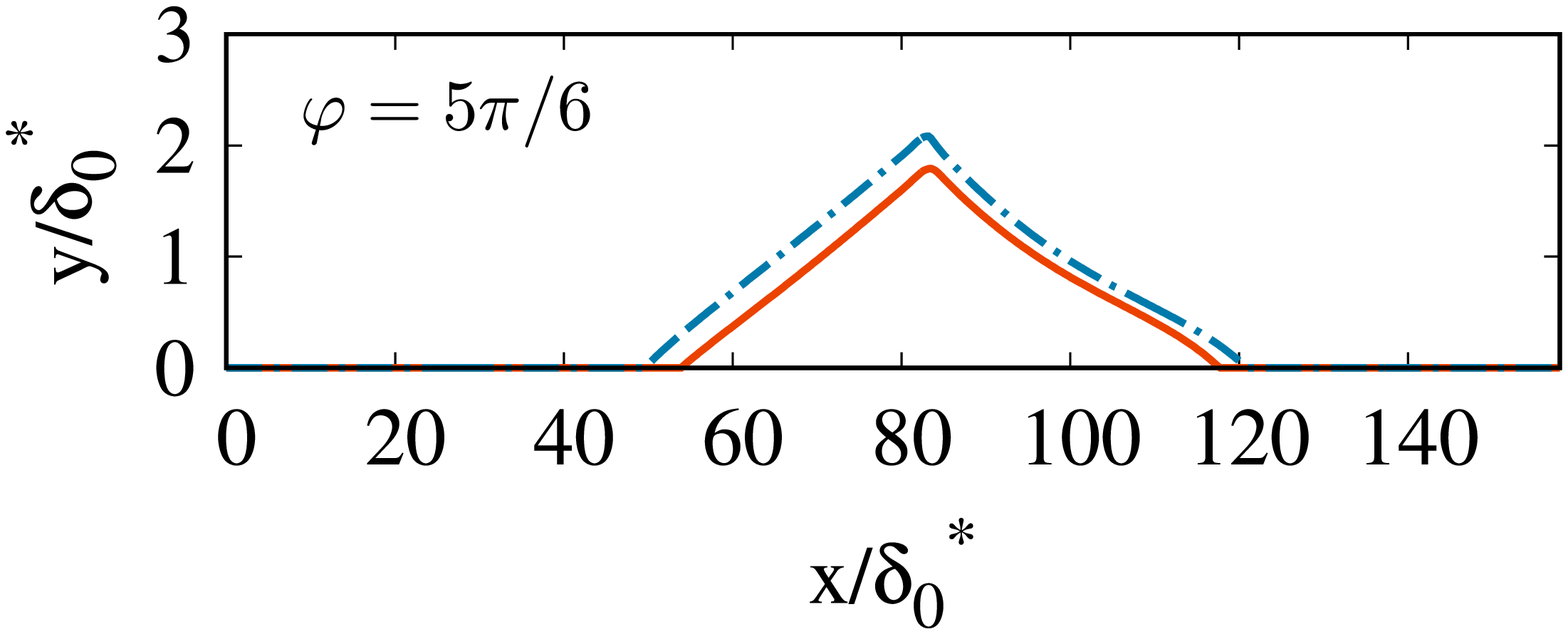}      
        \end{subfigure}
        \begin{subfigure}[]{0.49\textwidth}  
          \includegraphics[angle=-90,trim=123 63 120 25, clip,width=0.99\textwidth]{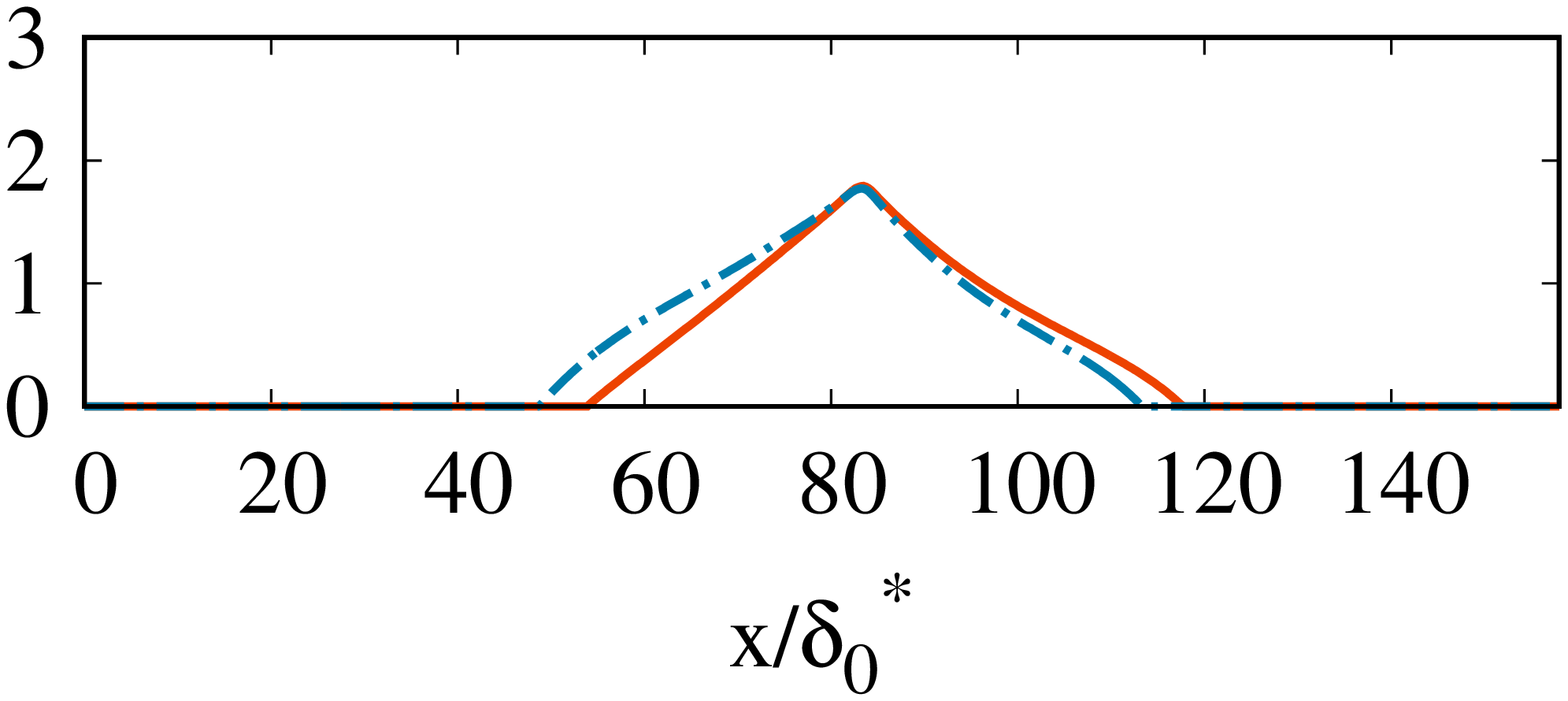}       
        \end{subfigure}
        \vspace{-0.15cm}
        \caption{Bubble dynamics at $M=2.20$ and $Re=1100$ resulting from the addition of the optimal (left) and sub-optimal (right) low-frequency resolvent modes to the base flow. The blue dashed lines is the distorted recirculation bubble corresponding to the location where $\bar{u}+A \hat{u}=0$ (using $A=0.3$). Its evolution at 6 different phases of a periodic cycle is shown from top to bottom. The red line is the recirculation bubble of the base flow ($\bar{u}=0)$.}
        \label{fig.bubble_dynamics}
    \end{figure}

    The first and second resolvent modes are now compared from the perspective of the bubble dynamics. 
    In order to observe its motion, one response mode at one chosen frequency is linearly added to the base flow.
    The frequency considered here is $\StL=10^{-3}$ but its value is actually not important since, as previously mentioned, the resolvent modes do not depend on $\StL$ at low frequency. 
    Snapshots at six different time steps of a periodic cycle, i.e. six different phases, are shown in figure \ref{fig.bubble_dynamics}.
    In each case, the distorted recirculation bubble is shown.
    It corresponds to the location in space where the total streamwise velocity is zero, calculated as $\bar{u} + A \hat{u} = 0$. 
    The scalar $A$ is an arbitrary amplitude chosen as $A=0.3$ to ease visualisation. 
    The undistorted bubble of the base flow ($\bar{u}=0$) is also plotted as a reference.
    The optimal resolvent mode, associated with $\sigma_1$, produces a breathing motion: the bubble periodically contracts and expands relatively to that of the basic state (figure \ref{fig.bubble_dynamics}-left).
    This dynamics is similar to the bubble motion found in the model of the low-frequency unsteadiness developed by \cite{PDDD09}.
    The dynamics produced by the sub-optimal mode, shown in figure \ref{fig.bubble_dynamics}-right, is different.
    The bubble is now periodically displaced upstream and downstream from the reference state.
    No displacement is observed in the wall-normal direction.

    \subsection{Discussion}

    In their review paper, \cite{clemens2014low} ended their discussion on whether the SWBLI behaved as a forced dynamical system at low-frequency.
    Our resolvent analysis supports the idea that, indeed, the low-frequency behaviour results from a forced dynamics, as opposed to an intrinsic dynamics such as that of a vortex street behind cylinder.
    From the resemblance between their DMD and stable global stability modes, \cite{nichols2017stability} suggested that the system could then be seen as a damped oscillator.
    Our study brings some new lights to this idea by showing that one stable global mode drives the resolvent analysis at low frequency.
    The fact that the optimal gain follows a first-order low-pass filter equation is furthermore consistent with the work of \cite{TS2011} who were the first to model the low-frequency dynamics of the SWBLI as a filter of this kind.
    From this perspective, the dynamics requires a range of low frequencies that force the system and are filtered by the transfer function which, in the resolvent framework, is related to the singular values of the resolvent operator.
    Therefore, while carried out in the laminar regime, our analysis provides insights that the SWBLI in the turbulent regime could possess a forced dynamics at low frequency, in which fluctuations over a wide range of scales would force the flow \citep{mckeon2010critical} and sustain its low-frequency behaviour.
    This would explain why low-frequency unsteadiness is observed in the turbulent regime rather that the laminar regime unless the later is externally forced as in \cite{SSH2014}.
    In the present study, because resolvent analysis assumes by nature the presence of a forcing field, the low-frequency behaviour of the system can be qualitatively recovered even though the flow is laminar.

    For both the optimal and sub-optimal gain, the damping rate of one stable global mode provides the time scale of the low-pass filter (the cut-off frequency).
    These frequencies scale as $(\Lsep / u_\infty)^{-1}$, which was expected as these modes are related to the bubble dynamic.
    This results in a constant cut-off Strouhal number $\StL$ for any Mach and Reynolds numbers.
    This scaling is indeed observed both experimentally and numerically for different configuration of SWBLI, with frequencies reported from $\StL \simeq 3 \times \tento{-2}$ to $\StL \simeq 6 \times \tento{-2}$ \citep{dussauge2006unsteadiness}.
    With values presently found around $\StL^{(1)} \simeq  1 \times \tento{-2}$ and $\StL^{(2)} \simeq 4 \times \tento{-2}$ for the optimal and sub-optimal gain respectively, $\StL^{(2)}$ compares more favourably to the literature.
    However, the dynamics of the bubble of the optimal mode is, on one hand, related to a breathing motion that strongly resembles that at play in the low-frequency unsteadiness of the SWBLI \citep{PDDD09,clemens2014low}.
    It is also strikingly similar to that observed experimentally in a turbulent incompressible recirculation bubble \citep{WMTS2015}, also occurring at Strouhal numbers of the order of $10^{-2}$.
    On the other hand, the sub-optimal mode that produces a very different type of motion, which is expected given the orthogonality of the resolvent modes.
    The resulting bubble motion is similar to that obtained by \cite{nichols2017stability} in their global stability analysis; this is indeed the same unsteady stable mode.
    However, no conclusive element allows us to link it to the low-frequency behaviour reported in the literature.
    This leads us to hypothesize that the optimal (rather than sub-optimal) gain and modes, associated with the bubble breathing, indeed describe the low-frequency dynamics of the SWBLI.
    The fact that $\StL^{(1)}$ is underestimated in the present study may be due to the laminar nature of the SWBLI understudied while results in the literature usually deals with turbulent SWBLI.
    The shape of the bubbles being different between laminar and turbulent regimes this could account for the quantitative mismatch.
    Finally, while the range of values of separation length studied in the present work is large enough
    to characterise the scaling of the low-frequency time scale, the Mach number is restricted to a small
    range around M = 2.20. 
    Similar analyses carried out at much different Mach
    numbers are then still needed to assess the universality of these findings.

  The larger the gain separation $\sigma_1 / \sigma_2$ between the optimal and sub-optimal gain, the more likely the optimal response obtained through resolvent analysis is to model the system.
  More rigorously, this also depends on the projection of the effective forcing of the system (whether it is from external perturbations or intrinsic, turbulent non-linear interactions) onto the forcing resolvent modes \citep{beneddine2016conditions}.
  Here, $\sigma_1 / \sigma_2 \simeq 3$ is found at low frequency.
  This value not being significantly greater than 1, if we assume that the optimal response drives the system (which we do based on the previous discussions on the bubble dynamics), then the actual forcing can be assumed to project reasonably well onto the optimal forcing mode.
  This gives us some insight on the actual forcing in the flow.
  The optimal forcing is indeed maximum around the separation point and along the separation line on the downstream part of the bubble (figure \ref{fig.resMode1_1Em4}).
  Thus, some forcing is required in these regions in order to maintain the low-frequency bubble dynamics.
  The limit of the resolvent analysis is eventually reached when seeking for the origin of the forcing. 
  Because of its linear nature, this framework does not allow us to discuss how the low-frequency structures can be produced by non-linear interactions.
  Thus, both boundary layer structures convected to the interaction region \citep{GCD2007} exciting the separation point and upstream travelling mechanisms \citep{SSH2016,BBGBMSJ2019} forcing the downstream part of the recirculation bubble could be valid candidates to sustain the low-frequency dynamics of the SWBLI.

%%%%%%%%%%%%%%%%%%%%%%%%%%%%%%%%%%%%%%%%%%%%%%%%%%%%%%%%%%%%%%%%%%%%%%%%%%%%%%%%%%%%%%%%%%%%%%%%%%%%%%%%%%%%%%%%%%%%%%%%%%%%%%%%
%%%%%%%%%%%%%%%%%%%%%%%%%%%%%%%%%%%%%%%%%%%%%%%%%%%%%%%%%%%%%%%%%%%%%%%%%%%%%%%%%%%%%%%%%%%%%%%%%%%%%%%%%%%%%%%%%%%%%%%%%%%%%%%%

\section{Linear dynamics of 3D perturbation at low frequency} \label{sec.3Dperturbations}

    \subsection{Resolvent analysis} \label{sec.resolvent3D}

\begin{figure}
        \centering
        \includegraphics[angle=-90,trim=0 0 0 0, clip,width=0.5\textwidth]{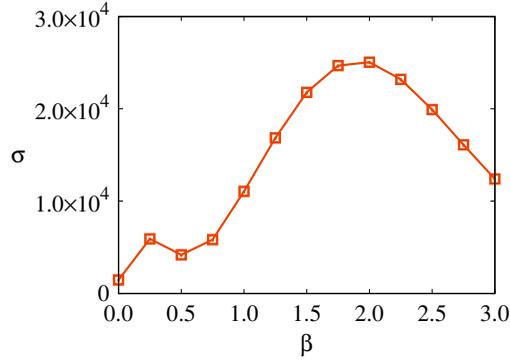}
        % \vspace*{0.3cm}
        \caption{Optimal gain as a function of the spanwise wave number at $St=10^{-4}$ ($Re=1100$, $M=2.20$).}
        \label{fig.gain3D_fBeta}
    \end{figure}

\begin{figure}
        \centering
        \begin{subfigure}[]{0.495\textwidth}  
        \includegraphics[angle=-0,trim=0 0 0 0, clip,width=1\textwidth]{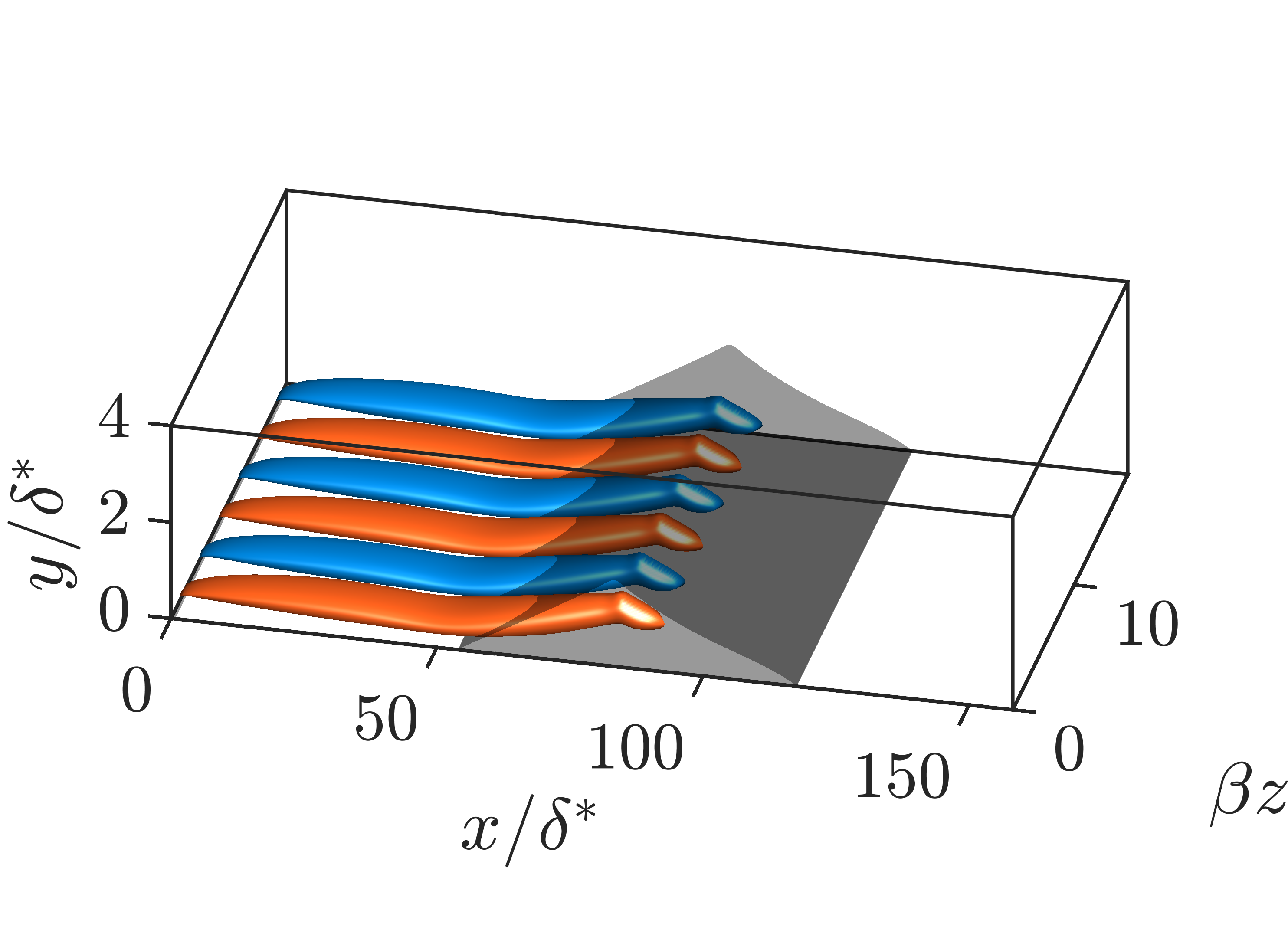}      
        \end{subfigure}
        \begin{subfigure}[]{0.495\textwidth}  
        \includegraphics[angle=-0,trim=0 0 0 0, clip,width=1\textwidth]{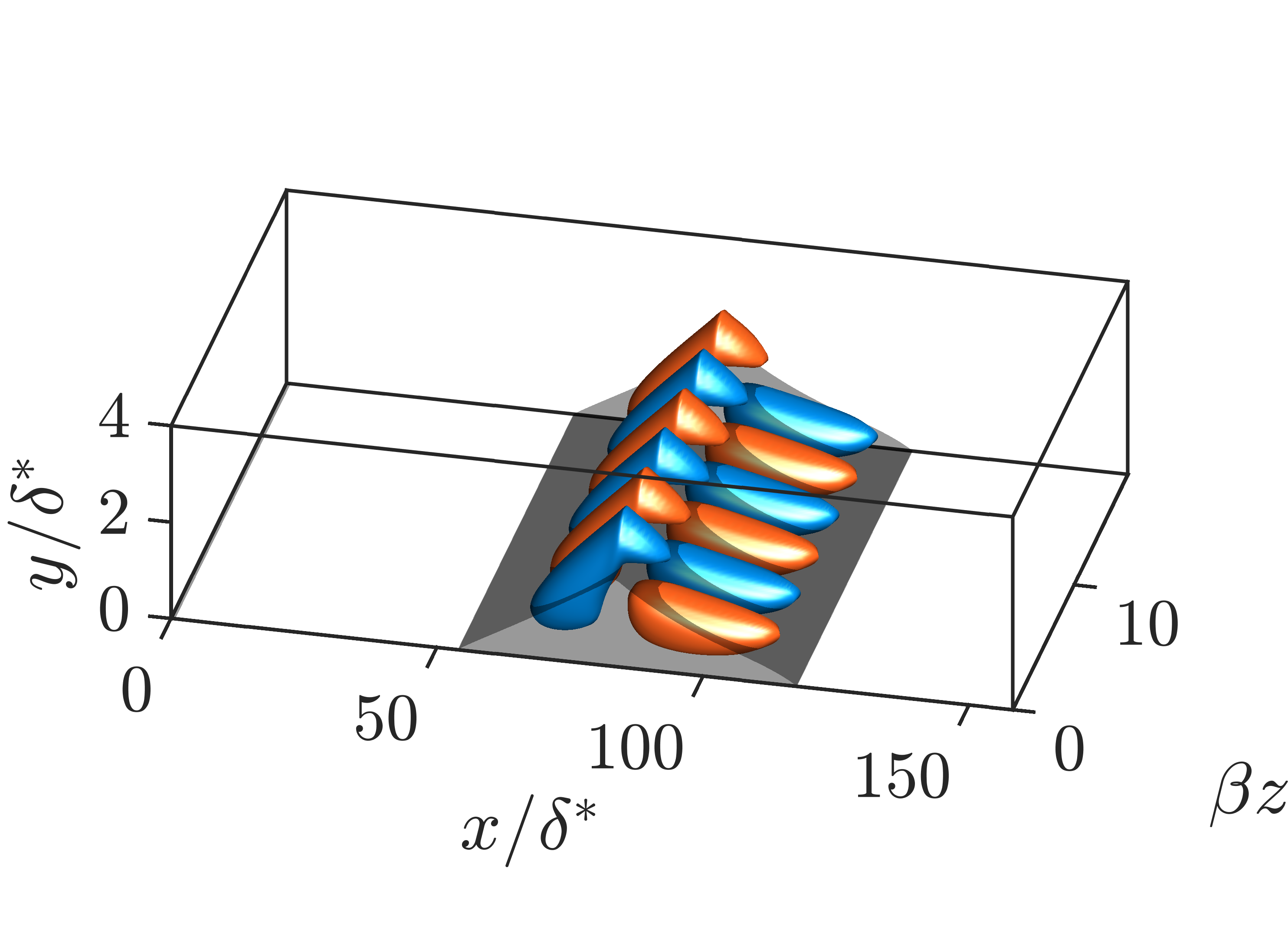}      
        \end{subfigure}
        \caption{Optimal resolvent mode at $\beta=0.25$ and $\StL=10^{-4}$, real part of $w$. Iso-surfaces at $40 \%$ and $-40 \%$ the maximum absolute value are plotted in red and blue. The dark grey surface is the recirculation bubble of the base flow (homogeneous in $z$).}
        \label{fig.resMode3D_beta025}
    \end{figure}

\begin{figure}
        \centering
        \begin{subfigure}[]{0.495\textwidth}  
          \includegraphics[angle=-90,trim=0 0 0 0, clip,width=1\textwidth]{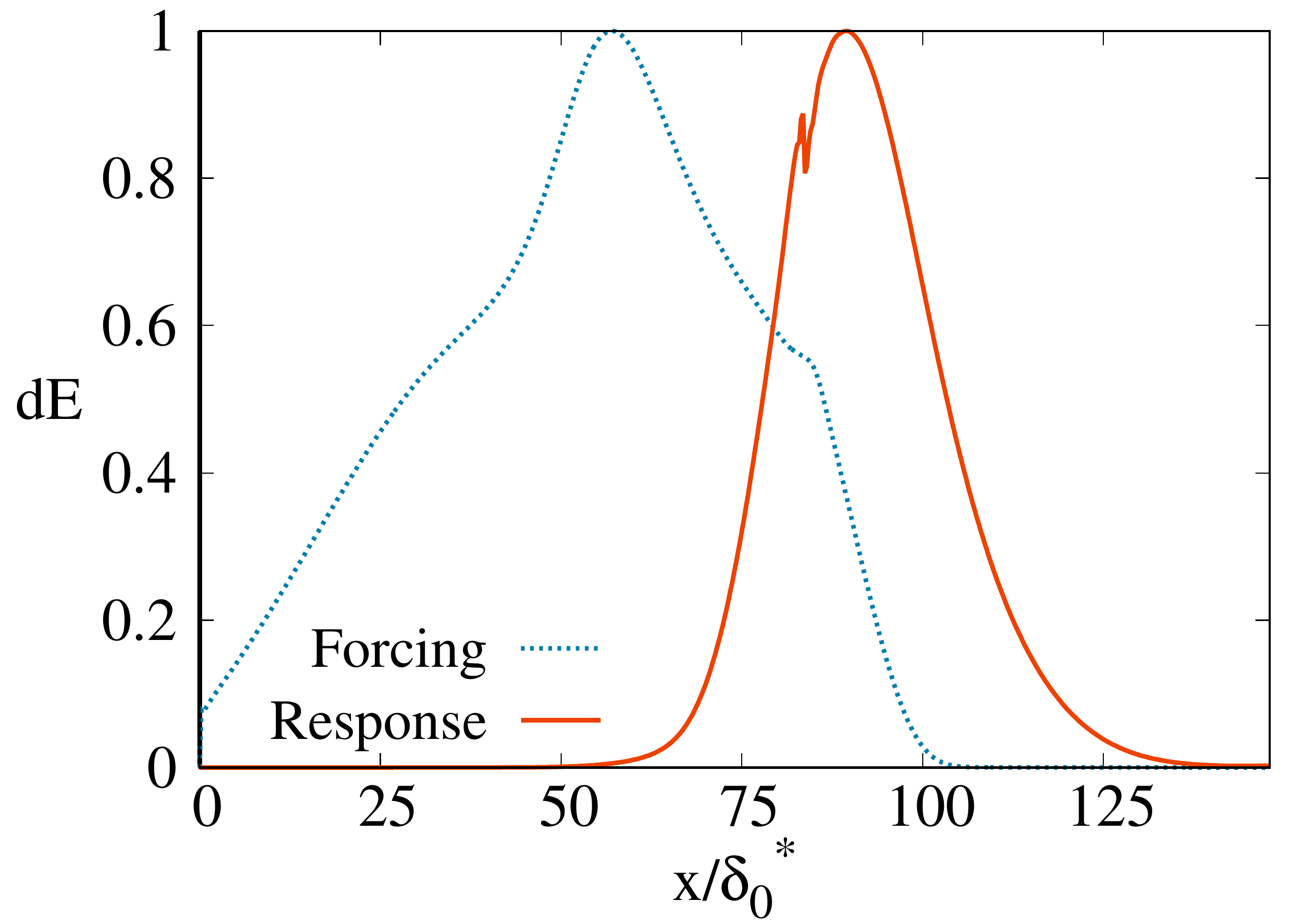}      
        \end{subfigure}
         \begin{subfigure}[]{0.495\textwidth}  
          \includegraphics[angle=-90,trim=0 0 0 0, clip,width=1\textwidth]{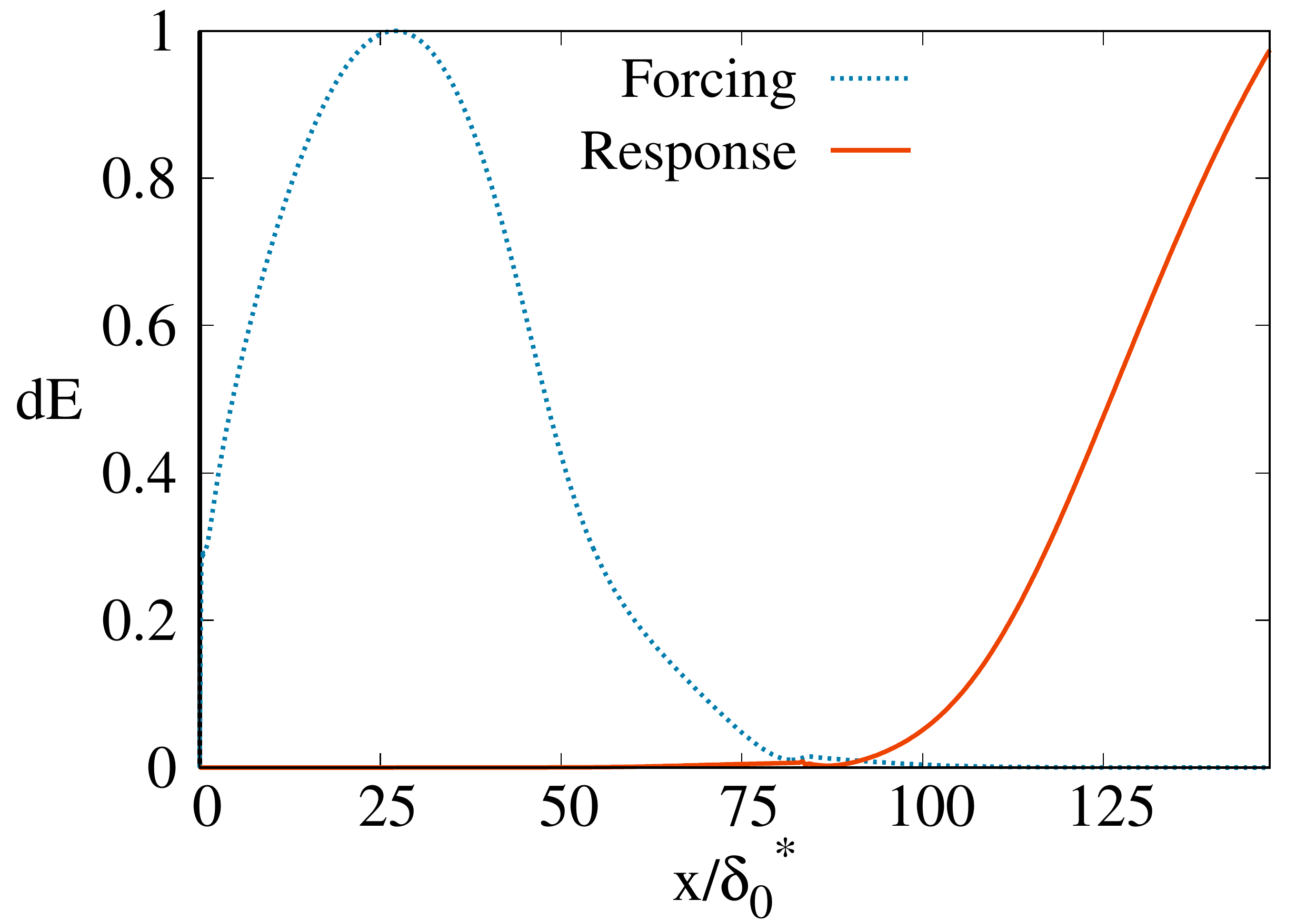}      
        \end{subfigure}
        \caption{Energy density profiles of the optimal resolvent modes at $\beta=0.25$ (left) and $\beta=2$ (right), $\StL=10^{-4}$. The profiles are normalised by their maximum value.}
         \label{fig.resMode3D_energyProfiles}
    \end{figure}

    The low-frequency dynamics of three-dimensional perturbations, characterised by their non-zero spanwise wave number $\beta$, is studied in this section. 
    The analysis is here limited to the base flow calculated at $M=2.20$ and $\Reynd=1100$.
    The influence of $\beta$ on the optimal gain at $\StL = 10^{-4}$ is shown in figure \ref{fig.gain3D_fBeta}.
    A first peak is observed for low wave numbers around $\beta = 0.25$.
    The corresponding wavelength is $\lambda_z = 25 \deltazero$, which is of the order of the recirculation length (approximately half of it, see table \ref{tab.baseflows}).
    The associated optimal forcing and response can be visualised by plotting the isosurface of a selected amplitude in the 3D physical space (figure \ref{fig.resMode3D_beta025}).
    The optimal response is located in the recirculation region and features a phase shift in the streamwise direction.
    A strong resemblance can be noted with the 3D global mode found by \cite{R07}.
    The possibility of a modal resonance to model the optimal gain, as developed previously for 2D perturbation, will be explored in the next section.
    The optimal forcing field is made of elongated structures that lay upstream from the recirculation bubble until the separation point.
    The energy density $\mathrm{d}E(x)$ of the response can be defined at each streamwise location as $\mathrm{d}E(x) = \int 0.5 \rho \left( u u^* + v v^* + w w^* \right) \mathrm{d}y$.
    An analogous quantity can be defined for the forcing field.
    The profiles of these energy densities confirm that the response is localised in the separation region as it experiences a rapid decay downstream from it (figure \ref{fig.resMode3D_energyProfiles}).
    The forcing energy is shown to spread over a larger portion of the domain and reaches a peak in the vicinity of the separation point.

\begin{figure}
        \centering
        \begin{subfigure}[]{0.495\textwidth}  
          \includegraphics[angle=-0,trim=0 0 0 0, clip,width=1\textwidth]{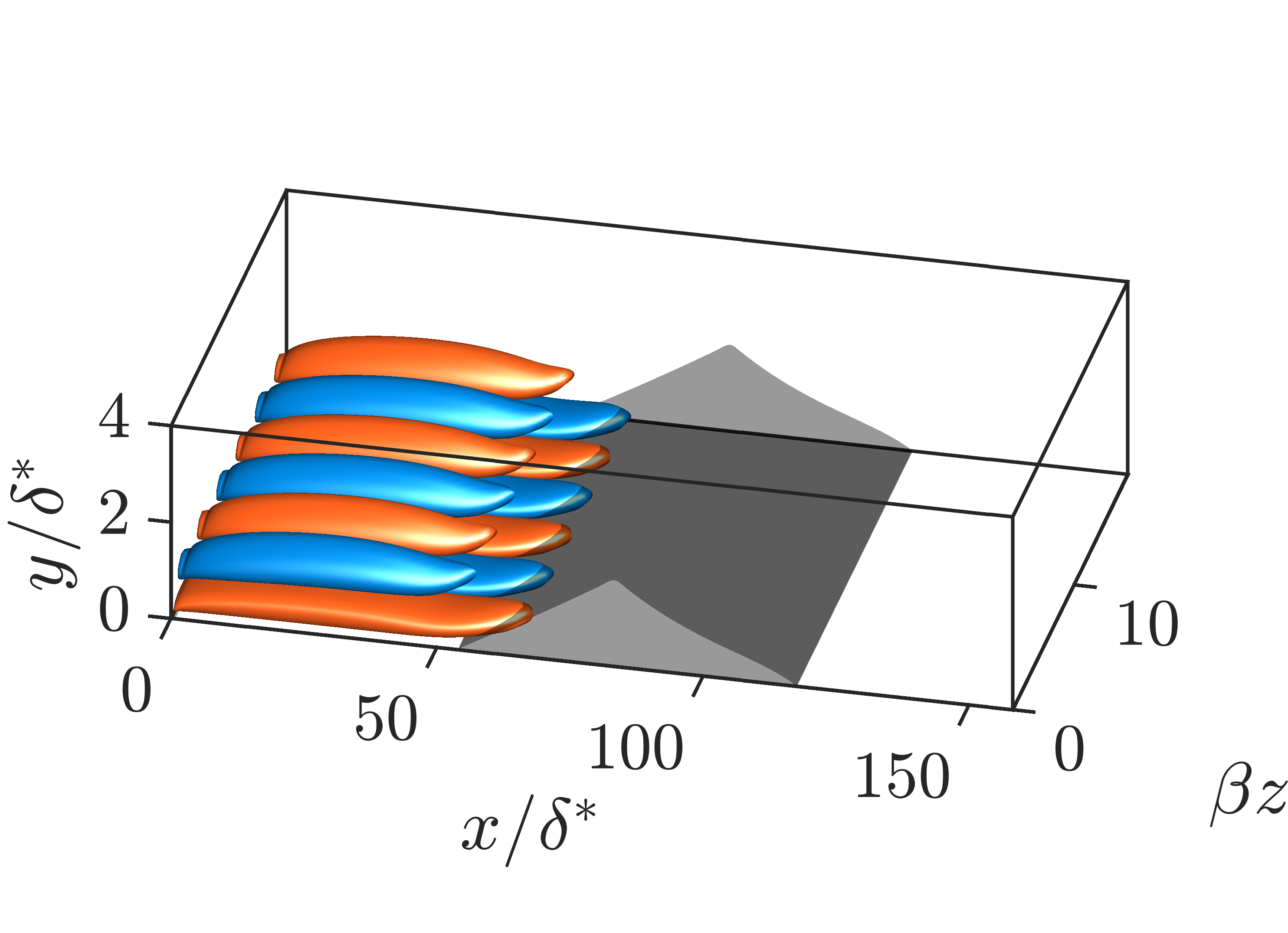}      
        \end{subfigure}
         \begin{subfigure}[]{0.495\textwidth}  
          \includegraphics[angle=-0,trim=0 0 0 0, clip,width=1\textwidth]{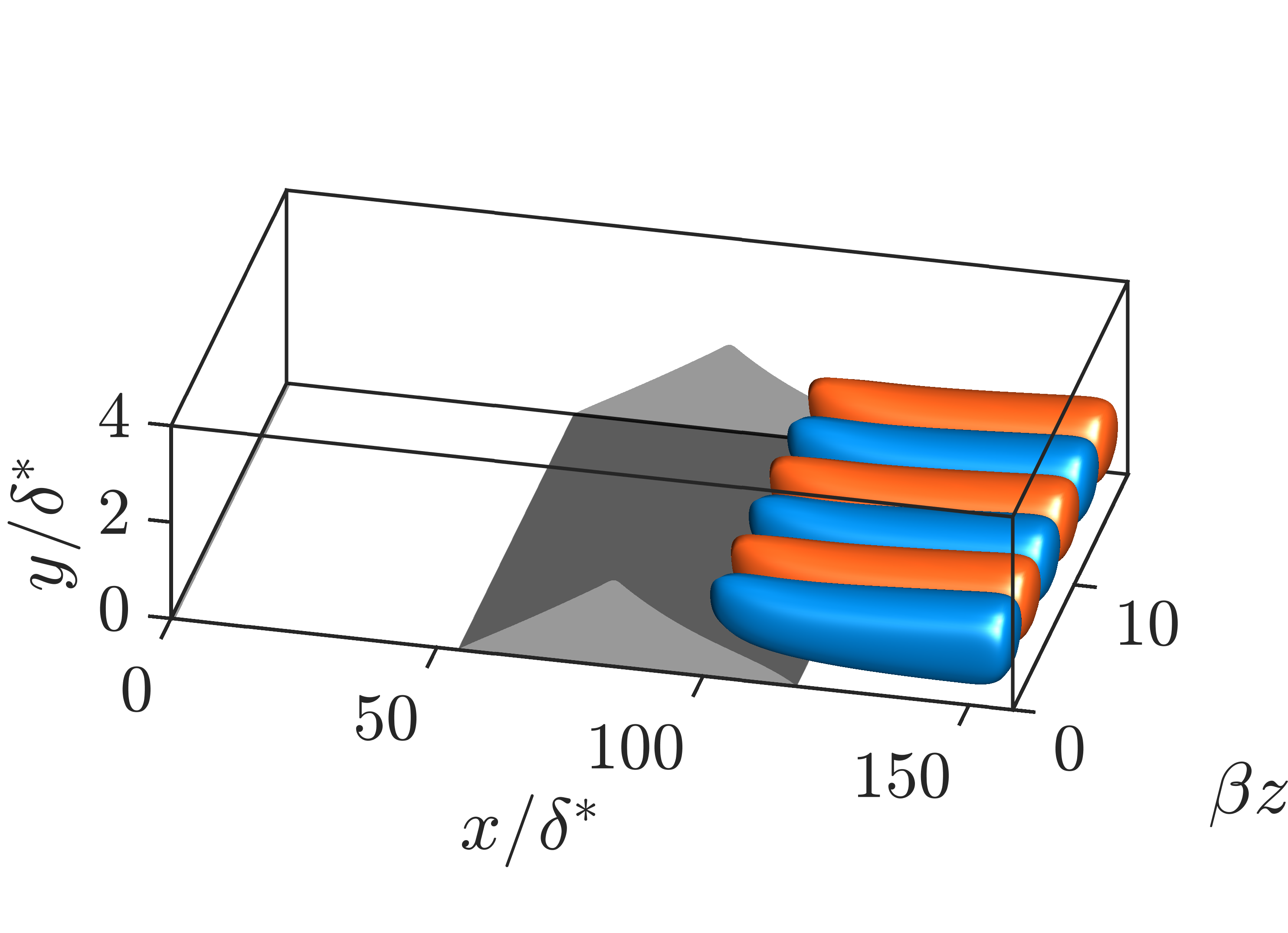}      
        \end{subfigure}
        \caption{Optimal resolvent mode at $\beta=2$ and $\StL=10^{-4}$, real part of $w$ of the forcing (left) and real part of $u$ of the response (right). Isosurfaces at $20 \%$ and $-20 \%$ of the maximum absolute value are plotted.}
         \label{fig.resMode3D_beta2}
    \end{figure}

    In addition of the local maximum at $\beta=0.25$, the optimal gain features a peak of larger magnitude around $\beta=2$.
    The corresponding wavelength is $\lambda_z = 3 \deltazero$ and is now of the order of the boundary layer thickness.
    Elongated streamwise vortices located upstream of the recirculation bubble are found in the optimal forcing field (figure \ref{fig.resMode3D_beta2}).
    A further inspection of the wall normal organisation of its components (available in appendix \ref{app.Profiles3D}, figure \ref{fig.resMode3D_profileBeta2}) shows that they correspond to streamwise vortices.
    Downstream of the recirculation bubble, streaks of streamwise velocities are observed in the boundary layer, for which profiles reveal that $u>>v$ and $u>>w$ (figure \ref{fig.resMode3D_profileBeta2}).
    These features indicate the action of the lift-up mechanism  that is ubiquitous in shear flows at low frequency and non-zero spanwise wave numbers \citep{landahl1980note}.
    Besides, these results are very similar to the 3D optimal response in a supersonic boundary layer without impinging shock \citep{bugeat20193d}.
    The energy profiles reveal the convective nature of this instability as perturbations keep growing until the downstream end of the domain (figure \ref{fig.resMode3D_energyProfiles}). 
    Most of the energy is contained downstream from the recirculation area whereas most of the forcing energy is localised upstream from it.
    These results hence show that the growth of streaks does not interact significantly with the recirculation region. 
    
    \subsection{Link with global stability}

    Eigenvalue spectra of the global stability problem are given in figure \ref{fig.spectrum3D} for the wave numbers $\beta=0.25$ and $\beta=2$ that correspond to maxima of the optimal gain.
    In both cases, the flow is found to be globally stable.
    Similarly to the analysis developed for 2D perturbations in section \ref{sec.2Dperturbations}, the relation between the optimal gain behaviour at low frequency and the global stability features is explored for these two spanwise wave numbers.

    \subsubsection{Low $\beta$: excitation of a steady global mode}

\begin{figure}
        \centering
        \begin{subfigure}[]{0.495\textwidth}  
          \includegraphics[angle=-0,trim=0 0 0 0, clip,width=1\textwidth]{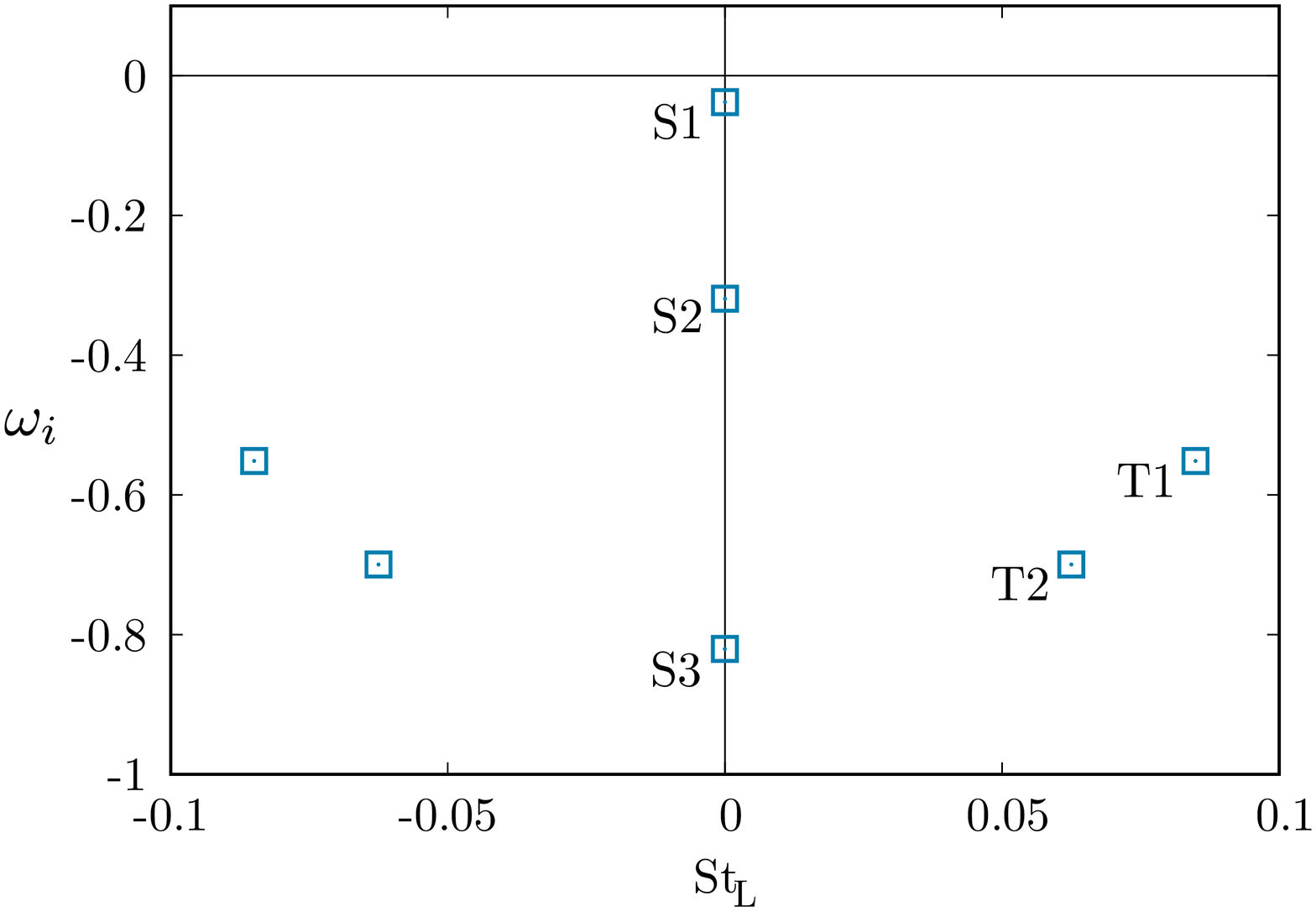}
        \end{subfigure}
         \begin{subfigure}[]{0.495\textwidth}  
          \includegraphics[angle=-0,trim=0 0 0 0, clip,width=1\textwidth]{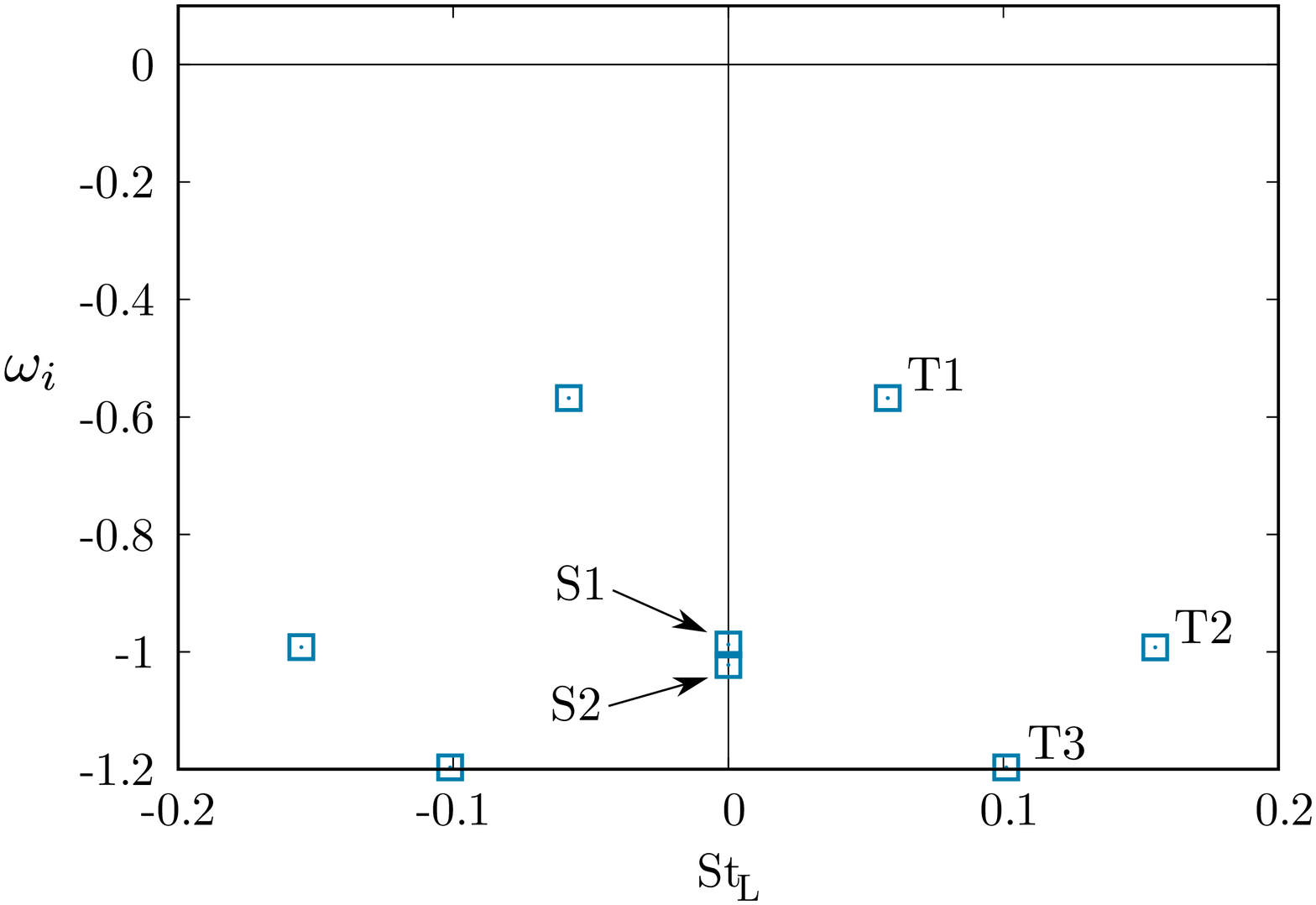}
        \end{subfigure}
        \caption{Eigenvalue spectrum of the global stability problem at $M=2.20$ and $\Reynd=1100$. Left: $\beta=0.25$. Right:  $\beta=2$.}
         \label{fig.spectrum3D}
      \end{figure}

\begin{figure}
        \centering
        \includegraphics[angle=0,trim=0 0 0 0, clip,width=0.51\textwidth]{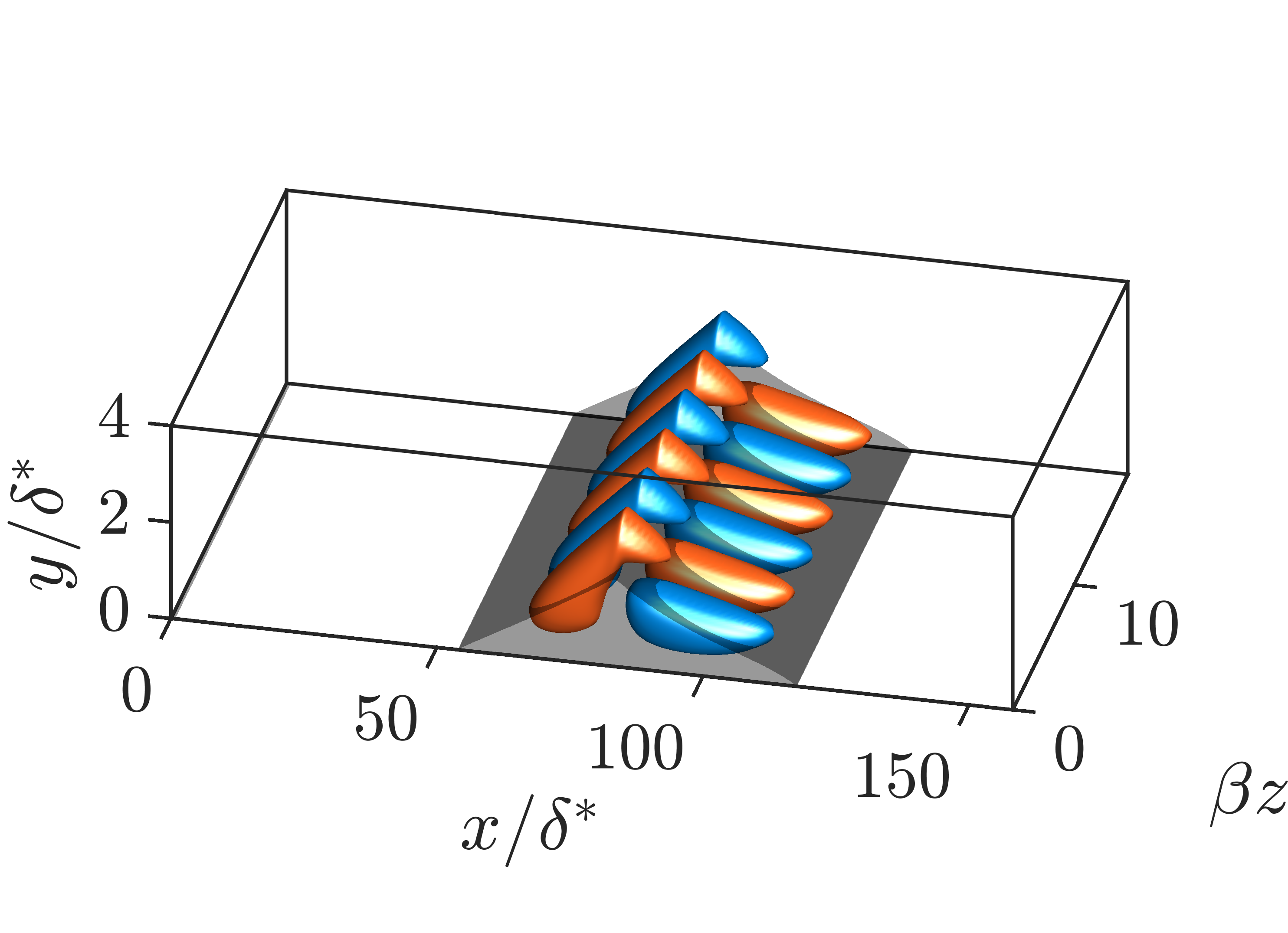}
        \caption{Global mode S1 at $\beta=0.25$, $M=2.20$ and $Re=1100$. Isosurfaces of $40 \%$ and $-40 \%$ of the maximum absolute value of $w$ are plotted in red and blue.}
        \label{fig.3Dfield_mode1_W}
    \end{figure}

\begin{figure}
        \centering
        \includegraphics[angle=-90,trim=0 0 0 0, clip,width=0.53\textwidth]{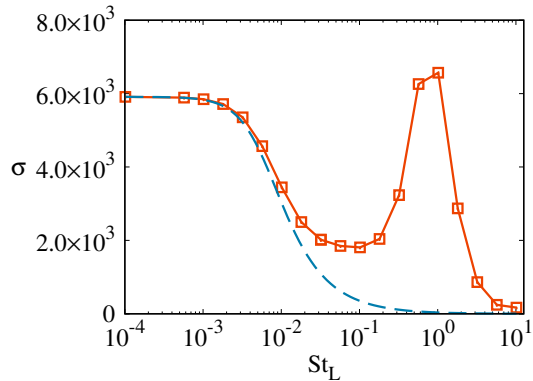}
        \caption{Optimal gain at $\beta=0.25$, $M=2.20$, $Re=1100$ (red line with squares). The low-pass filter model of equation \eqref{eq.lowpassfilter}, calculated using the damping rate of the 3D global S1 at $\beta=0.25$, is shown in dashed blue line.} 
        \label{fig.gain3D_beta025}
    \end{figure}

    At $\beta=0.25$, the least stable global mode is steady (mode S1) as shown in figure \ref{fig.spectrum3D}-left.
    Its structure is shown in figure \ref{fig.3Dfield_mode1_W}.
    It strongly resembles that of the optimal response previously observed.
    For 2D perturbations, these observations led us to formulate the hypothesis of a purely modal dynamics at low frequency, driven by a single mode.
    The first-order low-pass filter model (equation \eqref{eq.lowpassfilter}) that results from this assumption and that is based on the damping rate of the 3D global mode S1 is tested in figure \ref{fig.gain3D_beta025}.
    Once again, the model correctly predicts the cut-off frequency of the optimal gain.
    The slope of the gain drop is also reasonably captured until $\StL \simeq 10^{-2}$ where additional modes participate to the forced dynamics. 
    The low-frequency dynamics of 3D perturbations at low wave numbers thus behave the same way as that of 2D perturbations, which was expected as it is the limit when $\beta \rightarrow 0$.
    Finally, another peak of optimal gain can be observed at higher frequency in figure \ref{fig.gain3D_beta025}.
    This maximum  is associated with the convective instabilities that are known to exhibit larger growth rates for $\beta \ne 0$ in supersonic flows \citep{mack1984boundary}, resulting in larger values of the optimal gain \citep{bugeat20193d}.
    
    \subsubsection{Larger $\beta$: pseudo-resonance}

\begin{figure}
        \centering
        \begin{subfigure}[]{0.493\textwidth}  
        \includegraphics[angle=-0,trim=0 0 0 0, clip,width=0.99\textwidth]{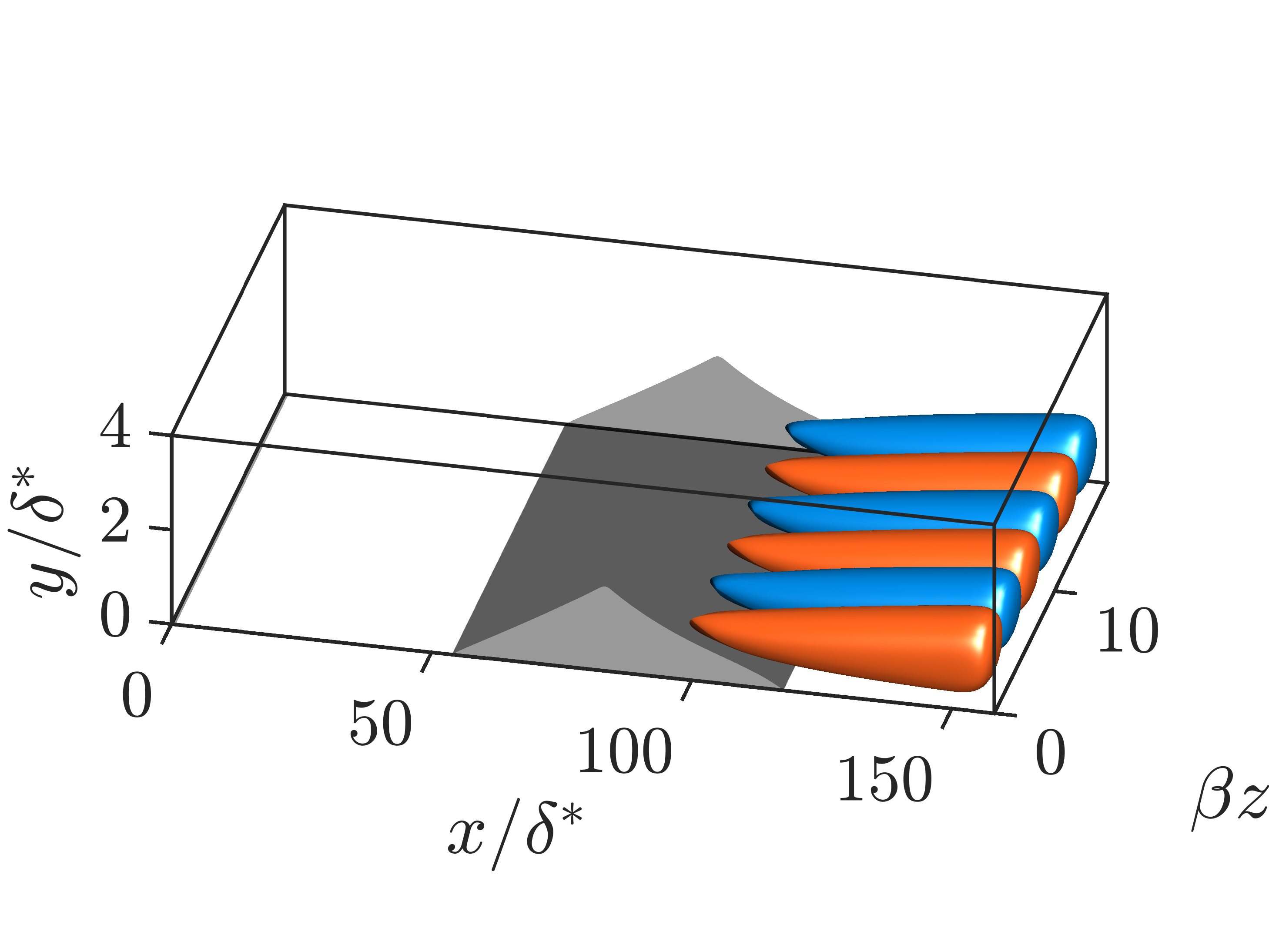}
        \end{subfigure}
        \begin{subfigure}[]{0.493\textwidth}  
        \includegraphics[angle=-0,trim=0 0 0 0, clip,width=0.99\textwidth]{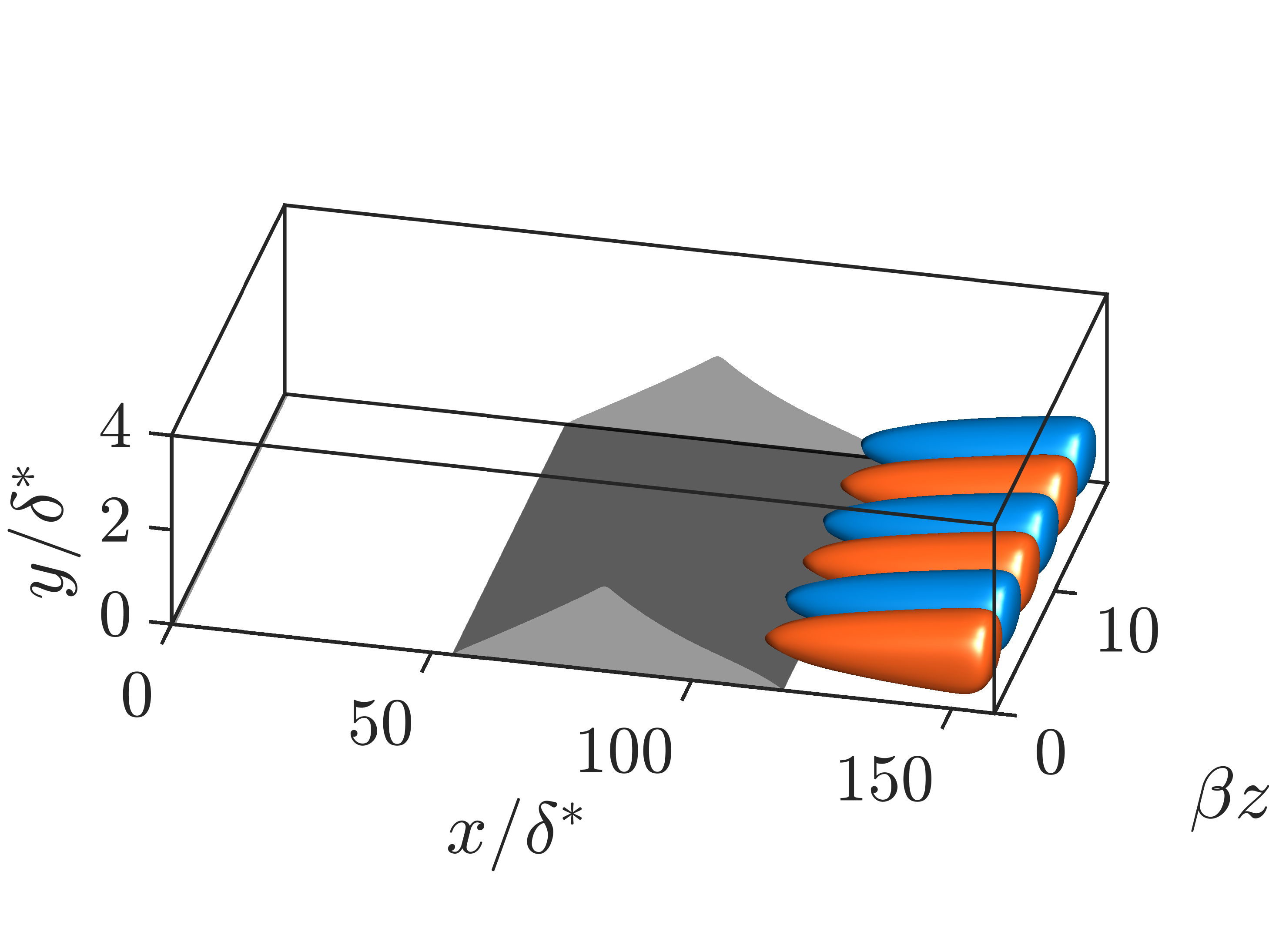}
        \end{subfigure}
        \caption{Global modes S1 (left) and T1 (right) at $\beta=2$ and $\StL=10^{-4}$, real part of $u$. Isosurfaces of $20 \%$ and $-20 \%$ of the maximum absolute value amplitude are plotted in red and blue.}
        \label{fig.modeBeta2}
    \end{figure}

\begin{figure}
        \centering
        \begin{subfigure}[]{0.495\textwidth}  
        \includegraphics[angle=-90,trim=0 0 0 0, clip,width=1\textwidth]{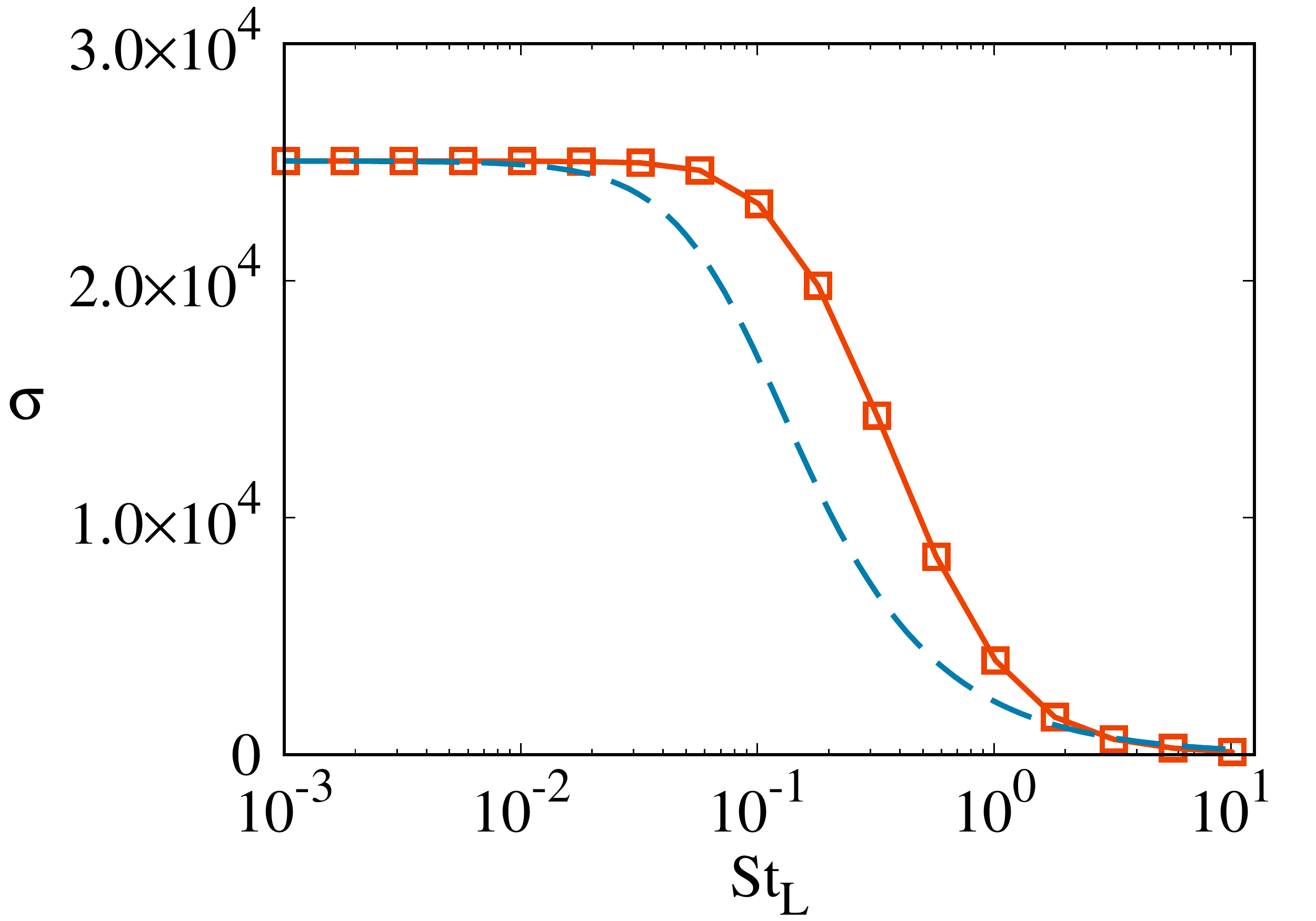}
        \end{subfigure}
        \begin{subfigure}[]{0.495\textwidth}  
        \includegraphics[angle=-90,trim=0 0 0 0, clip,width=1\textwidth]{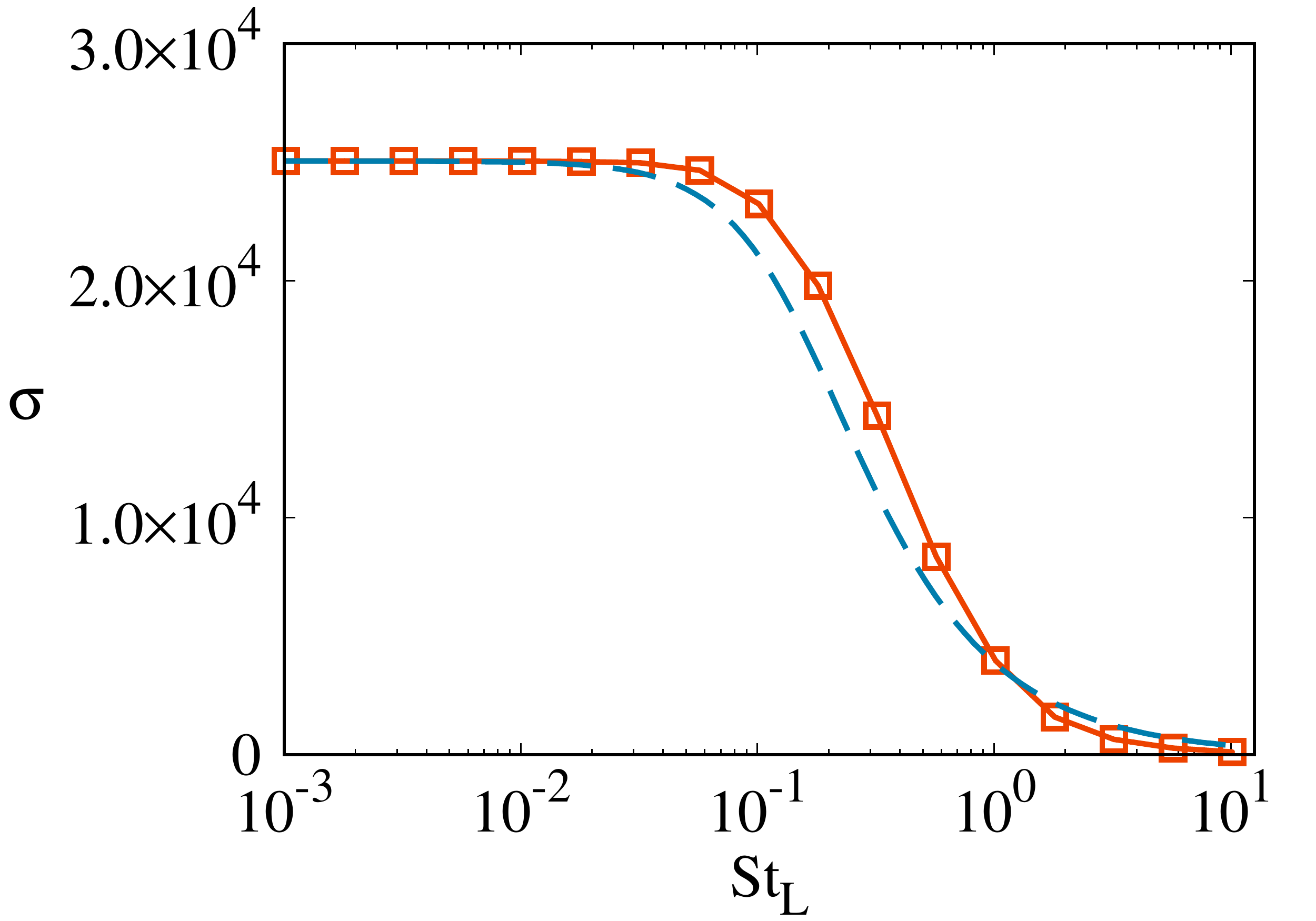}
        \end{subfigure}
        \caption{Optimal gain at $\beta=2$, $M=2.20$, $Re=1100$. The dashed blue line is the the  low-pass filter model based on global mode S1 (left) and T1 (right), which both fail due to the non-modal nature of the streaks.}
        \label{fig.gainBeta2}
    \end{figure}

    The lift-up mechanism leading to the formation of streaks is a linear algebraic instability, as theorised by \cite{ellingsen1975stability}.
    As such, linear stability theories based on a modal approach cannot capture it.
    Conversely, the resolvent analysis is particularly suited for their calculation as non-modal effects are taken into account \citep{schmid2007nonmodal}.
    We now suggest to show the \textit{failure} of the first-order low-pass filter model (based the excitation of a single mode) as a way to point out the non-modal character, or pseudo-resonance, of the optimal gain at wave numbers of the order of $\beta=2$.
    The least damped mode of the spectrum is the unsteady modes T1, followed by the steady mode S1 (figure \ref{fig.spectrum3D}-right).
    Their structure is shown in figure \ref{fig.modeBeta2}.
    They share similarities with the optimal response already presented in figure \ref{fig.resMode3D_beta2}-right as their growth takes place in the downstream region of the domain.
    Nevertheless, noticeable differences remain, which suggests that the optimal response does not proceed from the excitation of a unique global mode. 
    Two low-pass filters model based on the modes S1 and T1 are tested in figure \ref{fig.gainBeta2}.
    Both models underestimate the cut-off frequency and underpredict the optimal gain, confirming the non-modal nature of the optimal response at these wave numbers.
  
    \subsection{Discussion}

    Depending on their spanwise wave number, the 3D optimal response at low frequency can result from two different mechanisms, underpinned by a modal or non-modal nature.
    At low wave numbers, whose length scales are of the order of $\Lsep$, optimal perturbations interact with the recirculation bubble. 
    Their energy is localised in this region similarly to what was observed for 2D perturbations.
    Additionally, the low-pass filter model based on the excitation of a unique steady global mode reasonably describes the low-frequency behaviour of the optimal gain $\sigma$.
    Low wave numbers hence behave in continuity to the 2D results ($\beta=0$) presented in section \ref{sec.2Dperturbations}.
    Noticeably larger values of $\sigma$ are however reached for 3D perturbations.
    This raises the question of whether the dynamics of the system at low frequency and low wave number is dominated by 2D or 3D perturbations.
    On the one hand, larger values of the gain indeed indicates that more energetic 3D perturbations are expected in the flow, assuming the same forcing energy and the same projection of this forcing onto the resolvent modes for each wave number.
    These two assumptions are, at this point, impossible to verify.
    For example, if the forcing generated by non-linear interactions in a turbulent flow is preferentially 2D, then one may still expect the 2D response to be dominant even though it is associated with a lower gain value.
    Thus, no definitive conclusion can be drawn.
    On the other hand, the cut-off frequency based on the 3D global is $\StL = 6 \times 10^{-3}$. 
    This underestimates the usual value of $\StL = 3 \times 10^{-2}$ even more than the 2D perturbations, for which a value of $\StL = 1 \times 10^{-2}$ was found.
    This is in favour of the prevalence of 2D perturbations at low frequency even though a perfect quantitative agreement between our laminar analysis and the experimental data on turbulent flow is not expected.
    This is also backed up by previous work which restricted their analysis to 2D perturbations \citep{TS2011,SSH2014}

    The possibility that the low-frequency behaviour of the SWBLI may be a 3D phenomenon still remains a promising lead to explore further given some arguments found in our resolvent analysis.
    It could be interesting to investigate whether the optimal spanwise wave number would then scale with the recirculation bubble.
    A space-Fourier transform along the spanwise direction of the time-Fourier modes from experimental or numerical data would be required to explore this phenomenon.
    This, however, may not be manageable experimentally if the spanwise width of the facility is not wide enough compare to recirculation length scale.
    Recent studies have besides pointed out the influence of the side-walls in SWBLI experiments \citep{xiang2019corner,rabey2019two}.
    From a computational point of view, the same challenge of ensuring sufficiently wide domain is expected, additionally coupled with the costly computational time necessary to statistically converge the low-frequency time-Fourier transform.
    Such a study, if achievable, could help to characterise the 3D organisation of the forcing field and eventually understand its role in the low-frequency dynamics.

    While low wave numbers exhibit a \textit{modal} and \textit{localised} (around the bubble) dynamics, optimal responses at larger wave numbers have been shown to ignore the recirculation region as they are driven by a \textit{convective} instability growing downstream in the boundary layer, as a result of \textit{non-modal} effects.
    These findings are consistent with those of \cite{dwivedi2020transient} who recently carried out a transient growth analysis in a SWBLI at $M=5.92$.
    Such analysis also embeds non-modal effect and can in fact be seen as the temporal counterpart of the resolvent analysis \citep{sipp2010dynamics}.
    The authors detected streaky structures evolving in time and convected in space, similar to those obtained in the present study.
    % The optimal wave number is 
    Our resolvent analysis showed that streaks are associated with larger optimal gain values than the low-wave number perturbations.
    This is not surprising as it results from a convective instability which efficiently extract energy from the base flow.
    The role played by the streaks in the low-frequency dynamics ($\StL \simeq 3 \times 10^{-2}$) is not obvious.
    Contrary to low wave number perturbations, streaks are not directly involved in this dynamics since they do not interact with the recirculation region and are expected to scale with $\deltazero$ rather than $L$. 
    At first, it then seems that they could only be of interest regarding the transition to turbulence of the system.
    However, the large energy growth they trigger could quickly lead to non-linear interactions that may play a role in the forcing of the bubble through upstream feedback \citep{BBGBMSJ2019}.
    Studying the development of streaks in a turbulent SWBLI, seen as coherent structures, would then be of interest.
    Following the work of \cite{sartor2015unsteadiness}, this could be achieved by performing this 3D resolvent analysis on a turbulent mean flow.

%%%%%%%%%%%%%%%%%%%%%%%%%%%%%%%%%%%%%%%%%%%%%%%%%%%%%%%%%%%%%%%%%%%%%%%%%%%%%%%%%%%%%%%%%%%%%%%%%%%%%%%%%%%%%%%%%%%%%%%%%%%%%%%%
%%%%%%%%%%%%%%%%%%%%%%%%%%%%%%%%%%%%%%%%%%%%%%%%%%%%%%%%%%%%%%%%%%%%%%%%%%%%%%%%%%%%%%%%%%%%%%%%%%%%%%%%%%%%%%%%%%%%%%%%%%%%%%%%

\section{Conclusion} \label{sec.conclusion}

A resolvent analysis was carried out on a laminar SWBLI in order to study the low-frequency behaviour of the system for both 2D and 3D perturbations.
The optimal gain, that can be seen as a transfer function of the system, was computed over several decades of Strouhal numbers.
A model of optimal gain based on the excitation of a single global mode was developed, leading to the equation of a first-order low pass-filter in agreement with the work of \cite{TS2011}.
Our analysis was shown to be valid for 2D perturbations over different base flows calculated from different pairs of Reynolds and Mach numbers, but was not relevant at the largest Reynolds numbers tested. %($Re \simeq$ 1900).
The time scale of the model, i.e. the cut-off frequency of the filter, was found to be determined by the damping rate of the least damped global mode. 
This damping rate scales as $u_\infty/L$.
Thus, the low-frequency dynamics is associated with a constant Strouhal number based on the recirculation length, as observed experimentally and numerically.
The value obtained here is $\StL \simeq 1 \times 10^{-2}$, which slightly underestimates the classical values of the literature.
This mismatch could be attributed to the laminar nature of the flow studied in this paper.
Ultimately, this work carried out in the laminar regime suggests that the low-frequency dynamics of the SWBLI in the turbulent regime could be a forced dynamics, sustained by the background turbulent fluctuations.
The resulting bubble dynamics, which would be sustained by a background forcing produced by non-linear interactions in a turbulent flow, is then found to exhibit a breathing motion. 

Assuming the spanwise direction as homogeneous, a 3D resolvent analysis showed that two regimes of spanwise wave numbers $\beta$ could be identified.
At low $\beta$, the same modal mechanism as that of 2D perturbations, leading to a first-order low-pass filter, was observed.
Larger values of optimal gain were obtained for wave numbers of the order of the recirculation length.
The possibility of a prevailing 3D dynamics was then discussed.
At larger wave numbers, streaks generated by the non-modal lift-up effect were detected.
Their growth does not interact with the recirculation region, suggesting that streaks are not directly involved in the low-frequency dynamics of the SWBLI.
Finally, from the perspective of transition to turbulence, 3D resolvent analysis remains an efficient tool to characterise the energy growth of streaks and provides their optimal wave number and forcing location.

Future work investigating the forcing terms (both 2D and 3D) produced by non-linear interactions in a turbulent SWBLI, either through DNS or experimental observations, would be a decisive follow-up to this work.
While the low-frequency behaviour of the SWBLI was suggested to be understood as a forced dynamics, the physical origin of the forcing is missing. 
Ultimately, this could also provide some insights to assess the prevalence of 3D effects in the dynamics of the system.

%%%%%%%%%%%%%%%%%%%%%%%%%%%%%%%%%%%%%%%%%%%%%%%%%%%%%%%%%%%%%%%%%%%%%%%%%%%%%%%%%%%%%%
%%%%%%%%%%%%%%%%%%%%%%%%%%%%%%%%%%%%%%%%%%%%%%%%%%%%%%%%%%%%%%%%%%%%%%%%%%%%%%%%%%%%%%%%%%%%%%%%%%%%%%%%%%%%%%%%%%%%%%%%%%%%%%%%

\section*{Declaration of interests}
The authors report no conflict of Interest.

%%%%%%%%%%%%%%%%%%%%%%%%%%%%%%%%%%%%%%%%%%%%%%%%%%%%%%%%%%%%%%%%%%%%%%%%%%%%%%%%%%%%%%%%%%%%%%%%%%%%%%%%%%%%%%%%%%%%%%%%%%%%%%%%
%%%%%%%%%%%%%%%%%%%%%%%%%%%%%%%%%%%%%%%%%%%%%%%%%%%%%%%%%%%%%%%%%%%%%%%%%%%%%%%%%%%%%%%%%%%%%%%%%%%%%%%%%%%%%%%%%%%%%%%%%%%%%%%%

\appendix

\section{Mesh convergence}\label{sec.mesh}

The influence of the main numerical parameters on the eigenvalues computation of the global stability problem is tested.
Eigenvalues associated with the first steady (S1) and unsteady (T1) global modes, that appear in the low-pass filter models in section \ref{sec.model} and \ref{sec.modelsubopt}, are considered.
The reference set of parameters used throughout the paper is given on the first line of table \ref{tab.meshconv}. 
Independency with respect to the number of grid points $N_x$ and $N_y$ is verified in the cases A and B.
The influence of the height of the domain $L_y/\deltazero$ and the length of the domain \textit{downstream} from the impinging shock $L_x^{DS} / \deltazero$ are assessed in the cases C and D.
Overall, the unsteady mode is the most easily converged mode: relative differences of eigenvalues between the reference case and the other ones are equal to or below than $0.1 \%$ for the range of tested parameters.
As for the steady mode, relative difference of eigenvalues lower than $1 \%$ are found, which is deemed acceptable.

  \begin{table}
  \centering
  \begin{tabular}[t]{c|c|c|c|c|c|c|c|c}
      % \hline
      Name & $N_x$ & $N_y$ & $L_x^{DS} / \deltazero $ & $L_y / \deltazero$ & $\omega^{(S1)} \times 10^{4}$ & $\frac{| \Delta \omega_{\mathrm{ref}}^{(S1)}|}{| \omega_{\mathrm{ref}}^{(S1)}|}$ & $\omega^{(T1)} \times 10^{3}$ & $\frac{| \Delta \omega_{\mathrm{ref}}^{(T1)}|}{| \omega_{\mathrm{ref}}^{(T1)}|}$  \\
      % \hline
      
      \rule{0pt}{4ex}Ref & 600 & $270$ & $70$ & $50$ & $6.673 i$ & - & $3.272 + 2.589 i$ & -  \\     
      A  & 900  & $270$ & $70$ & $50$ & $6.731 i$ & $0.87 \%$ & $3.271 + 2.589 i$ & $0.02 \%$  \\     
      B  & 600  & $400$ & $70$ & $50$ & $6.651 i$ & $0.33 \%$ & $3.272 + 2.585 i$ & $0.10 \%$  \\   
      C  & 680  & $270$ & $90$ & $50$ & $6.737 i$ & $0.96 \%$ & $3.271 + 2.589 i$ & $0.02 \%$  \\  
      D  & 600  & $343$ & $70$ & $75$ & $6.631 i$ & $0.63\%$ & $3.272 + 2.589 i$ & $<0.01\%$   \\     

    \end{tabular}
%    \captionsetup{justification=centering}
      \caption{Mesh convergence of the most unstable mode at $M=2.30$, $\Reynd=1100$. The reference set of numerical parameters, used in all the calculation of the paper, is the first one of the table. $\Delta \omega_\mathrm{ref}$ corresponds to the difference between the eigenvalue of the reference case and that obtained in each other case.}
      \label{tab.meshconv}
  \end{table}

\section{Profiles of the 3D resolvent modes} \label{app.Profiles3D}

\begin{figure}
        \centering
        \begin{subfigure}[]{0.49\textwidth}  
          \includegraphics[angle=-90,trim=0 0 0 0, clip,width=0.99\textwidth]{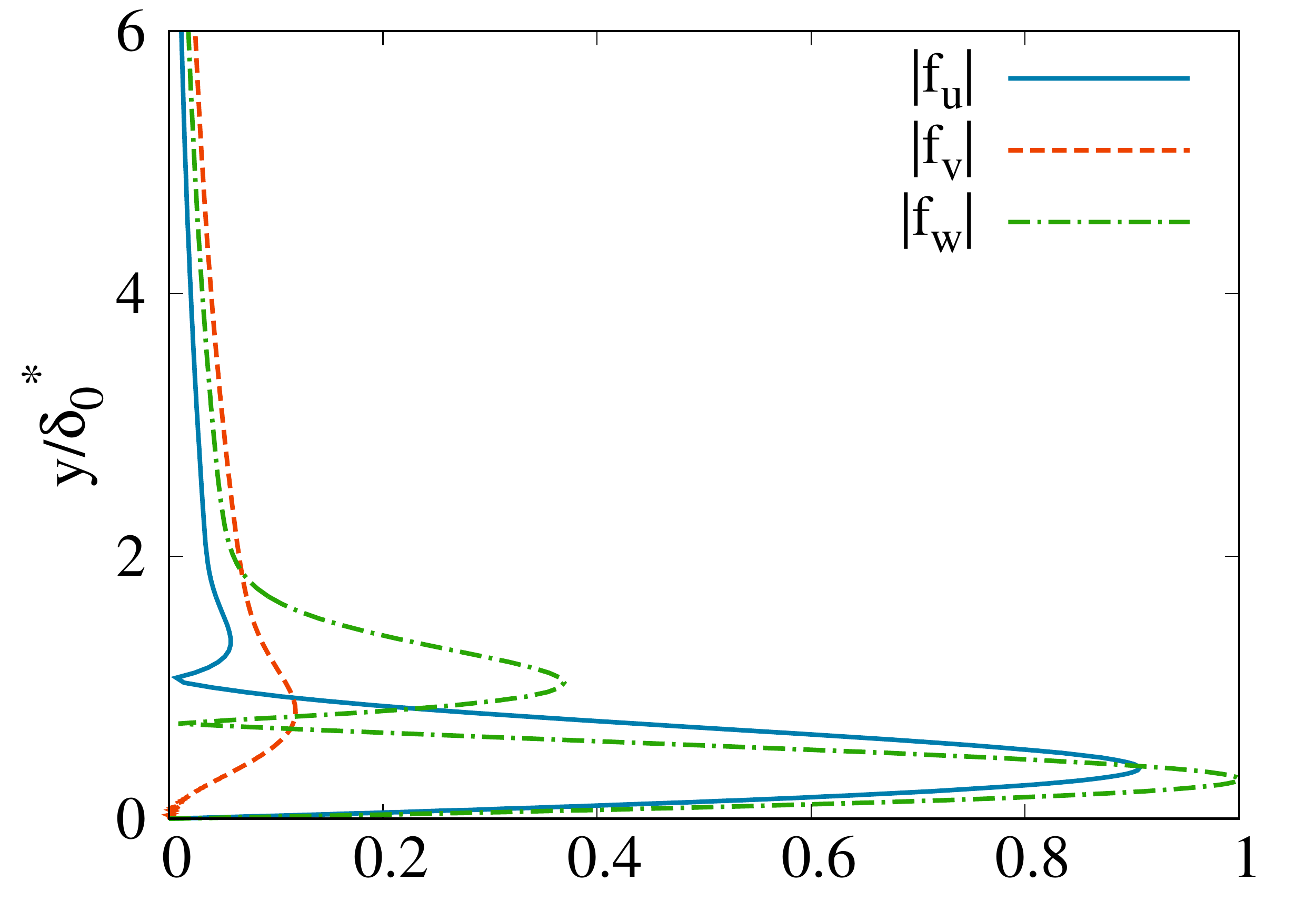}      
        \end{subfigure}
         \begin{subfigure}[]{0.49\textwidth}  
          \includegraphics[angle=-90,trim=0 0 0 0, clip,width=0.99\textwidth]{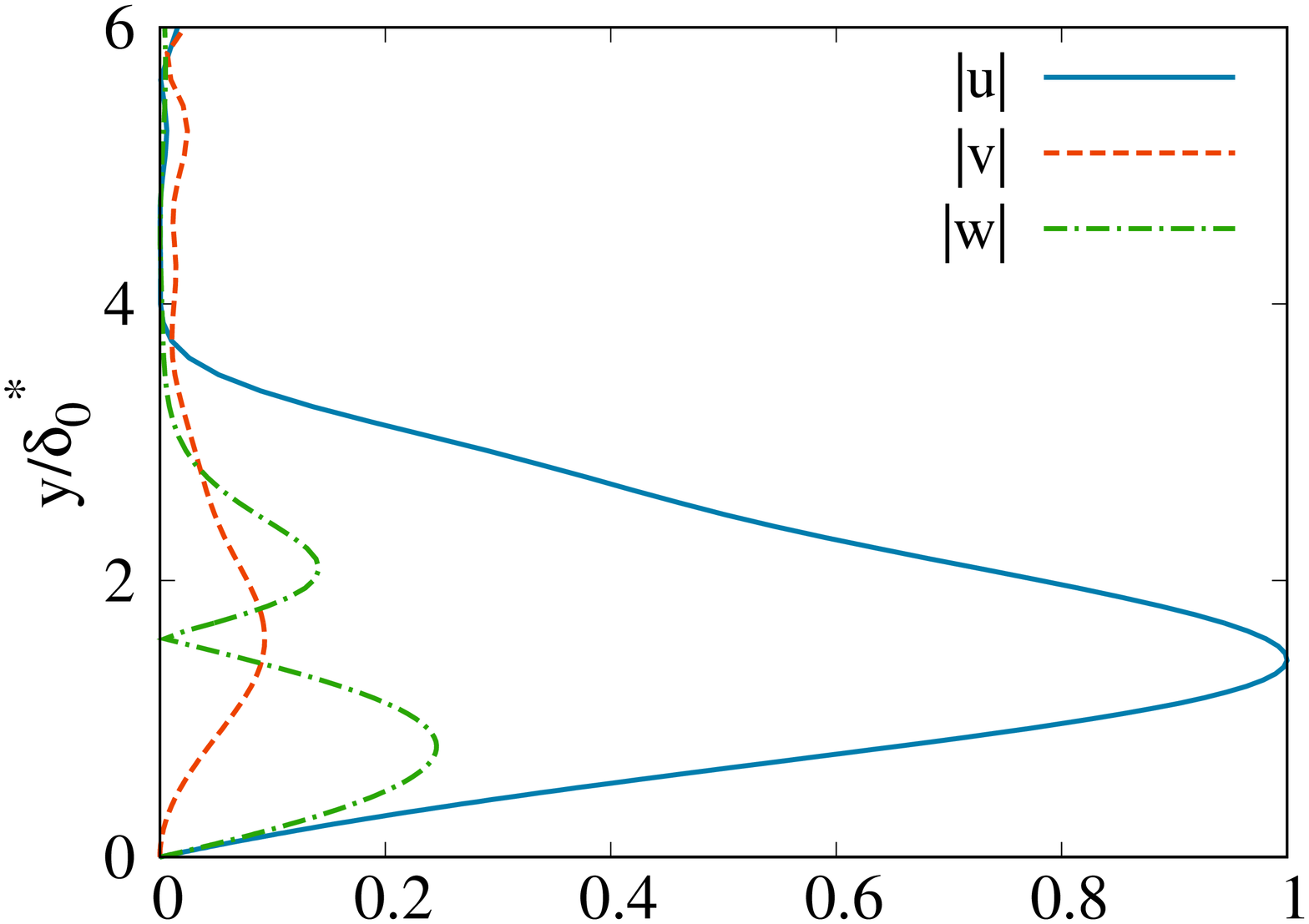}      
        \end{subfigure}
        \caption{Profiles of the optimal resolvent mode at $\beta=0.25$, $M=2.20$, $Re=1100$, $\StL=10^{-4}$.}
         \label{fig.resMode3D_profileBeta025}
    \end{figure}

\begin{figure}
        \centering
        \begin{subfigure}[]{0.49\textwidth}  
          \includegraphics[angle=-90,trim=0 0 0 0, clip,width=0.99\textwidth]{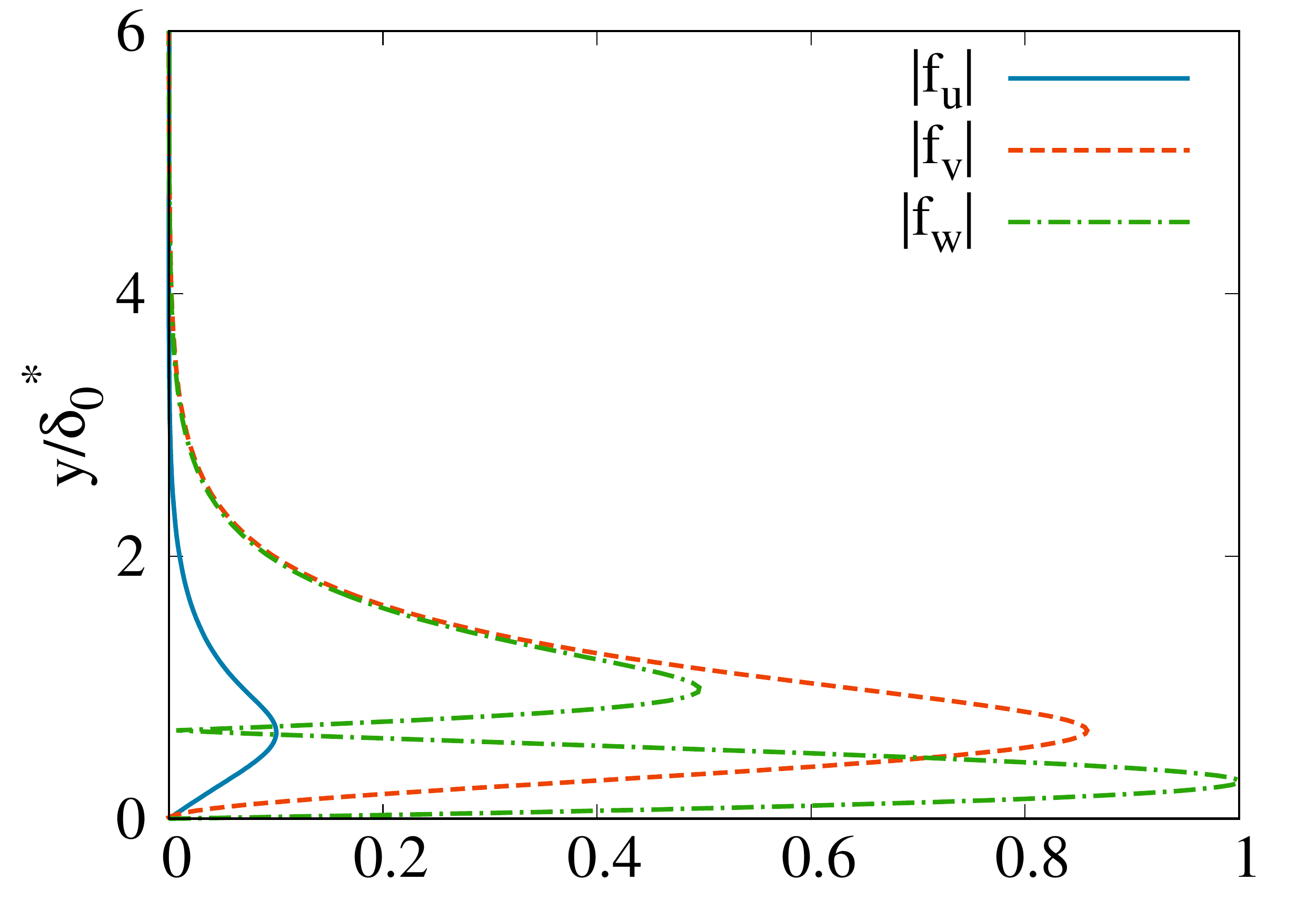}      
        \end{subfigure}
         \begin{subfigure}[]{0.49\textwidth}  
          \includegraphics[angle=-90,trim=0 0 0 0, clip,width=0.99\textwidth]{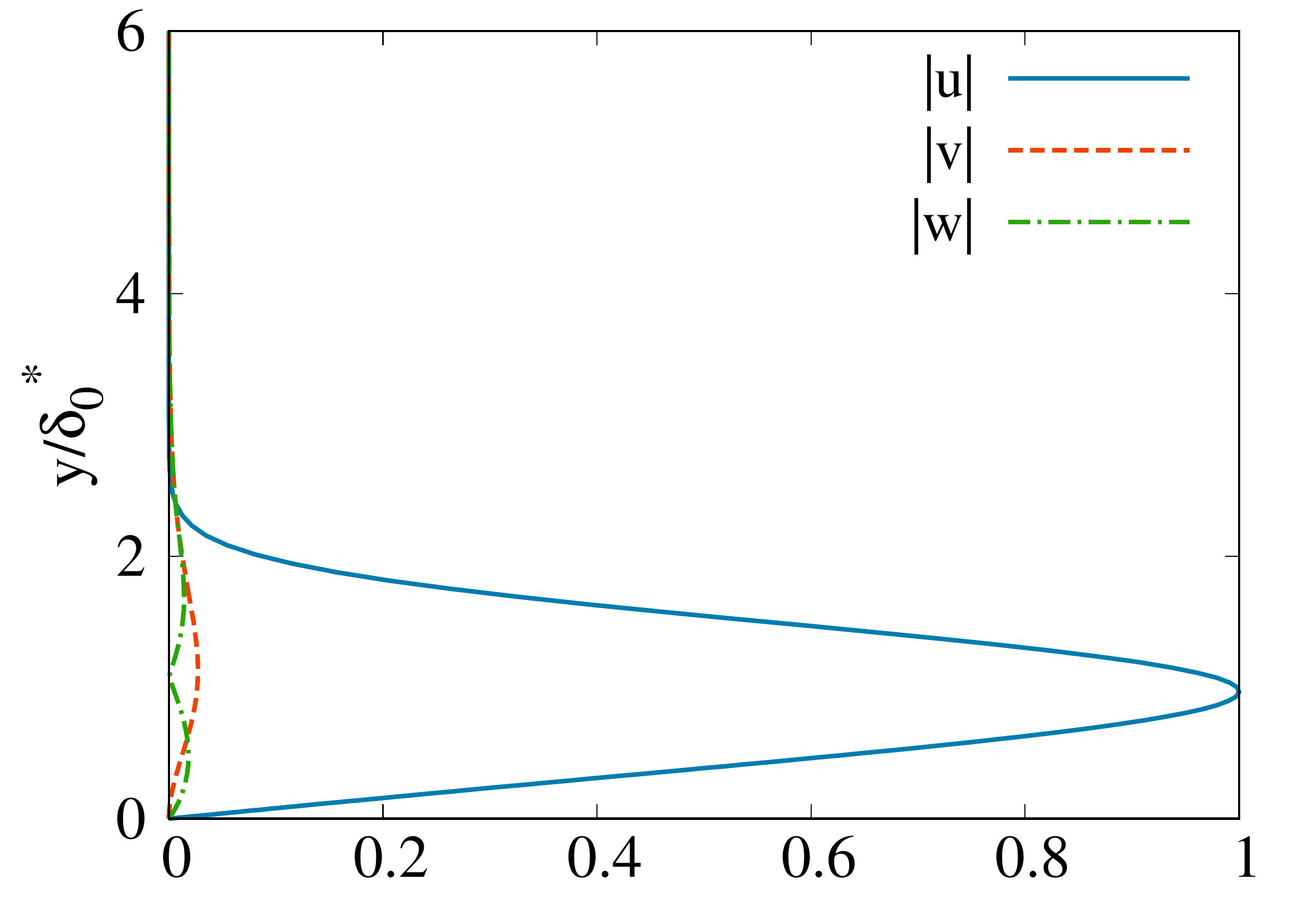}      
        \end{subfigure}
        \caption{Profiles of the optimal resolvent mode at $\beta=2$, $M=2.20$, $Re=1100$,\textbf{} $\StL=10^{-4}$.}
         \label{fig.resMode3D_profileBeta2}
    \end{figure}

Profiles of the forcing and response of the 3D low-frequency resolvent modes at $\beta=0.25$ and $\beta=2$ are shown in figure \ref{fig.resMode3D_profileBeta025} and \ref{fig.resMode3D_profileBeta2} respectively.
The $x$-location of each profile corresponds to that of the maximum of energy density as defined in section $\ref{sec.resolvent3D}$.
Note that this location is different for the forcing and the response (see figure \ref{fig.resMode3D_energyProfiles}).
As a result, the boundary layer thickness is different on each plot.

\bibliographystyle{jfm}
\bibliography{jfm-Bugeat.bib}

\end{document}